\documentclass[fleqn,usenatbib]{mnras}

\usepackage{times} %Use times new roman
\usepackage{newtxtext,newtxmath}
\usepackage{lscape}

\usepackage[T1]{fontenc}

\DeclareRobustCommand{\VAN}[3]{#2}
\let\VANthebibliography\thebibliography
\def\thebibliography{\DeclareRobustCommand{\VAN}[3]{##3}\VANthebibliography}

\usepackage{graphicx}	% Including figure files
\usepackage{amsmath}	% Advanced maths commands
\def\cm2{cm$^2$ }
\def\se1{s$^{-1}$ }

\newcommand{\ha}{H$\alpha$ }
\newcommand{\hb}{H$\beta$ }

\usepackage{xfrac}
\newcommand{\altfrac}[2]{\ifmmode\def\tmp{$}\else\def\tmp{}\fi\mbox{%
    {\raisebox{.34\ht\strutbox}{\tmp#1\tmp}}%
    \kern-2.2pt\scalebox{1.5}[1.5]{/}\kern-0.0pt%
    {\tmp#2\tmp}%
    }}

\usepackage{threeparttable}
\usepackage{tablefootnote}
\usepackage{multirow}
\usepackage{booktabs}
\usepackage{wrapfig}
\usepackage{float}
\newcommand{\ii}{~\textsc{ii}}
\newcommand{\iii}{~\textsc{iii}}

\usepackage{float}
\usepackage[font=footnotesize]{subfig}

\newcommand{\kmsmpc} {$\rm {km~s^{-1}}~Mpc^{-1}$}

\newcommand{\kms} {$\rm {km~s^{-1}}$}

\newcommand{\oiii}{[O\,{\sc iii}]}
\newcommand{\oii}{[O\,{\sc ii}]}
\newcommand{\oi}{[O\,{\sc i}]}
\newcommand{\nii}{[N\,{\sc ii}]}
\newcommand{\sii}{[S\,{\sc ii}]}
\newcommand{\siii}{[S\,{\sc iii}]}

\def\hbeta {H$\beta$}
\def\halpha {H$\alpha$}

\def\oh{12 + $\log$(O/H)}

\title[Spatially--resolved metallicity in Seyferts]{Spatially--resolved gas-phase metallicity in Seyfert galaxies}

\author[M. Armah et al.]{Mark Armah,$^1$\thanks{E-mail: armah@ufrgs.br} Rog\'erio Riffel,$^1$
L. G. Dahmer-Hahn,$^2$
R. I. Davies,$^3$
 O.L. Dors,$^4$
Darshan Kakkad,$^5$
\newauthor
Rogemar A. Riffel,$^6$
A. Rodr\'iguez-Ardila,$^{7,8,9}$
D. Ruschel-Dutra$^{10}$ and
T. Storchi-Bergmann$^1$
\\
$^1$Departamento de Astronomia, Instituto de Física, Universidade Federal do Rio Grande do Sul, CP 15051, 91501-970, Porto Alegre, RS, Brazil \\
$^{2}$Shanghai Astronomical Observatory, Chinese Academy of Sciences, 80 Nandan Road, Shanghai 200030, China 
\\
$^{3}$Max Planck Institute for Extraterrestrial Physics, Giessenbachstrasse 1, Garching, 85748, Germany
\\
$^4$Universidade do Vale do Para\'iba. Av. Shishima Hifumi, 2911, CEP: 12244-000, São José dos Campos, SP, Brazil\\
$^{5}$Space Telescope Science Institute, 3700 San Martin Drive, Baltimore, 21210 MD, USA.
\\
$^6$Departamento de F\'isica, CCNE, Universidade Federal de Santa Maria, Av. Roraima 1000, 97105-900,  Santa Maria, RS, Brazil
\\
$^7$Laborat\'orio Nacional de Astrof\'isica/MCTIC, Rua dos Estados Unidos, 154, Bairro das Nac¸oes, Itajub\'a, MG 37504-364, Brazil
\\
$^8$Instituto Nacional de Pesquisas Espaciais. Divis\~ao de Astrof\'isica. Avenida dos Astronautas 1758. S\~ao Jos\'e dos Campos, 12227-010, SP, Brazil
\\
$^9$Observatório Nacional, Rua General José Cristino 77, CEP 20921-400, São Cristóvão, Rio de Janeiro, RJ, Brazil\\
$^{10}$Departamento de F\'isica, Universidade Federal de Santa Catarina, P.O. Box 5064, 88035-972, Florianop\'olis, SC, Brazil
}
\date{Accepted XXX. Received YYY; in original form ZZZ}

\pubyear{2024}

\begin{document}

\label{firstpage}
\pagerange{\pageref{firstpage}--\pageref{lastpage}}
\maketitle

\begin{abstract}
We explore the relations between the gas-phase metallicity radial profiles (few hundred inner parsec) and multiple galaxy properties for 15 Seyfert galaxies from the AGNIFS (Active Galactic Nuclei Integral Field Spectroscopy) sample using optical Integral Field Unit (IFU) observations from Gemini Multi-Object Spectrographs (GMOS) and Multi Unit Spectroscopic Explorer (MUSE) processed archival data. The data were selected at $z \lesssim 0.013$ within black hole mass range $\left[6<\log \left(M_{\rm BH}/{\rm M_\odot} \right)<9\right]$ with moderate 14--150\,keV  X-ray luminosities $\left[42\,\lesssim\,\log L_X (\rm erg\,s^{-1})\,\lesssim\,44\right]$. 
We estimated the gas-phase metallicity using the strong-line methods and found mean values for the oxygen dependent ($Z \sim 0.75Z_\odot$) and  nitrogen dependent ($Z \sim 1.14Z_\odot$) calibrations. These estimates show excellent agreement  with $\Delta Z \approx 0.19$ dex and $\Delta Z \approx 0.18$ dex between the mean values from the two strong-line calibrations for GMOS and MUSE respectively, consistent with  the order of metallicity uncertainty via the strong-line methods.  
We contend that our findings align with a scenario wherein local Seyferts have undergone seamless gas accretion histories, resulting in positive metallicity profile over an extended period of time, thereby providing insights into galaxy evolution and the chemical enrichment or depletion of the universe. Additionally, we argue that metal-poor gas inflow from the local interstellar medium (ISM) and accreted through the circumgalactic medium (CGM) onto the galaxy systems regulates the star formation processes by diluting their central metallicity and inverting their metallicity gradients, producing a more prominent anti-correlation between gas-phase metallicity and Eddington ratio.
\end{abstract}


%%%%%%%%%%%%%%%%%%%%%%%%%%%%%%%%%%%%%%%%%%%%%%%%%%

%%%%%%%%%%%%%%%%% BODY OF PAPER %%%%%%%%%%%%%%%%%%

\section{Introduction}
Active galactic nuclei (AGNs) present a unique challenge in terms of gas-phase metallicity estimation, primarily because their powerful central engines—supermassive black holes (SMBHs) accreting matter at relatively $\sim$10$\rm ^{-4}-10\:M_{\odot}yr^{-1}$ 
 \citep[e.g.][]{Diniz+19,Riffel-Rogemar+23}—create environments distinct from typical H{\ii} regions in star-forming galaxies. 
 Therefore, AGNs are indicative of growth periods for massive compact objects, believed to be ubiquitously located at the centres of massive galaxies \citep[e.g.][]{Rees+84,Magorrian+98}. The concept of co-evolution between AGNs and their host-galaxies has been invoked through a range of observational studies that reveal correlations between their respective physical properties \citep[e.g. a tight correlation between the black hole mass and velocity dispersion of the stellar bulge, i.e. $M_{\rm BH}-\sigma_\star$ relation; ][]{Gebhardt+2000, Ferrarese+2000, Gultekin+09, Kormendy+13, Bosch+16, Caglar+20, Caglar+23}.  AGN feedback processes \citep[e.g.][]{Fabian+12} are crucial mechanisms in our current understanding of galaxy evolution and are deeply integrated into cosmological models and simulations to reproduce observed galaxy properties \citep[e.g.][]{Marulli+08, Ward+22}.  Negative feedback is the primary mechanism for suppressing star formation and shaping the overall morphology in massive galaxies  \citep[e.g.][]{Zubovas+17, Man+18, Piotrowska+22}, while positive feedback enhances star formation processes \citep[e.g.][]{Bieri+16,Zhuang+20}. Moreover, observational studies have occasionally presented instances where AGN-driven outflows or jets trigger enhanced star formation, and in some remarkable cases, both suppression and enhancement of star formation occur concurrently within the same galaxy \citep[e.g.][]{Elbaz+09,Cresci+15}. However, some studies, particularly for moderate-luminosity AGNs, posit that star formation is independent of AGN activity \citep[e.g.][]{Azadi+15,Stanley+15,Suh+17,Forster+20}. Additionally, direct observational evidence that AGN feedback causes quenching on a large scale across galaxy populations remains elusive.  

AGN feedback plays a key role in the distribution of metals to the outskirts of galaxies, as evidenced by the metallicity content out to circumgalactic medium (CGM) scales \citep[e.g.][]{Choi+20,Jecmen+23,Yang+24}. 
The first two moments of the abundance distribution—the mean metal abundances of discs and their radial gradients—have so far been observed in several studies on the chemical abundance of extragalactic nebulae
(e.g.  \citealt{Belfiore+17, Sanchez+19,  Berg+20, Nascimento+22}). 
Due to the need for high spatial resolution (typically below 2~arcsec), the spatially resolved physical properties
(e.g. electron temperature,  \citealt{Revalski+18a, Revalski+18b,Revalski+21,Revalski+22}) and metal distribution
(e.g. \citealt{Guo+20})
have only been studied in a limited number of AGNs.

Several studies have demonstrated that AGN hosts generally exhibit young stellar populations \citep[][and references therein]{CidFernandes+04,Riffel+09,Riffel+07,FalconBarroso+14, Ruschel-Dutra+17, Burtscher+21,DahmerHahn+22,Riffel+21,Riffel+22,RiffelRA+23,Ni+23}. Moreover, the age of the stellar population has been found to correlate with AGN luminosity \citep{Mallmann+18,Riffel+22,RiffelRA+23}. Furthermore, the stellar metallicity in AGNs is observed to be typically lower compared to non-active galaxies with similar mass \citep{CampsFarina+21,RiffelRA+23}. Additionally, investigations have highlighted the significant role that recycled gas may play in AGN feeding \citep{Choi+23, Riffel+24}.

Some aspects of the intricate interactions between the chemical abundances of galaxies and their physical properties have been revealed by previous spectroscopic studies \citep[e.g.][]{Maiolino+19,Sanchez+22, Mainieri+24}.  Over  decades, nebular emission lines from bright galaxies have been our primary sources for measuring the gas-phase metallicity at specific spatial positions in galaxies. To constrain galactic and extra-galactic chemical evolution, a good observational depiction of the distribution of elemental abundances across the surfaces of adjacent galaxies is required. Similarly, precise determination of star formation histories (SFHs) improves our understanding of stellar nucleosynthesis in typical spiral galaxies, which are inseparable important pieces of information about galaxy evolution processes.

The primordial composition, the amount and distribution of molecular and neutral gas, the AGN and stellar feedback effects (on the SFH), outflows and inflows of gas, the transport and mixing of gas, the initial mass function (IMF), and other factors all affect the chemical evolution of galaxies \citep[][]{Pagel+09,Matteucci+12}. 
Over the years, several relations between metallicity and galaxy features have been proposed, for instance, luminosity-metallicity ($L$-$Z$), mass-metallicity (MZR), fundamental-metallicity (FMR) and surface brightness versus metallicity relations, effective yield versus luminosity and circular velocity relations; abundance gradients and the effective radius of disks; systematic differences in the gas-phase abundance gradients between the inner and outer regions of the galaxy \citep[see][for a review]{Maiolino+19,Sanchez+22}.

The gas-phase  metallicity\footnote{The gas-phase  metallicity ($Z)$ and oxygen abundance [\oh]  are used interchangeably in this work.} can be 
derived in both star-forming regions (SFs, i.e. H{\ii} regions and star-forming galaxies) and AGNs through direct estimation of the electron temperature of the ionized gas using the intensity of one or more temperature-sensitive auroral lines (e.g. \oiii$\lambda$4363, \nii$\lambda$5755, \siii$\lambda$6312 and \oii$\lambda$7325), usually known as $T_{\rm e}$-method.\footnote{For a review of the $T_{\rm e}$-method for SFs see \citet{Peimbert+17}, \citet{PerezMonteiro+17} and for AGNs  see \citet{Dors+20b}.} However, these auroral lines are intrinsically faint ($\sim$100 times weaker than H$\beta$), especially in low ionization and/or metal-rich SFs (e.g. \citealt{vanZee+98,Diaz+07,Dors+08}) and  AGNs (e.g. \citealt{Dors+20a}).
Therefore, when these lines are undetectable or not measured, calibrations between strong emission line intensity ratios and metallicity, known as the strong-line method \citep{Pagel+79}, are used to estimate the metallicity of the ionized gas.\footnote{For details of strong-line methods for SFs  \citep[see][for a review]{LopezSanchez+10}
and for AGNs \citep[see][for a review]{Dors+20a}.} 

Employing the strong line method, recently, \citet{Armah+23} studied a large sample (i.e. 561 objects) of Seyfert galaxies and found an anti-correlation between X-ray emission and oxygen abundance. This is interpreted as evidence that the brightest sources are experiencing a stronger inflow of pristine gas from the outskirts of the host galaxy towards its center, feeding the AGN. Such inflow would dilute the metal-rich gas content in the central region of the galaxy. Additionally,
\citet{Nascimento+22}  used observational data from the SDSS-IV MaNGA survey together with the strong-line calibrations by \citet[][]{sb98}   
and \citet[][]{Carvalho+20}, and found that the nuclear region (AGN-dominated) has lower metallicity than the outer parts, indicating the possible effect of the inflow of pristine metal-poor gas towards the centre of the AGN (see also \citealt{Dors+15}). 
 
Traditionally, the physical conditions (e.g. metallicity, electron density and temperature) of AGNs have been derived from long-slit spectroscopy centred on the nuclear regions of galaxies. Presently, integral field spectroscopy (IFS)  or Integral Field Unit (IFU) spectrograph is a powerful technique to map the physical properties of central parts
of galaxies. In fact, \citet{RiffelRA+21b}, by using
Gemini GMOS-IFU observations of three luminous nearby Seyfert galaxies (Mrk\,79, Mrk\,348, and Mrk\,607) and through direct electron temperature measurements ($T_{\rm e}$-method), for the first time, found the presence of temperature fluctuations in
AGNs. Similar studies, also based on IFU spectroscopy, produce available information on electron density structures in this object class (e.g. \citealt{Freitas+18, Kakkad+18, RuschelDutra+21}). Furthermore, deep observations of 80 quasars 
at $z\sim3$ based on MUSE-VLT data by \citet{Guo+20}, found a radial decline of ultraviolet (UV) emission lines in about 20\,\% of the sample, indicating that central parts of AGNs can present a higher metallicity than gas regions located in outerstick parts, i.e. the existence of radial gradients in quasars. Therefore, the presence of temperature fluctuations  \citep{RiffelRA+21b}  and radial metallicity gradients  \citep{Guo+20,Amiri+24} observed in some AGNs suggest that a more comprehensive understanding of the spatial distribution of metals within these objects is crucial.
In this context, we focused on the inner $5^{\prime\prime}\times5^{\prime \prime}$ of nearby galaxies with a significantly greater spatial resolution ($\sim$1$^{\prime \prime}\approx$ 78 to 259\,pc), in contrast to prior Sloan Digital Sky Survey (SDSS)  program via Mapping Nearby Galaxies at Apache Point Observatory (MaNGA) and CALIFA Survey (Calar Alto Large
Integral Field Area Survey) results, which are concentrated on a larger field of view (FoV) but with lower spatial resolution.  We combine GMOS and MUSE IFU data to derive the resolved gas-phase metallicity maps and radial profiles using multiple dust-corrected diagnostic emission line-ratios at physical scales of only a few hundred parsecs.

The paper is organized as follows: In \S~\ref{chapt2}, we present a complete description of the observational data. In \S~\ref{chapt3}, we present the adopted metallicity calibrations together with flux measurements and optical emission line intensity maps, as well as the maps of the properties of the ionized gas for the individual targets.  In \S~\ref{chapt4}, we present the results and discuss the gas-phase metallicity distribution and its implications from individual sources and general standpoints.  Finally, we present a summary of this work and our main findings in \S~\ref{concl}. 
Throughout this paper, we adopt a spatially flat standard $\Lambda$CDM cosmology with the following parameters: $\rm \Omega_M$ = 0.315, $\Omega_\Lambda$ = 0.685 and $\rm H_0$ = 67.4 \kmsmpc \citep{Planck+20}.

\section{Sample selection}
\label{chapt2}
The data used here were obtained through cross-correlations between the Gemini Near-Infrared Integral-Field Spectrograph (NIFS) sample \citep[AGNIFS sample][]{Riffel+18,Riffel+22,Riffel-Rogemar+23} and archived data from the Gemini Multi-Object Spectrographs (GMOS) and Multi Unit Spectroscopic Explorer (MUSE). Our final sample (see Table~\ref{tab1}) is made up of 9 GMOS and 6 MUSE data cubes after duplicates and galaxies with bad data quality have been eliminated. These optical data-cubes have been presented in \citet{DahmerHahn+22} where more details can be found. The AGN nuclear activity type is based on the classification by \citet{Osterbrock+81} where we define Seyfert~1 galaxies (Sy~1) as the sources with Sy~1-1.5 classifications and Seyfert~2 galaxies (Sy~2) as those with Sy~1.8-2 classifications, since Sy~1.8 shows very weak broad H$\beta$ and H$\alpha$ in its optical spectrum while in Sy~1.9 only the weak broad component is detected at H$\alpha$, thus the overall spectra of Sy~1.8 and Sy~1.9 are closer to a type~2 source than to a true type~1.

The GMOS data cubes were obtained using the B600 grating, resulting in a spectral resolution $R = \lambda/\Delta\lambda$ of 1688, and with a 0\farcs1$\times$0\farcs1  spaxel sampling. The MUSE data were obtained with the wide field mode, at an average spectral resolution of $R$=1770 at 4800 {\AA} and $R$=3590 at 9300  {\AA}, and with a 0\farcs2$\times$0\farcs2  spaxel sampling.  Since the FoV of MUSE (60\farcs0$\times$60\farcs0) is considerably larger than that of GMOS (5\farcs0$\times$7\farcs5), we used the inner $5^{\prime \prime}\times5^{\prime \prime}$ of MUSE data cube extraction, centred on the location of the continuum peak, in order to properly compare their data for the whole sample. 

Both sets of data-cubes observations (GMOS and MUSE) were seeing limited, with angular resolution varying between 0.6 and 1.1 arcsec (spatial point spread function), estimated as the full width at half maximum (FWHM) of the flux distributions of point sources in the acquisition images. The archival MUSE datacubes consist of already reduced data, whereas we reduced the GMOS data with the standard {\sc iraf} pipeline. Both reduction processes consisted of bias subtraction, cosmic ray detection, flat-field correction, wavelength calibration, sky subtraction, and finally flux calibration through a standard star \citep[for details, see][]{DahmerHahn+22}.

\subsection{Literature data}
We estimate the Eddington ratio ($\lambda_{\rm Edd}$) from the  Eddington luminosity: $L_{\rm Edd}/{\rm (erg/s)} = {4\pi G M_{\rm BH} m_{\rm p}c}/{\sigma_{\rm T}}$, where $G$ is the gravitational constant, $M_{\rm BH}$ is the black hole mass, $m_{\rm p}$ is the mass of the proton, $c$ is the speed of light, and $\sigma_{\rm T}$ is the Thomson cross-section. The bolometric luminosity ($L_\mathrm{Bol}$) was calculated from the intrinsic
luminosity in the 14--150\,keV range translating hard X-ray luminosity to bolometric luminosity  by  adopting the 14--150\,keV bolometric correction factor of $\kappa_{14-150\,{\rm keV}}=8.0$, i.e., $L_{\rm Bol}=\kappa_{14-150\,{\rm keV}}\times L_{14-150\,{\rm keV}}$. This is equivalent to a 2--10\,keV bolometric correction considering $\kappa_{2-10\,{\rm keV}}= 20$  \citep{Vasudevan+09}, following $\kappa_{2-10\,{\rm keV}}\propto \lambda_{\rm Edd}$ over the range $\lambda_{\rm Edd}\leq 0.04$ for an X-ray photon index of $\Gamma=1.8$ \citep[e.g.][]{Tozzi+06}, consistent with the typical value of {\it Swift}/BAT AGN \citep{Ricci+17}. The Eddington ratio estimates were derived from the Eddington and bolometric luminosities presented by \citet{Koss+22a}, defined as 
$\lambda_{\rm Edd} = \frac{L_\mathrm{Bol}}{L_\mathrm{Edd}} = \frac{\kappa_{14-150\,{\rm keV}}\times L_{14-150\,{\rm keV}}}{1.5 \times 10^{38} \, M_{\rm BH}\,\rm erg\,s^{-1}}$.

\begin{table*}
 \centering
\addtolength{\tabcolsep}{-0.8pt}
\caption{The sample and main properties: (1) galaxy name; (2) redshift taken from NED (NASA/IPAC Extragalactic Database: \href{https://ned.ipac.caltech.edu/}{https://ned.ipac.caltech.edu/}); (3)  AGN classification type;  (4) galaxy morphological classification; (5) telescope used; (6) Black hole mass derived from  velocity dispersions \citep{Koss+22b}  and broad Balmer lines \citep{MejaRestrepo+16}; (7)  AGN bolometric luminosity estimated from the intrinsic {\it Swift}-BAT hard 14--150\,keV  X-ray luminosity \citep{Ricci+17}; (8) Eddington ratio estimated from Eddington and bolometric luminosities; (9) median electron density; (10) \& (11) mean metallicity via strong-line calibrations by \citet[][]{sb98}   and \citet[][]{Carvalho+20}, respectively. }
\begin{tabular}{@{}lc c c c c c c c c cl@{}}
 \toprule
Galaxy 	 &$z$        &  Activity   & Morph.  &  Instrument/     & log $M_{\rm BH}$ & log $L_{\rm Bol}$ & $\lambda_{\rm Edd}$ &  $N_{\rm e}$ & \oh$_{\rm SB98F1}$ & \oh$_{\rm C20}$ \\
--- & --- & --- & --- & Telescope & M$_{\odot}$ & erg s$^{-1}$ &  --- &  cm$^{-3}$ & --- &  ---\\
(1) 	 & (2)        &  (3) & (4)   & (5)  & (6)       & (7) & (8) & (9) & (10) & (11)  \\
 \midrule
MRK\,607    &  0.009    &      Sy\,2   &  SAa   &    GMOS &  6.78   & 43.12  & $1.67\times10^{-2}$ &  $775\pm177$   &  $8.45 \pm 0.06$ &  $8.67 \pm 0.06$   \\   

MRK\,1066   &  0.012    &      Sy\,2   &  SB0  &    GMOS &  7.41 &  43.35  &  $6.75\times10^{-3}$  & $1015\pm175$ & $8.53\pm0.04$ & $8.73 \pm 0.08$     \\

NGC\,1052   &  0.005    &      Sy\,2   &  E4   &    GMOS & 8.82  &  42.93 &  $9.89\times10^{-5}$  &  $668\pm171$  & $8.57\pm0.05$ & $8.68\pm0.08$    \\

NGC\,2110   &  0.008    &      Sy\,2   &  SAB0 &    GMOS &  8.78   &  44.50   &  $3.99\times10^{-3}$  &  $857\pm155$   &  $8.64\pm0.04$ & $ 8.83\pm0.11$\\

NGC\,3516   &  0.009    &      Sy\,1.5 &  SB0  &    GMOS &  7.39 &  44.16   &   $4.53\times10^{-2}$   & $910\pm207$ &  $8.52 \pm 0.04$ & $8.73 \pm 0.08$ \\

NGC\,3786   &  0.009    &      Sy\,1.8 &  SABa &    GMOS & 7.82  &  43.42 &    $3.05\times10^{-3}$  & $771\pm176$  & $8.61 \pm 0.05$ & $8.77\pm0.08$  \\

NGC\,4235  &  0.008    &      Sy\,1.2   & SAa       &   GMOS  &  7.28  &  43.26    &  $7.30\times10^{-3}$   & $478\pm96$  & $8.68\pm0.04$ & $8.85\pm0.09$ \\

NGC\,4939   &  0.010    &      Sy\,2   &  SAbc &    GMOS &  7.75  &  43.61  &   $5.59\times10^{-3}$    & $1024\pm119$ & $8.61 \pm 0.04$ &  $8.83\pm0.08$\\

NGC\,5899   &  0.009    &      Sy\,2   &  SABc &    GMOS &  7.96  &  43.56  &  $3.03\times10^{-3}$    & $720\pm144$ & $8.73\pm0.06$ & $8.92\pm0.10$\\

NGC\,1068   &  0.004    &      Sy\,2   &  SAb  &    MUSE & 6.93  &   43.55  &   $3.23\times10^{-2}$  & $835\pm181$ & $8.63\pm0.06$ & $8.83\pm0.08$\\

NGC\,1194   &  0.013    &      Sy\,2   &  SA0  &   MUSE &  7.83   &   44.52     &  $3.76\times10^{-2}$    & $344\pm121$ & $8.43\pm0.08$ & $8.59\pm0.08$\\

NGC\,2992   &  0.008    &      Sy\,2   &  Sa   &   MUSE &  7.97  &  43.48   &  $2.49\times10^{-3}$   & $1022\pm 217$  &$8.51 \pm 0.05$ & $8.71 \pm 0.09$\\

NGC\,3081   &  0.008    &      Sy\,2   &  SAB0 &    MUSE &   7.67  &  44.08 &  $1.98\times10^{-2}$  & $611\pm194$  & $8.51\pm0.07$ & $8.71\pm0.10$  \\

NGC\,3393   &  0.013    &      Sy\,2   &  SBa &    MUSE &  7.52  &   43.96      &   $2.11\times10^{-2}$  & $1126\pm284$  & $8.53 \pm 0.05$ & $8.75\pm0.10$\\

NGC\,5728   &  0.009    &      Sy\,2   &  SABa &   MUSE &  8.25  &   44.14     &   $6.02\times10^{-3}$  & $422\pm110$ & $8.61\pm0.08$ & $8.80 \pm 0.10$\\
 \bottomrule
\end{tabular}
\label{tab1}
\end{table*}

\section{Methodology}
\label{chapt3}
Our method for estimating the gas-phase metallicity from AGN data cube spaxels involves several key steps. First, we identify and isolate spaxels dominated by AGN emission based on specific emission line ratios or ionization states. Next, we measure the fluxes of relevant emission lines  (e.g. \oiii$\lambda$5007, \hb and \nii$\lambda$6583) within these spaxels.  AGN strong-line methods from the literature are then applied to these measured line ratios to derive the gas-phase metallicities.
By combining these individual spaxel estimates, we construct detailed metallicity maps of the AGN-dominated regions. These maps are then analyzed to investigate spatial variations and gradients in metallicity. A detailed description of our methodology is provided below.

\subsection{Spectral fitting}
To accurately measure the emission line fluxes, the stellar absorption spectrum needs to be subtracted since the stellar absorption lines can introduce biases in the measurement of the emission line fluxes. Therefore, the emission-line fluxes were measured in a pure emission-line
spectrum using the fits performed by \citet{DahmerHahn+22}, free from the underlying stellar flux contributions.
The emission lines were then fitted using the
\texttt{IFSCUBE} package\footnote{\href{https://github.com/danielrd6/ifscube/}{https://github.com/danielrd6/ifscube/}}  \citep[][]{RuschelDutra+21}. 
Multiple Gaussian components were employed to model the emission lines, while a single
Gaussian component was sufficient for most targets without outflows in all emission line fits. Initially, 
we fit the emission lines with the same component of each line, forced to have the
same profile for a narrow component modelled with a single Gaussian centred at the wavelengths \halpha, \hbeta, \oiii$\lambda\lambda$4959,5007, \oi$\lambda$6300, \nii$\lambda\lambda$6548,6584, \sii$\lambda\lambda$6717,6731. A subsequent additional broad component was included as a second Gaussian for the Balmer lines (\halpha\, and \hbeta) and \sii$\lambda\lambda$6717,6731, as well as a wing component modelled with one Gaussian for \oiii$\lambda\lambda$4959,5007 when necessary.
The Gaussian components were classified as narrow (FWHM $\lesssim$ 200 \kms) or broad (FWHM $\gtrsim$ 500 \kms). 
We added the following constraints to our fits in order to obtain physically plausible solutions: 
(i) the centroid velocities were restricted to within $\lesssim\pm700$ \kms~ relative to the rest frame velocity of each galaxy;
(ii) the velocity offset of the Balmer and forbidden lines were
each tied, i.e. a single velocity offset for the Balmer lines (\halpha\ and \hbeta) was fitted and a separate offset was chosen for the other lines;
(iii) the observed velocity dispersion were restricted to
$\lesssim500$ \kms and pairs of lines for \oiii$\lambda\lambda$4959, 5007, \nii$\lambda\lambda$, 6548, 6584 and \sii$\lambda\lambda$6717, 6731 had tied
values for the velocity dispersion;
(iv) the flux ratios of \oiii$\lambda\lambda$4959, 5007 and \nii$\lambda\lambda$6584, 6548 were fixed to the theoretical values of 3.06 and 2.96,
respectively \citep[e.g.][]{Osterbrock+06}. As an integral part of the fitting process, \texttt{IFSCUBE} also estimates the uncertainties associated with each of the fitted model parameters.
These uncertainties represent the range of parameter values that are statistically consistent with the observed data, considering the noise and other sources of error such as the uncertainty in the peak flux. The emission line flux errors are estimated following the empirical relations by \citet{1992PASP..104.1104L} and \citet{2016MNRAS.456.3774W} based on the signal-to-noise ratio at the centre of each line. This empirical relation links the fractional uncertainty in flux ($\Delta F/F$) to the signal-to-noise ratio at the line centre: $\Delta F/F =  \frac{({\rm FWHM}/\Delta\lambda)^{-0.5}}{C\times{\rm SNR}}$, where $\Delta F$ represents the uncertainty or error in the measured flux of the emission line, $F$ denotes the measured flux of the emission line,  $\Delta\lambda$ is the wavelength sampling interval, the width of the Gaussian profile relative to the velocity dispersion ($\sigma$), usually measured in angstroms via FWHM $=2\sqrt{2ln2}\times\sigma \approx 2.35\sigma$, $C$ is an empirical constant  \citep[i.e. $0.5\,\lesssim\,C\,\lesssim\,1.5$;][]{1992PASP..104.1104L} based on the line shape and SNR is the signal-to-noise ratio calculated at the central peak of the emission line. The typical computed relative flux errors are in the range $\sim$10 -- 40\,\% of the strong and weak emission lines, respectively. The individual flux uncertainties are then propagated by performing 100 Monte Carlo iterations for the subsequent calculation of the physical properties that rely on these line fluxes. The errors for each parameter are then determined as the 1$\sigma$ dispersion obtained from the 100 Monte Carlo simulations.

\begin{figure*}
\includegraphics[width=2.1\columnwidth]{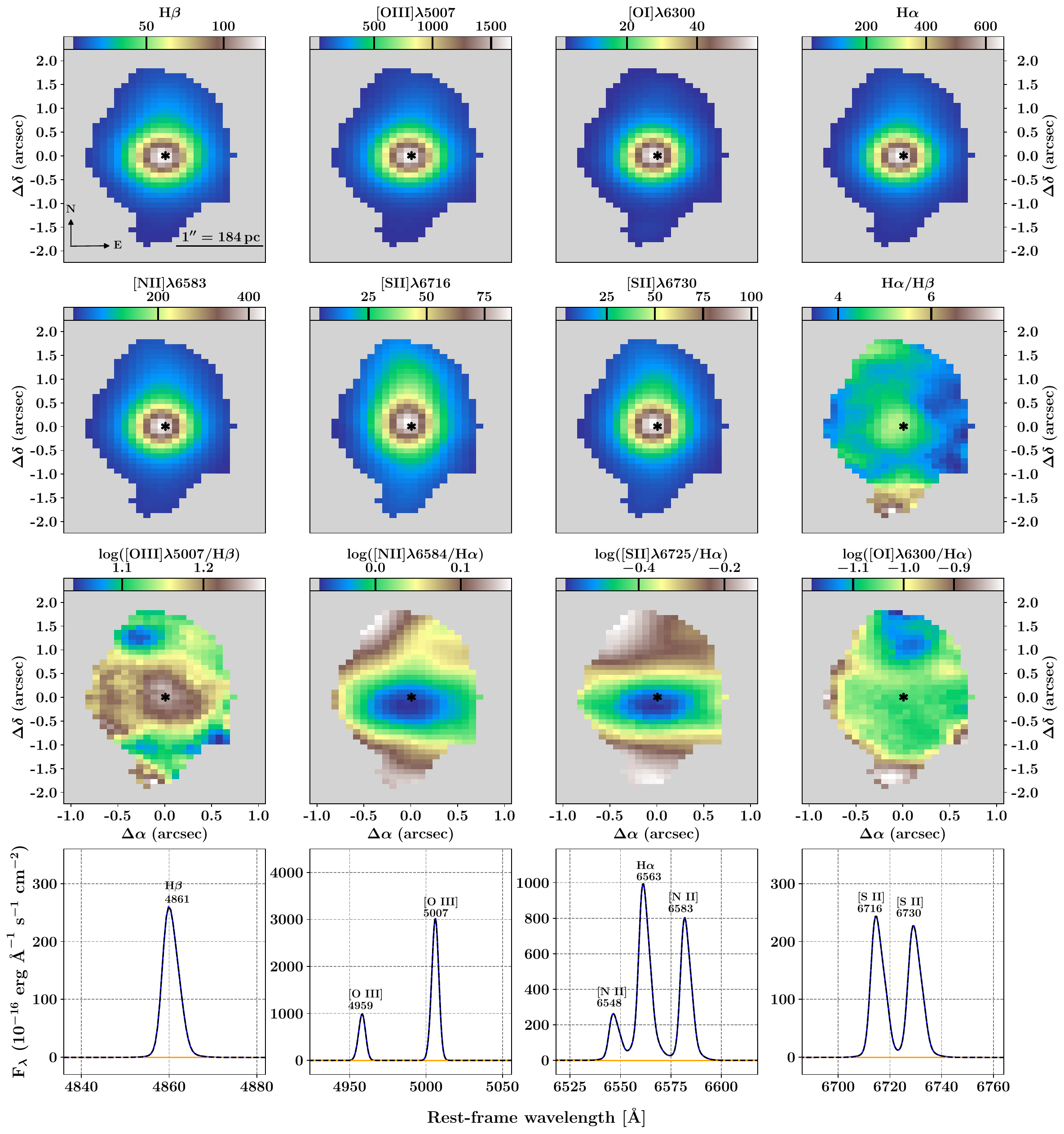}
\caption[]{Upper panels: emission line flux maps of GMOS Mrk\,607 (NGC\,1320):  \hbeta, \halpha, \oiii$\lambda$4959, \oiii$\lambda$5007,  \nii$\lambda$6548, \nii$\lambda$6583, \sii$\lambda$6716 and \sii$\lambda$6730 in units  $10^{-17}$ erg \AA$^{-1}$  s$^{-1}$ cm$^{-2}$  spaxel$^{-1}$.  Lower panels: maps of the diagnostic line ratios in logarithmic scale. The inset  indicates the Balmer lines intensity ratio    \halpha/\hbeta\ and the line ratios  \oiii$\lambda$5007/\hbeta,  \nii$\lambda$6584/\halpha\ and \sii$\lambda$6725 represents the
sum of the lines \sii$\lambda$6717 and \sii$\lambda$6731. The bottom plots show the pure nebular emission line spectrum for the nuclear spaxel of Mrk\,607 after subtracting the contribution from the stellar population continuum. The solid blue, dotted black and solid orange lines correspond to the nebular emission, modelled spectrum fits and 
the stellar continuum, respectively. The emission lines used in our analysis have been labelled, as shown. The grey regions correspond to locations where the spaxels have no available measurements or are below the signal-to-noise ratio limit. }
\label{ratios}
\end{figure*}

\begin{figure*}
\includegraphics[width=2.1\columnwidth]{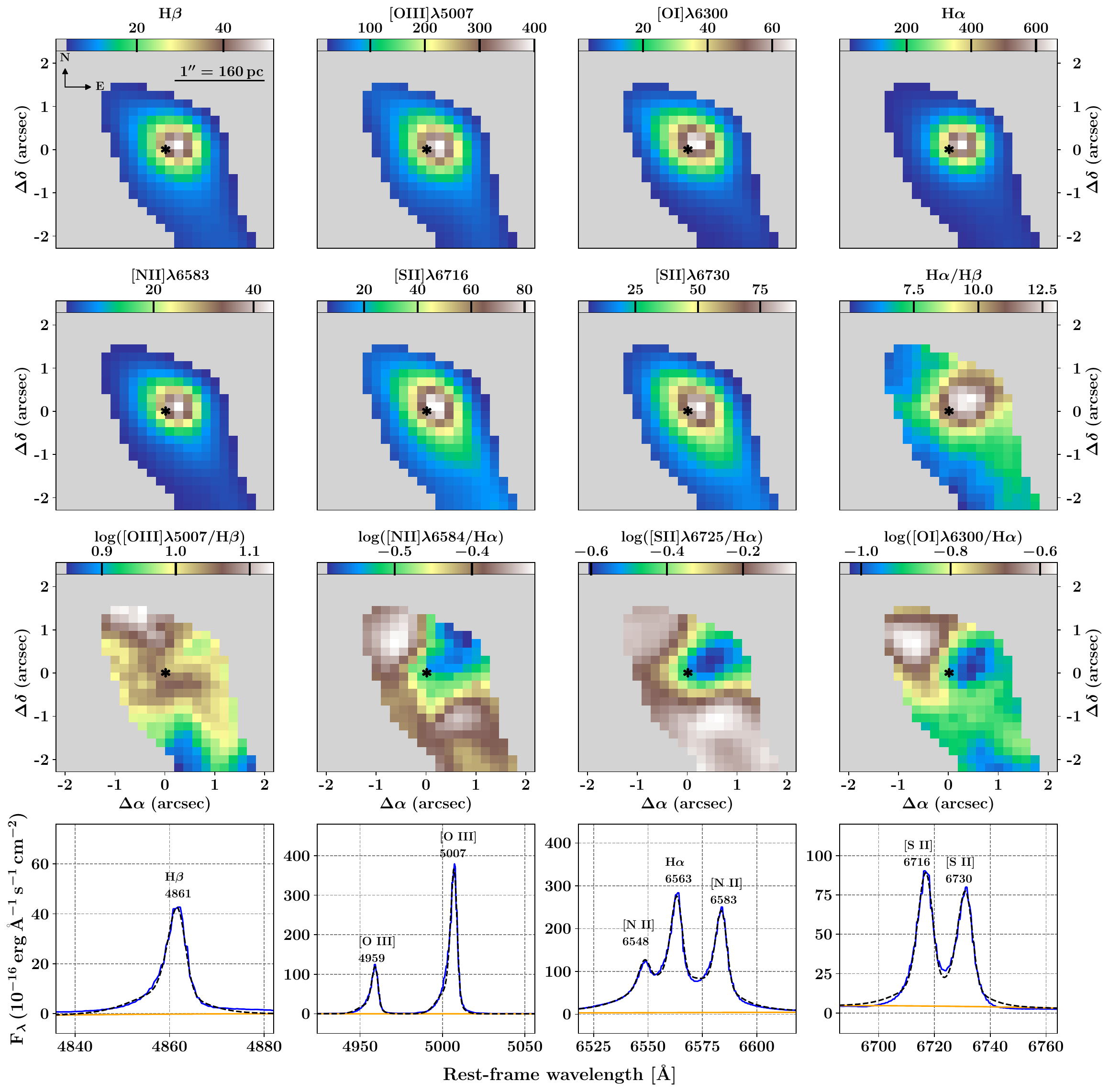}
\caption[]{Same as Fig.~\ref{ratios} but for NGC\,2992 from MUSE.}
\label{ratios2}
\end{figure*}

The emission line flux maps for Mrk\,607 (GMOS) and NGC\,2992 (MUSE) are shown in Figures~\ref{ratios} and \ref{ratios2}, respectively. The emission lines \hbeta, \oiii$\lambda$5007, \oi$\lambda$6300,  \halpha, \nii$\lambda$6583, \sii$\lambda$6716 and \sii$\lambda$6730 are  predominantly produced in the central regions of the nebulae. The fitted spectra for the individual objects of the present sample are continuum-subtracted spectra, i.e. the stellar population contribution and the AGN contribution, represented by a power-law function, were removed before the line fitting procedure. The integrated spectrum was obtained by summing up all the spectra of the IFU data cube from the central spaxel. The emission line flux maps for the remaining 13 targets are available in appendix \ref{appendix}.

\subsection{Diagnostic diagram}
\begin{figure*}
\includegraphics[width=2.1\columnwidth]{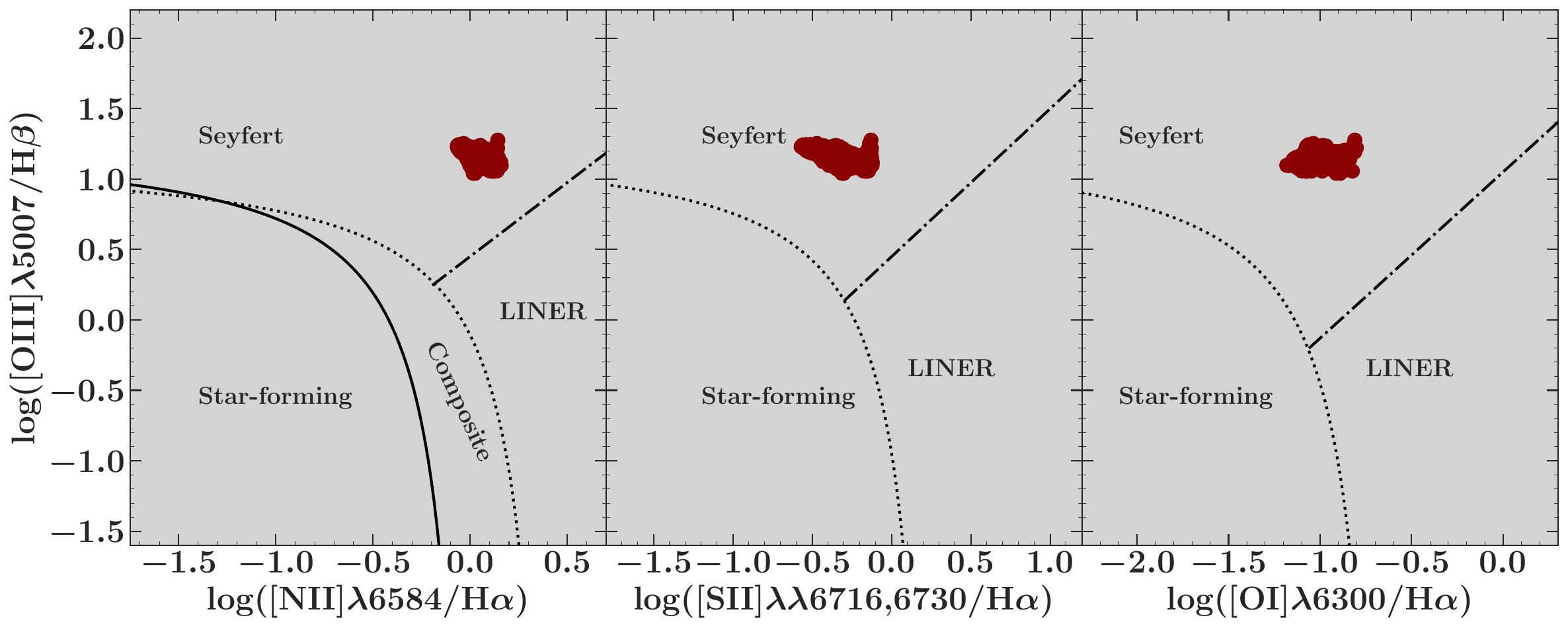}
 \caption[]{The BPT \citep[]{Baldwin+81}   diagnostic diagrams for MRK\,607 from the emission-line ratios of log([O\iii]$\lambda5007$/H$\beta$) versus log([N\ii]$\lambda6584$/H$\alpha$), log([S\ii]$\lambda$6725/H$\alpha$) and log([O\i]$\lambda6300$/H$\beta$). The black continuous solid and dotted lines, proposed by  \citet{2003MNRAS.346.1055K} and \citet{Kewley+01} respectively, separate objects ionized by massive stars from those ionized by AGN-like objects, as indicated. The dot-dashed lines represent the criterion proposed by \citet{CidFernandes+11} to separate AGN-like and low-ionization (nuclear) emission line regions [LI(N)ERs] objects.}
\label{fig2}
\end{figure*}

We created the spatially resolved Baldwin-Phillips-Terlevich  \citep[BPT;][]{Baldwin+81, Veilleux+87} diagnostic diagrams by combining the flux maps of different emission lines.  Spatially-resolved BPT diagrams for Mrk\,607 are shown in Fig.~\ref{fig2}: \oiii$\lambda 5007$/\hbeta\ vs. \nii$\lambda 6584$/\halpha, \sii$\lambda 6717,31$/\halpha\,  and \oi$\lambda$6300/\halpha. Each point corresponds to one spaxel, with line detection threshold of $\rm S/N\gtrsim3$ for each line.  The three panels in Fig.~\ref{fig2} show that for every position in the galaxy, our emission-line ratios fall in the Seyfert region based on the empirical classification line by \citet{CidFernandes+11} and the  theoretical upper bound to pure star-formation given by \cite{Kewley+01}, i.e. above and to the left of the separation lines in all BPT diagrams, specifically, in the high excitation and metal-rich region. This indicates that photoionization from AGN is the dominant excitation mechanism within the nebular regions of Mrk 607 (GMOS). Similarly, for all the remaining 14 galaxies,  except for NGC\,1052, which has a dual Seyfert/LINER classification, implying that it harbours a low luminosity AGN in comparison with the other objects, all spaxels are observed in the Seyfert region
of the BPT diagram.
The emission-line ratios of individual spaxels occupy a narrow band on the BPT diagnostics diagram, consistent with  AGN excitation.  

We followed a similar methodology for the dust extinction correction as that used by  \citet{Armah+23}. Therefore,  the emission-line fluxes can be corrected for dust extinction  using the Balmer decrement and the \cite{Cardelli+89} extinction curve. We assumed an $R_V = A_V/E(B-V) =3.1$  and an intrinsic H$\alpha$/H$\beta$ ratio of  3.1 for galaxies dominated by AGN  \citep[the Balmer decrement for case B recombination at $T = 10^4$ K and $N_{\rm e} \sim 10^2-10^4$ cm$^{-3}$;][]{Osterbrock+89}. 

\subsection{AGN metalicity calibrations}
\label{abund}
We adopted the strong-line methods for optical emission lines to be applied in the estimation of the spatially resolved metallicity in the NLRs of Seyfert nuclei since the lines \oii$\lambda\lambda$3726,3729 and \oiii$\lambda$4363 (or/and other auroral lines) are very weak and undetectable in most of our GMOS targets and outside the MUSE wavelength coverage.  To address this, few studies have developed AGN-specific calibrations and have also suggested using multiple line ratios to better constrain the estimates.  Hitherto, the only AGN calibrations that solely rely on [N\ii]$\lambda$$\lambda$6548,6584/H$\alpha$ and/or [O\iii]$\lambda$$\lambda$4959,5007/H$\beta$ are the two calibrations proposed  by \citet[][]{sb98} 
 and \citet[][]{Carvalho+20}. 
 
 AGN strong-line calibrations based on comprehensive photoionization models are derived by simulating an entire AGN. Thus, strong-line methods, in principle, can be used to interpret integrated NLR emission.
 AGN strong-line calibrations, while valuable for estimating gas-phase metallicity in the NLRs of AGNs using integrated spectra, face significant limitations when applied to individual spaxels within an AGN. This stems from several key factors, including but not limited to contamination from H{\ii} regions, shock excitation and spatial resolution limitations. On spatial scales, individual spaxels often capture a mixture of AGN-ionized gas and emission from H{\ii} regions. Star-forming galaxies exhibit distinct emission-line ratios compared to AGN-ionized gas; thus, metallicity estimates derived using any specific calibration method will be inherently biased. AGN outflows and jets can induce shocks in the surrounding gas, leading to additional excitation mechanisms beyond photoionization by the AGN \citep[e.g.][]{RiffelRA+21}. Shock-excited gas displays different emission-line ratios than purely photoionized gas, complicating the interpretation of strong-line diagnostics and potentially skewing metallicity estimates. Current integral field spectrographs often lack the spatial resolution needed to isolate the NLR from other emitting regions on sub-kiloparsec scales. This results in mixed spectra where the AGN contribution may be diluted, leading to inaccurate metallicity measurements when applying AGN-specific calibrations. Despite these aforementioned challenges, on smaller scales, AGN strong-line calibrations remain applicable when considering integrated spectra that encompass the inner hundreds of parsecs to $\sim$1 kpc of the galaxy. In this region, the influence of the  AGN is dominant, and the contribution from star-forming regions or shocks is minimized. Consequently, the observed emission-line ratios are primarily driven by AGN photoionization, allowing for reliable metallicity estimates using AGN-specific calibrations. 
In the following sections, we provide some details of the calibrations by \citet[][]{sb98} and   \citet[][]{Carvalho+20}, as well as the metallicity estimated from the data under consideration in this work. 

\subsection{\texorpdfstring{\citet{sb98}}~~calibration}
\label{sb}
\citet[][]{sb98}   used a grid of photoionization models  by assuming a typical AGN continuum, which were built with the {\sc Cloudy} code \citep{Ferland+17} and, for the first time, proposed two  AGN theoretical calibrations between  the NLRs emission line ratios
[N\ii]$\lambda\lambda$6548,6584/H$\alpha$, [O\ii]$\lambda3727$/[O\iii]$\lambda\lambda4959,5007$
as well as [O\iii]$\lambda\lambda4959,5007$/H$\beta$ and the metallicity (traced by the oxygen abundance). Three different approximations for the AGN ionizing continuum were considered: a typical AGN continuum with a combination of power laws \citep[e.g.][]{Mathews+87} and two power laws $F_v\propto v^\alpha$, with $\alpha$ = 1 and 1.5, respectively. The ionization parameter ($\log U)$ varies over the interval $-4 \: \lesssim \:  \log U \: \lesssim \: -2$.  The models were constructed assuming a gas density value of $300\, \rm cm^{-3}$. These calibrations are valid for the range of $\rm 8.4 \: \lesssim \: [12+\log(O/H)] \: \lesssim \:  9.4$ and the oxygen abundances obtained from these calibrations  differ systematically by only $\sim0.1$ dex \citep{sb98,Dors+20a}.

In this work we used only one calibration proposed by \citet[][hereafter  \textcolor{blue}{SB98F1}]{sb98}
 because we are unable to use [O\ii]$\lambda3727$ line as a selection criterion 
due to observational limitations.  The  \textcolor{blue}{SB98F1}
 calibration is defined by:
 \begin{eqnarray}
       \begin{array}{l@{}l@{}l}
\rm 12+(O/H)\, & =\, &  8.34  + (0.212 \, x) - (0.012 \,  x^{2}) - (0.002 \,  y)  \\  
         & + & (0.007 \, xy) - (0.002  \, x^{2}y) +(6.52 \times 10^{-4} \, y^{2}) \\  
         & + & (2.27 \times 10^{-4} \, xy^{2}) + (8.87 \times 10^{-5} \, x^{2}y^{2}),   \\
     \end{array}
\label{cal_SB}
\end{eqnarray}
where $x$ = [N\ii]$\lambda$$\lambda$6548,6584/H$\alpha$ and 
$y$ = [O\iii]$\lambda$$\lambda$4959,5007/H$\beta$.

It is important to apply the correction proposed by these authors to  Eq.~\ref{cal_SB}  in order to account for the deviations from the assumed  gas density, therefore, the final calibration is given by 
\begin{equation}
\label{cal_SB2}
{\rm \log(O/H)_{SB98F1}=[\log(O/H)}]-\left[0.1 \: \times \: \log \frac{N_{\rm e}({\rm cm^{-3}})}{300 \: ({\rm cm^{-3}})}\right],
\end{equation}
 where $N_{\rm e}$ is the electron density and Eq.~\ref{cal_SB2} is valid for $10^2\lesssim N_{\rm e} \rm (cm^{-3})\lesssim 10^4$ \citep[also, see][]{Peterson+97}. We note that only high-density objects ($N_{\rm e} > 3 \times 10^3\, {\rm cm^{-3}}$) have values of this correction that surpass 0.1 dex. Additionally, assuming  $N_{\rm e} =100 \: \rm cm^{-3}$ if \sii$\lambda6717$/\sii$\lambda6731 \gtrsim1.45$ and $N_{\rm e} =10\,000 \: \rm cm^{-3}$ if \sii$\lambda6717$/\sii$\lambda6731 \lesssim0.45$ in order to avoid collisional de-excitation effect  will produce a maximum correction of $\sim 0.15$ dex
\citep[e.g.][]{Denicol+02, Marino+13}. We used the \sii-doublet to estimate the electron density as these lines are strong in the spectra of all galaxies, however, we acknowledge the limitations and potential biases associated with using \sii-based measurements in AGN hosts \citep[e.g.][]{Shimizu+19, Davies+20, Nicholls+20}.   
Therefore, the metallicity derived from \textcolor{blue}{SB98F1} calibration is not significantly influenced by the electron density \citep[e.g., see \textcolor{blue}{Figure 5} by][]{Armah+23}.

\subsection{\texorpdfstring{\citet{Carvalho+20}}~~calibration} 
\citet[][hereafter \textcolor{blue}{C20}]{Carvalho+20} presented a comparison using  photoionization model predictions
 built with the {\sc Cloudy} code \citep{Ferland+17},  considering a wide range of nebular parameters, and   [O\iii]$\lambda5007$/[O\ii]$\lambda3727$ versus [N\ii]$\lambda$6584/H$\alpha$ diagram obtained from observational data of 
 463 Seyfert~2 nuclei ($z\lesssim0.4$).
 From this comparison, these authors derived a semi-empirical calibration 
 between the $N2$=log([N\ii]$\lambda$6584/H$\alpha$) line ratio and the metallicity $Z$, given 
 by  
\begin{equation}
Z/Z_{\odot} = x^{N2} - y, 
\label{c20}
\end{equation}
where $x = 4.01\pm0.08$ and $y = 0.07\pm0.01$, which is valid for $0.3 \: \lesssim \: Z/Z_{\odot} \:\lesssim  \: 2.0$. 
The metallicity results obtained from Eq.~\ref{c20} can be converted to oxygen abundance via the relation:
\begin{equation}
12+\log({\rm O/H})_{\rm C20}=12+\log[(Z/Z_{\odot}) \: \times \: 10^{\log(\rm O/H)_{\odot}}], 
\label{c20b}
\end{equation}
where $\log(\rm O/H)_{\odot}=-3.31$ is the solar oxygen abundance value taken from \citet{Asplund+21}. The $N2$ calibration index relies on an assumed relation between nitrogen-to-oxygen (N/O) abundance ratio and oxygen-to-hydrogen (O/H) abundance ratio. The relation $\log({\rm N/O}) = 1.29 \times [12+\log({\rm O/H})] - 11.84$ was adopted in the derivation of Eq.~\ref{c20}, which differs slightly from that assumed in calibration by \citet[][]{sb98}. Uncertainties or variations in this relation can introduce errors in the metallicity estimates. Potential
discrepancy between metallicity estimates from  \textcolor{blue}{SB98F1} and \textcolor{blue}{C20} calibrations can be attributed to the assumed (N/O) - (O/H) relation.  The $N2$ index is primarily calibrated for high-metallicity regimes 
($N2$ index also saturates at $Z/{\rm Z_{\odot}}<0.3$ and $Z/{\rm Z_{\odot}}>2$) due to the secondary nature of nitrogen production in stars. It may not be reliable for low-metallicity AGNs where nitrogen is predominantly produced through primary nucleosynthesis.

However, the $N2$ index has advantages  over other metallicity indicators such as [N\ii]$\lambda6584$/[O\ii]$\lambda3727$ 
because  it involves emission lines with very close wavelengths, i.e., the proximity of the {\nii} and \ha lines in the spectrum minimizes the impact of dust extinction and flux calibration errors in the derived metallicity \citep[e.g.][]{Marino+13,Castro+17}.  Unlike some other metallicity indicators (e.g. $O3N2$ or $R_{23}$), the $N2$ index is relatively insensitive to variations in the ionization parameter, making it more robust in diverse AGN environments. 

\subsection{Electron density and temperature}
The electron density ($N_{\rm e}$) and temperature ($T_{\rm e}$) were estimated from the \sii$\lambda6717$/\sii$\lambda6731$ and [O\iii]$\lambda 4363/\lambda 5007$ line ratios, respectively, using the Python package \texttt{PyNeb}\footnote{\texttt{PyNeb} is an innovative code for analysing emission lines and computing the physical properties of both theoretical and observational diagnostic plots; we used version 1.1.19 \url{https://github.com/Morisset/PyNeb_devel/tree/master/docs}} \citep{Luridiana+15}.  For each spaxel, we propagated the error in the line intensity using Monte Carlo simulations from the extinction-corrected line intensity ratios via \texttt{PyNeb}, assuming an electron temperature of 10 000 K. However, for the targets (Mrk\,607, Mrk\,1066, NGC\,1052, NGC\,2110 and NGC\,4235) where we measured the temperature sensitive auroral line [O\iii]$\lambda 4363$, we computed the electron
temperature values for each spaxel using the [O\iii] line ratio, and then we calculate the electron density assuming the $T_{\rm e}$ values obtained for each spaxel. This process is
repeated 100 times per spaxel. We obtain a distribution of density for each spaxel taking the mean as the value of the spaxel and the standard deviation as a measure of the associated uncertainties dominated by the systematic uncertainties in the scaling relations rather than the emission line fitting measurements, which correspond to
typical 0.1 dex metallicity correction. 

\begin{figure*}
\includegraphics[width=1.03\columnwidth]{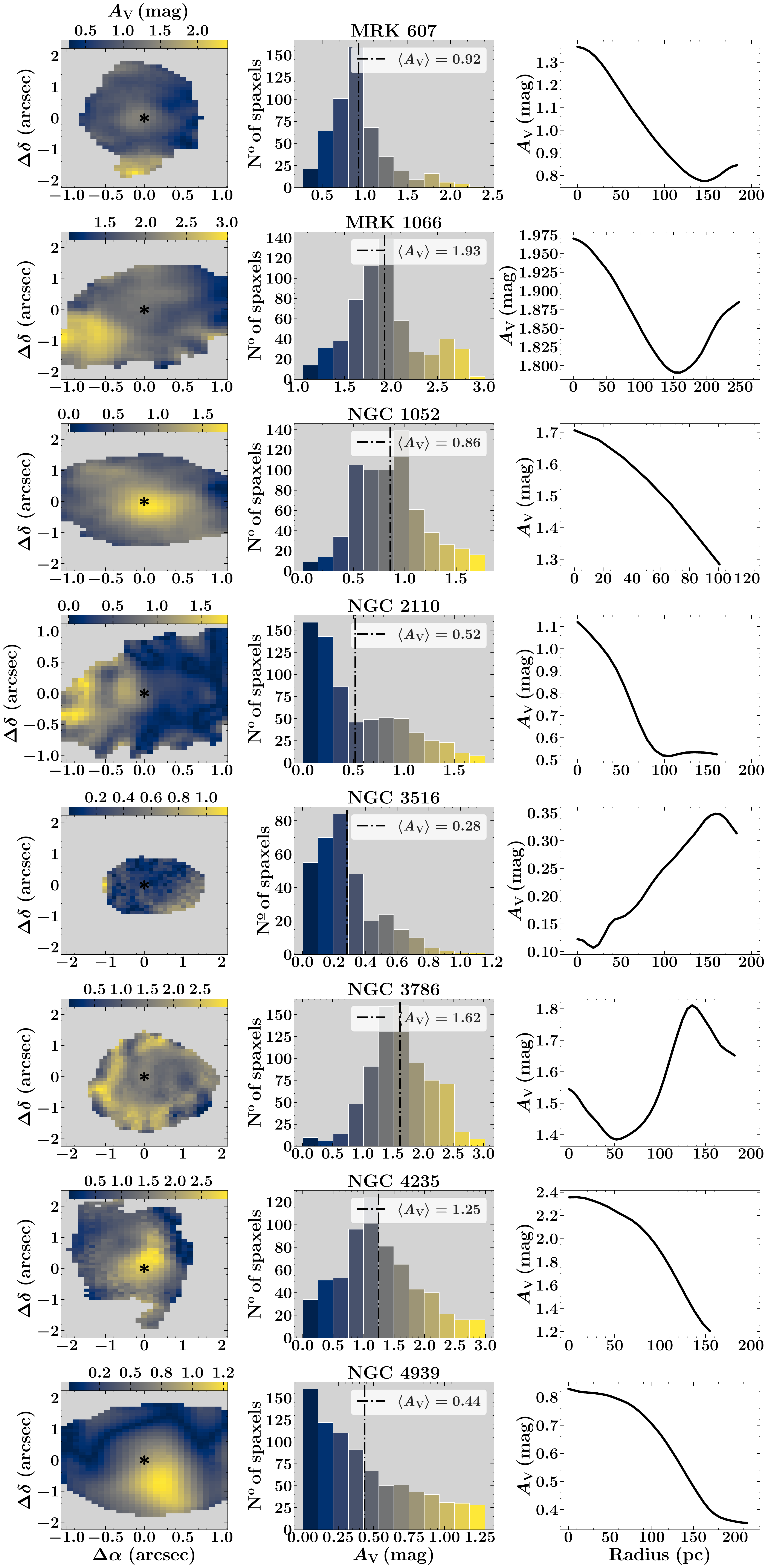}
\includegraphics[width=1.03\columnwidth]{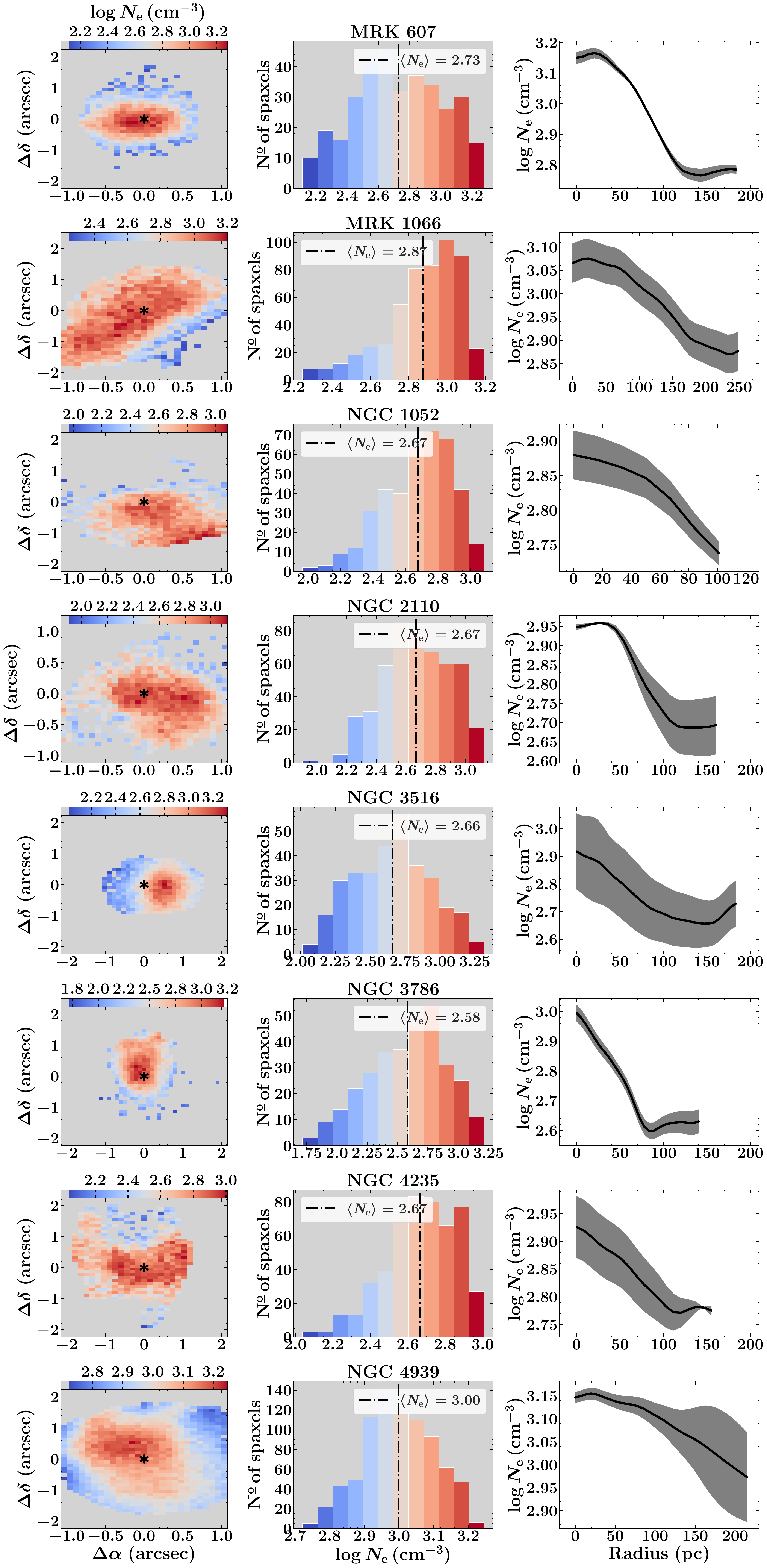}
\caption{The left three panels are maps, histograms, and radial distributions of the visual extinction ($A_{\rm V}$) derived from the Balmer decrement H$\alpha$/H$\beta$.  The right three panels are the same as in the left panel, but for the electron density ($N_{\rm e}$) derived from the \sii$\lambda6717$/\sii$\lambda6731$ line intensity ratio. The black solid curves correspond to the average profiles. The shaded regions in the radial profiles represent 1$\sigma$ uncertainties.
}
\label{fig4}
\end{figure*}

\begin{figure*}
\includegraphics[width=1.03\columnwidth]{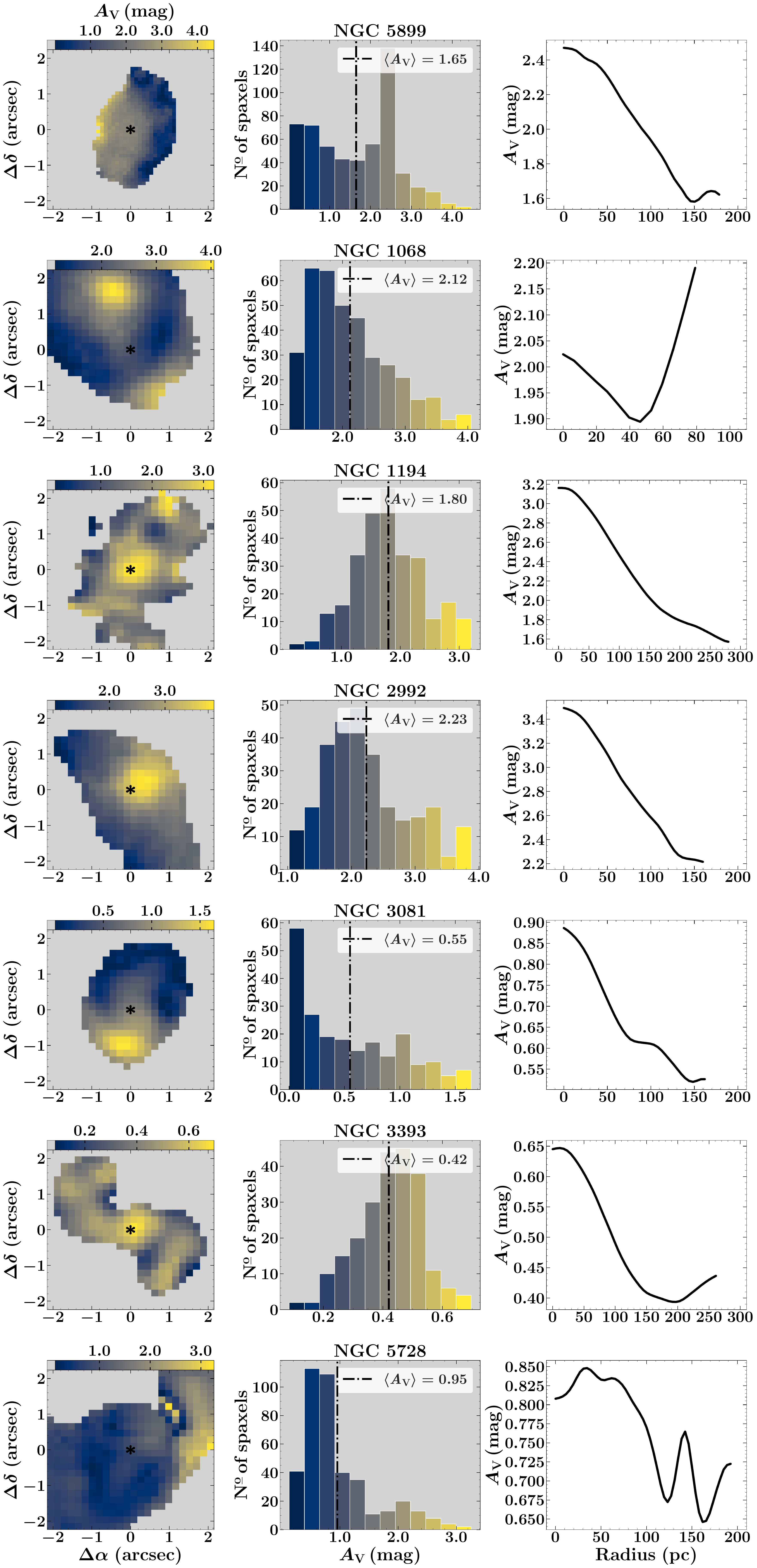}
\includegraphics[width=1.03\columnwidth]{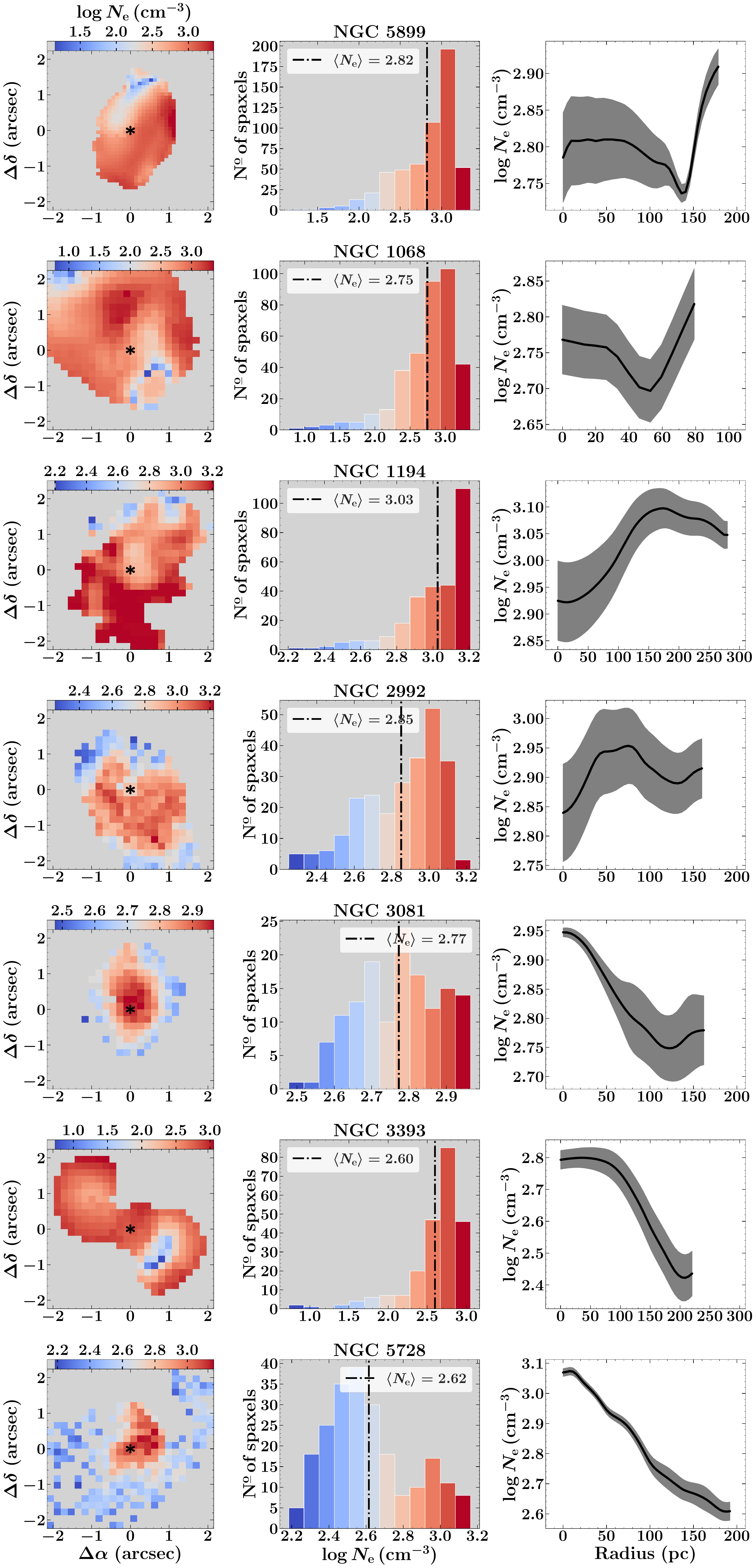}
\caption{Same as Fig.~\ref{fig4} but for the remaining objects. }
\label{fig5}
\end{figure*}

Figs.~\ref{fig4} and \ref{fig5} show the maps and radial profiles (see description in \S~\ref{res_1}) of the electron density derived from the \sii$\lambda6717$/\sii$\lambda6731$ line intensity ratio for individual sources. The density tends to follow a somewhat homogeneous distribution trend, either  from a central increase to an outskirt decrease or vice versa. The density distribution varies depending on the morphology and AGN activity of the galaxy. The density distribution in the spiral galaxies exhibits higher densities concentrated in the central bulge and along the spiral arms \citep[e.g.][]{Boselli+22}. While the general trend of decreasing $N_{\rm e}$ with radius in the disk and spiral arms is still present, the influence of the AGN can create localized variations due to ionization cones, outflows, or shocks. For instance, AGN feedback, in the form of jets or outflows, can drive gas outwards, creating cavities or bubbles of low $N_{\rm e}$ in the surrounding ISM \citep[][]{Harrison+14}. For barred spiral galaxies, the density distribution follows a similar pattern, with higher densities along the central bar and in the spiral arms. The presence of a bar structure can lead to enhanced star formation activity and gas dynamics within the galaxy \citep[e.g.][]{Crenshaw+03b, Galloway+15}. The density decreases gradually away from the centre towards the outskirt regions. On the other hand, the density distribution in elliptical galaxy is typically smooth and symmetric \citep[e.g.][]{Voit+15}. Furthermore,  elliptical galaxies lack the distinct spiral arms seen in spiral galaxies, so their density distribution tends to be more uniform. 

\begin{figure*}
\includegraphics[width=1.03\columnwidth]{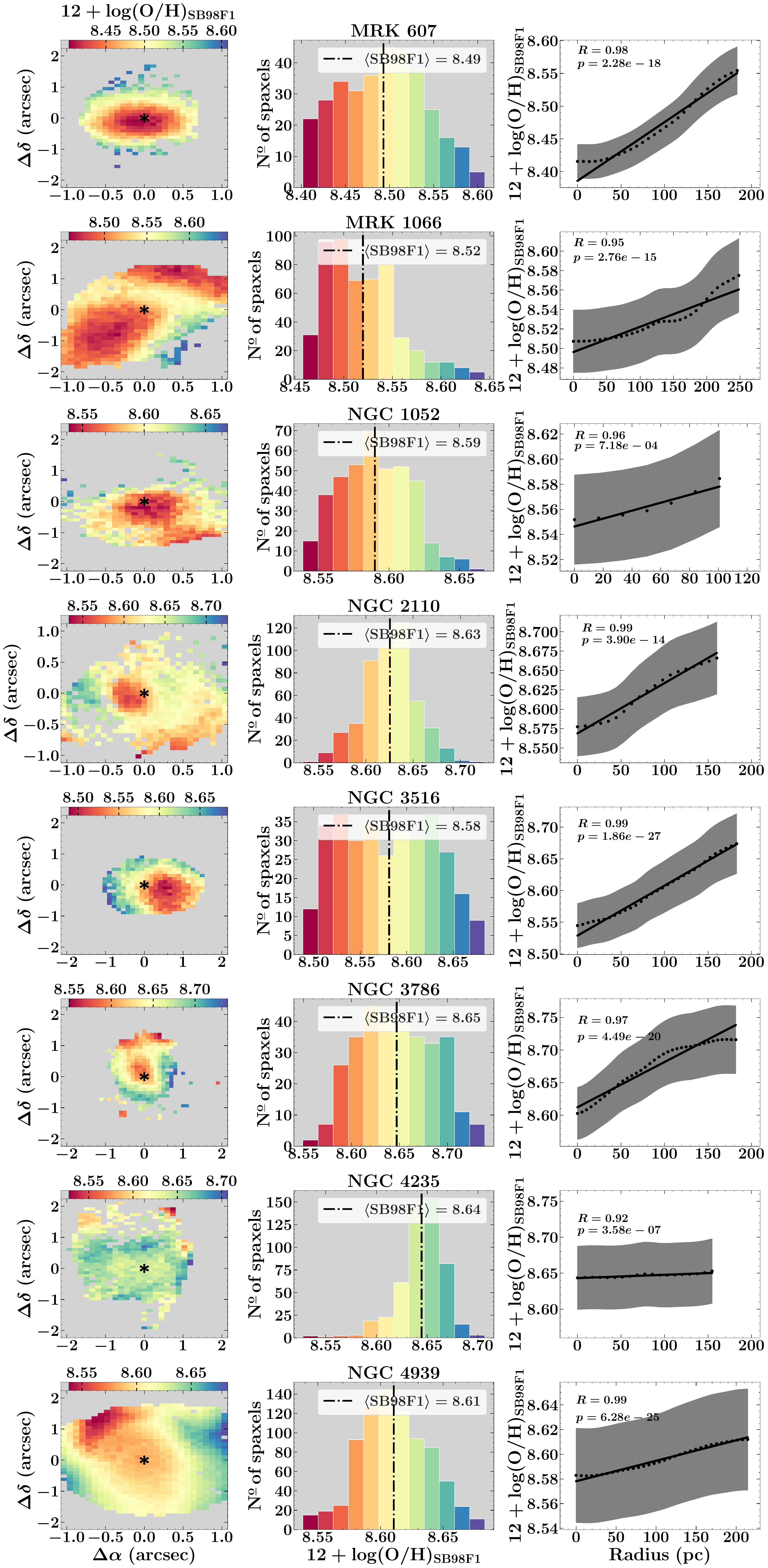}
\includegraphics[width=1.03\columnwidth]{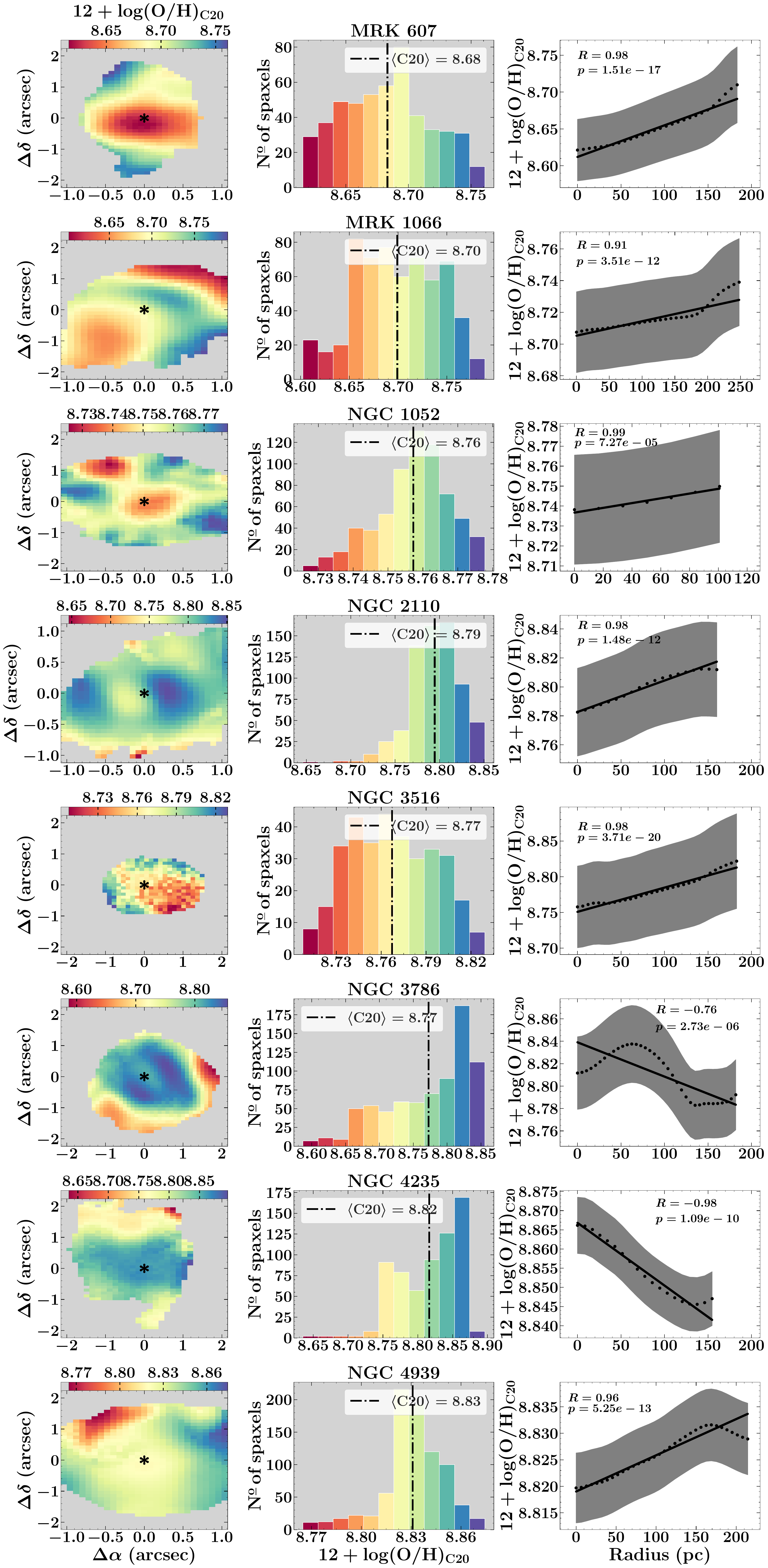}
\caption{The left three panels are the metallicity maps, histogram and radial distributions in [\oh] via the strong-line calibrations by  \textcolor{blue}{SB98F1}. The right three panels are the same as the left panels, but for \textcolor{blue}{C20} calibration.  Each data point represents the average value calculated within a specific annulus for each 10 pc radial bin. The solid line represents the best-fit regression over all data points using 1000 bootstrap realisations. The black solid curves  correspond to the average profiles. The shaded regions in the radial profiles represent 1$\sigma$ uncertainties.
}
\label{fig6}
\end{figure*}

\begin{figure*}
\includegraphics[width=1.03\columnwidth]{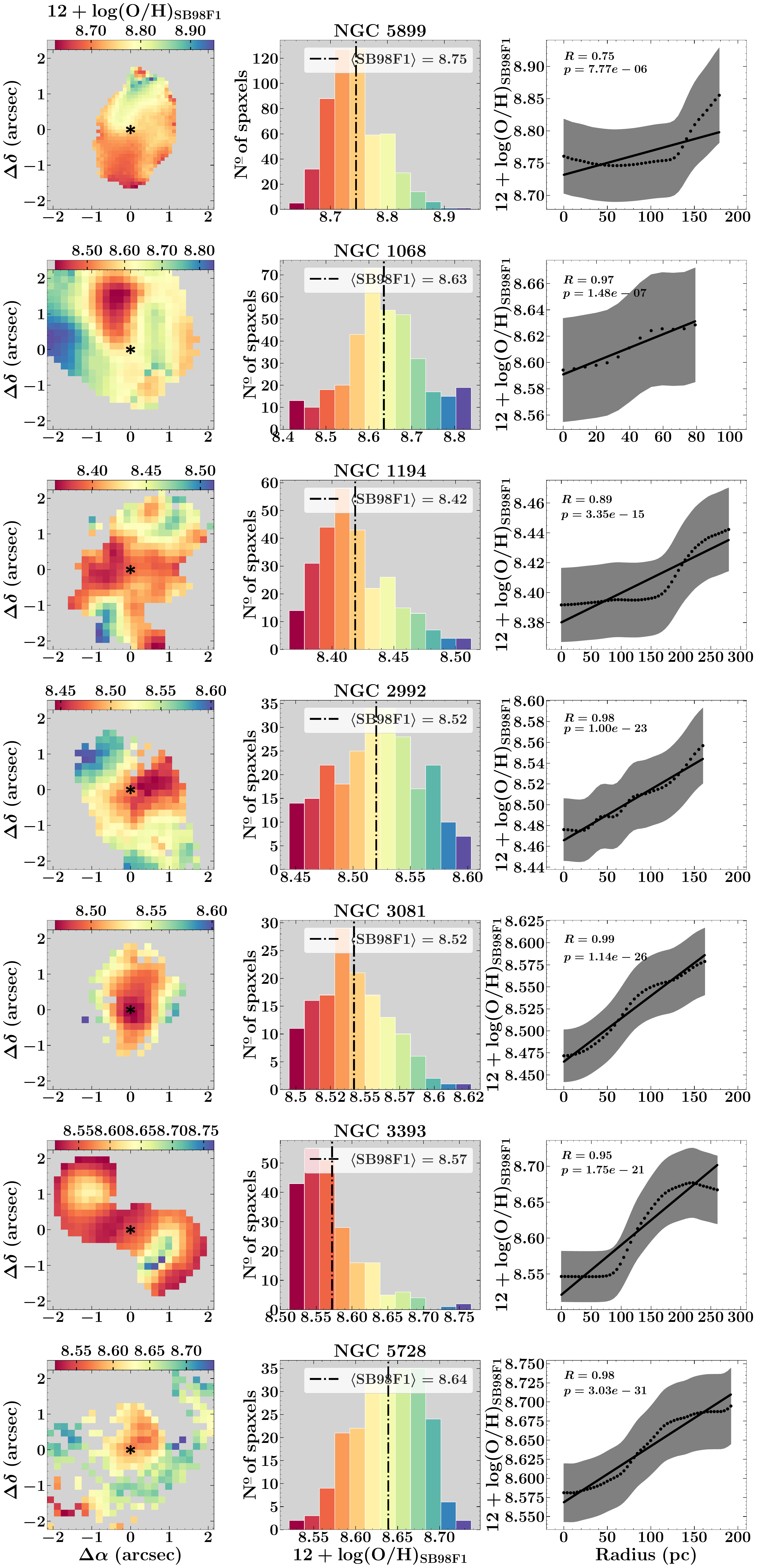}
\includegraphics[width=1.03\columnwidth]{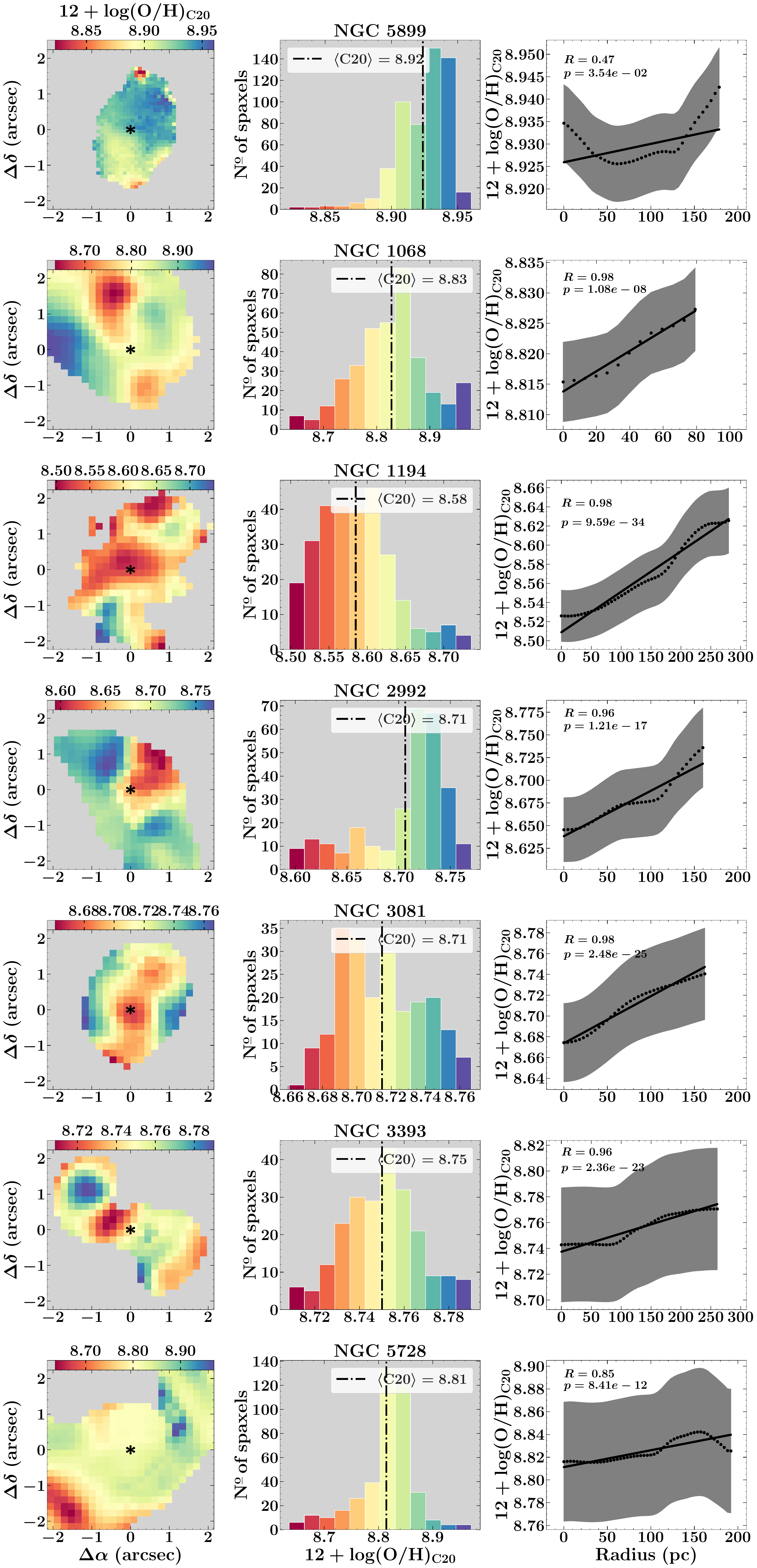}
\caption{Same as Fig.~\ref{fig6} but for the remaining objects. }
\label{fig7}
\end{figure*}

\section{Results and Discussion}
\label{chapt4}
\subsection{Individual result and discussion}
\label{res_1}
To maximize the spatial information from the IFU observations, we fit the physical property maps using a straightforward 2D axisymmetric gradient model. In this model, the physical property varies depending on the galactocentric distance, corrected for projection effects. We build our gradient model in the source plane by producing a de-projected galactocentric distance 2D map, using the centre of the galaxy, the ratio between the minor and major axis, and the position angle. We define 10 parsec annular apertures starting at radius, $r=0$, measuring the average physical properties of these annuli. 

The individual results (maps, histograms and radial profiles) from our sample are presented in Figs.~\ref{fig4} -- \ref{fig7} with each row representing one galaxy. In Figs.~\ref{fig4} and \ref{fig5}, we show the $V$-band dust extinction  ($A_{\rm V}$) derived from the Balmer decrement H$\alpha$/H$\beta$ and electron density ($N_{\rm e}$) derived from the  \sii$\lambda6717$/\sii$\lambda6731$ line intensity ratio, in  left three columns and right three columns, respectively.  The distribution of dust within these galaxies, depicted by the radial dust extinction profiles, vary significantly, with both negative (decreasing outward) and positive (increasing outward) gradients observed in our sample. The negative radial dust extinction profiles exhibited by most of the targets except NGC\,3516, NGC\,3786 and NGC\,1068, suggest that radiation of the AGN or outflow might be responsible for expelling dust from the BLR to the inner NLR (likewise from the inner NLR to the outer NLR). This indicates that the central regions are less obscured by dust compared to the outer regions.Though not as prevalent, positive dust extinction profiles have also been reported in some Seyfert galaxies, suggesting that dust extinction increases with distance from the nucleus. This phenomenon has been primarily associated with Seyfert 2 galaxies, which implies a higher concentration of dust in the inner NLRs. The revealed positive dust extinction gradient is possibly due to the presence of a dusty torus obscuring the central region or the influence of the structure of the host galaxy and environment \citep[e.g.][]{SchnorrMuller+16}.

All the targets except NGC\,5899, NGC\,1068, NGC\,1194 and NGC\,2992 show negative electron density profiles, suggesting a decrease in electron density as one moves away from the galactic centre   \citep[e.g.][]{Freitas+18, Kakkad+18,Prieto+21}. As one moves away from the nucleus (inner NLR), the gravitational potential weakens, causing the gas density to decrease and resulting in a negative $N_{\rm e}$ profile in the outer NLR. The gas becomes less tightly bound, allowing for expansion and cooling, which favours recombination of electrons and ions, thereby lowering the electron density in the outer NLRs \citep[e.g.][]{Crenshaw+03a, Osterbrock+06}. This expansion can also be driven by feedback mechanisms from the AGN, such as jets and outflows, that inject energy and momentum into the surrounding medium, further decreasing the outer NLR density \citep[e.g.][]{Kakkad+18}.
 Shock heating from these outflows or feedback from star formation could also contribute to the outward decrease in density. While less common, positive electron density profiles have also been observed in some Seyfert galaxies.  The infalling matter experiences shocks and heating, leading to enhanced collisional ionization and excitation, further augmenting $N_{\rm e}$ \citep{Peterson+97}. Additionally, radiative processes, powered by the accretion onto the SMBH, contribute to ionization state and the gas density  particularly in the outer regions of Seyfert galaxies \citep[e.g.][]{Bennert+06}.

We calculated the metallicity for each object by employing Eq.~\ref{cal_SB} (using the density distribution) and \ref{c20b}.
The third and sixth panels in Figs.~\ref{fig6} and \ref{fig7} show the typical three main classes of radial gas-phase metallicity gradients, i.e. negative (decreasing), flat (zero) and positive (increasing) gradients.  These variations reflect the complex interplay of processes such as inflow of metal-rich gas, outflows, mixing, and shocks within AGN environments. Understanding these different trends provides insights into the enrichment processes and dynamics within AGN environments. The observed negative metallicity gradients, where the central regions of galaxies are more metal-rich than the outskirts, are commonly found in local spiral galaxies \citep[e.g.][]{Carton+18,Sharda+21}. This suggests efficient metal dilution consistent with galaxy evolution closed system and form according to an inside-out galaxy formation scenario where star formation and subsequent chemical enrichment start earlier in the inner regions than the outskirts \citep[e.g.][]{Tissera+22}. Negative gradients have been observed in approximately 66\,\% of galaxies hosting AGNs \citep[][]{Nascimento+22}. Under certain mechanisms, such as metal-poor gas accretion, intense stellar feedback, galaxy merging and interactions, as well as strong shocks and efficient radial mixing of metals \citep[e.g.][]{Simons+21}, the metallicity gradient can become flat. In this scenario, the metallicity is uniform across the galaxy, with no significant variation with radial distance. Around 9\,\% of galaxies hosting AGNs exhibit flat gradients \citep[][]{Nascimento+22}. Positive metallicity gradients, where the outer regions of galaxies are more metal-rich than the centres, are less common in AGNs but have been observed in some high-redshift star-forming galaxies \citep[e.g.][]{Carton+18,Simons+21}. Positive gradients are thought to be more prevalent at higher redshifts due to the higher rates of gas accretion of external pristine gas towards the central regions of the galaxy \citep[e.g.][]{Ceverino+16,Molla+16, Curti+20} and mergers, which can redistribute metals and flatten or even reverse the gradients \citep[e.g.][]{Simons+21}. Strong galactic outflows enriched with metals may also play a role in creating a positive gradient by displacing metal-rich gas from the centre of starburst regions to their outskirts \citep[e.g.][]{Wang+19,Tissera+22}.  However, positive gradients have also been observed in some nearby AGNs, for instance NGC\,7130 \citep[][]{Amiri+24} and approximately 25\,\% of 108 Seyfert galaxies from SDSS-IV MaNGA survey  \citep[][]{Nascimento+22}.  
The majority of the galaxies ($15/15 - 100\,\%$ from \textcolor{blue}{SB98F1} and $13/15 \sim 87\,\%$ from \textcolor{blue}{C20}) in our sample show positive gas-phase metallicity gradients. The diversity of metallicity gradients observed in galaxies, including those hosting AGNs, highlights the intricate interactions between various processes such as star formation, gas accretion, mergers, and radial mixing of metals \citep[e.g][]{Carton+18,Sharda+21,Simons+21,Nascimento+22}. 
The specific dust, density and radial gas-phase metallicity gradients are presented, considering individual galaxies, in the appendix.

From Fig.~\ref{fig6}, NGC\,3786 and NGC\,4235 are the only objects in our sample with both positive and negative metallicity gradients from \textcolor{blue}{SB98F1} and \textcolor{blue}{C20} strong-line calibrations, respectively. 
The phenomenon of a galaxy exhibiting both positive and negative gas-phase metallicity radial gradients from different strong line metallicity calibrations can be attributed to several factors: (a) Sensitivity to physical conditions: different strong line metallicity calibrations are based on different emission line ratios, which are sensitive to various physical conditions such as ionization parameter, metallicity, and gas density. These calibrations may have different dependencies on these parameters, leading to variations in the inferred metallicity gradients. Models that rely on assumptions about gas inflows, outflows, and metal equilibrium processes \citep[e.g.][]{Sharda+21} may oversimplify the complex interplay of these factors in real galaxies. (b) Assumptions and degeneracies: strong line metallicity calibrations often rely on assumptions about the ionization state of the gas and the shape of the ionizing spectrum. Differences in these assumptions can result in degeneracies, where multiple combinations of physical parameters can produce similar emission line ratios. As a result, different calibrations may yield different interpretations of the metallicity gradient. (c) Spatial variations: the metallicity of the gas in galaxies can vary spatially due to a variety of factors such as inflows, outflows, gas mixing, and interactions with neighboring galaxies. Different regions within the same galaxy may exhibit different metallicity gradients, depending on their specific evolutionary histories and environmental conditions. (d) Calibration uncertainties: strong line metallicity calibrations are often calibrated using samples of galaxies with known metallicities, but there can be uncertainties associated with these calibrations, including systematic biases and incompleteness in the calibration samples. These uncertainties can propagate into the inferred metallicity gradients and lead to discrepancies between different calibrations. (e) Evolutionary effects: the metallicity gradient of a galaxy may evolve over cosmic time due to processes such as gas accretion, star formation, and feedback from supernovae and AGNs. The observed metallicity gradient at a given epoch may depend on the evolutionary history of the galaxy and the timescale over which the metallicity measurements are made.

In summary, the presence of both positive and negative gas-phase metallicity radial gradients from different strong line metallicity calibrations in the same galaxy can arise from a combination of factors, including sensitivity to physical conditions, assumptions and degeneracies in the calibrations, spatial variations in metallicity, calibration uncertainties, and evolutionary effects. Understanding these factors is essential for accurately interpreting metallicity gradients in galaxies and elucidating their evolutionary histories.

\subsection{General results and discussion}
We note that all the spaxels show line ratios consistent with AGN-dominated regions. In order to provide an average map of the  properties we have obtained--dust, electron density and gas-phase metallicity distributions. We show in Fig.~\ref{avg}, average maps of these quantities in the top panels with corresponding average radial profiles in the bottom panels. They show that the visual dust extinction  and electron density peak at the nuclei, while the metallicities depict opposite trends.

\subsubsection{Dust extinction }
Seyfert galaxies often contain significant amounts of interstellar dust \citep[e.g.][]{Prieto+14,Lu+19,Prieto+21}, particularly in their central regions surrounding the AGN. This dust can obscure the central region from optical observations and contribute to the infrared emission. The distribution of dust can be complex, influenced by factors such as  galaxy interactions, star formation activity and overall morphology of the galaxy. AGN radiation can heat and evaporate dust grains in the vicinity of the central black hole, leading to a depletion of dust in the inner regions of the galaxy. Additionally, AGN-driven shock and outflows can compress and condense dust in the surrounding gas, leading to the formation of dense dust clouds and filaments. Furthermore, the presence of an AGN can lead to the formation of ionization cones and outflows that may affect the density distribution within the galaxy, compressing or dispersing gas clouds and triggering star formation in certain regions.
Interactions with neighboring galaxies or companions, if present, can also influence the density distribution, potentially leading to tidal effects and gas stripping. The three left panels in 
Figs.~\ref{fig4} and \ref{fig5} are the maps, histograms and radial profiles of the $A_{\rm V}$. On the one hand, the $A_{\rm V}$ distributions in GMOS and MUSE are nearly identical, with the nucleus exhibiting the highest values. 
The $A_{\rm V}$ values from both observations imply that the emission lines involved in the estimation of the metallicity are affected by dust extinction. Therefore, we note here that reddening correction was applied to all the emission lines used for the electron density and the metallicity estimates. 

The negative dust radial profile observed in Fig.~\ref{avg} refers to the trend where the amount of dust decreases with increasing distance from the galactic centre.  The negative dust radial profile in Seyfert galaxies reflects the combined effects of 
several factors including AGN activity, radiation processes, and galactic dynamics, affecting the distribution of dust within the galaxy. This radial profile is crucial for unraveling the physical processes occurring in Seyfert galaxies and their impact on galactic evolution.
(i) AGN activity: Seyfert galaxies are characterized by the presence of an active galactic nucleus, which emits intense radiation across a wide range of wavelengths, including X-rays and UV radiation. This high-energy radiation can heat the dust grains in the inner regions of the galaxy to temperatures where they evaporate or are destroyed. Consequently, the inner regions of Seyfert galaxies, where the AGN is located  (BLR), exhibit lower levels of dust content compared to the inner NLRs.
(ii) Dust destruction: the intense radiation from the AGN can directly heat and destroy dust grains in the central regions of Seyfert galaxies. This process reduces the overall abundance of dust in the inner regions and contributes to the negative radial profile observed. (iii) Radiation pressure: in addition to heating and destroying dust grains, the radiation from the AGN can exert pressure on the remaining dust particles, pushing them to the inner NLRs from the galactic centre. This radiation pressure can effectively drive dust out of the central regions of the galaxy, leading to a decrease in dust abundance with increasing distance from the galactic centre. (iv) Galactic dynamics: the distribution of dust in Seyfert galaxies can also be influenced by the galaxy's gravitational potential and dynamical processes such as gas inflows, outflows, and radial mixing. These processes can redistribute dust throughout the galaxy, leading to variations in the amount of dust as a function of radius. (v) Ionization front: the ionizing radiation emitted by the AGN can create an ionization front, where gas is ionized and electrons are liberated. However, as the distance from the AGN increases, the intensity of the radiation decreases, causing the ionization front to move outward. Consequently, the density of free electrons decreases with increasing radius from the galactic centre. (vi) Dust reddening: the presence of dust grains can absorb and scatter optical and UV radiation, resulting in dust reddening. Therefore, regions with higher dust abundance will exhibit higher levels of reddening compared to regions with lower dust abundance. The negative dust radial profile observed in Seyfert galaxies suggests that the inner regions of the galaxy have higher levels of dust reddening compared to the outer regions.

 In general, Seyfert 1 galaxies have less obscuring dust distribution compared to Seyfert 2s
\citep[e.g.][]{Prieto+14,SchnorrMuller+16,Lu+19,Prieto+21}. We observe moderate extinction $A_{\rm V}\sim0.9$ mag (see Fig.~\ref{avg}), which is insufficient to obscure or suppress any of the strong emission lines, even though nuclear filaments and lanes can produce distinct gas  kinematic and morphological properties observed in Seyferts 1 and 2, depending on their optical thickness and/or location in relation to the nucleus.  

\subsubsection{Electron density}
The density distribution in Seyfert galaxies can vary depending on their morphology and activity level. For instance, in barred spiral Seyfert galaxies, higher densities may be observed along the central bar and in regions of active star formation \citep[e.g.][]{Galloway+15}. The AGN itself can also play a role in influencing gas dynamics and density profiles, ultimately shaping the overall density distribution within the galaxy. The presence of an AGN can stabilize the gas disk against gravitational instabilities and enhance the formation of a central bar structure in the galaxy, affecting the overall density distribution. The $N_{\rm e}$ distributions in both observations (GMOS and MUSE) follow the same trends and the values are well within the limits of the NLRs of Seyfert galaxies  \citep[$N_{\rm e}\lesssim 10^4\, {\rm cm}^{-3}$; e.g.][]{Osterbrock+06,Ho+08}. 

The negative electron density radial profile observed in Fig.~\ref{avg}  refers to the trend where the density of free electrons decreases with increasing distance from the galactic centre. It can be attributed to several interconnected factors e.g. gravitational potential and AGN-driven outflows and winds (radiation pressure and mechanical feedback).  (i) Gravitational potential: as we move away from the central SMBH, the gravitational potential decreases. This leads to a decrease in the density of gas and dust, resulting in a negative electron density profile in the outer NLR. Additionally, the gas in the outer regions is less strongly bound by gravity and may experience expansion and cooling. This can further reduce the electron density due to recombination of electrons and ions \citep{Osterbrock+06}; (ii) Radiation pressure: the intense radiation emitted by the AGN exerts pressure on the surrounding gas. This radiation pressure can accelerate gas outwards, creating powerful outflows or winds. These outflows can sweep away gas from the central regions, leading to a decrease in electron density as one moves away from the centre. (iii) Mechanical feedback: AGN-driven jets and winds can also inject mechanical energy into the surrounding ISM \citep[e.g.][]{Fabian+12}. This energy input can heat the gas and drive turbulence, further promoting the outward movement of gas and contributing to the negative density profile in the outer NLR. (iv) Outflow interactions: the interaction of AGN-driven outflows with the surrounding ISM can create shocks. These shocks heat the gas, causing it to expand and resulting in a decrease in density outwards.

These aforementioned factors can act independently or in combination to produce the observed negative electron density profiles in Seyfert galaxies. The relative importance of each mechanism can vary depending on the specific properties of the AGN and its host galaxy. Thus, the radial density relation can be explained by the fact that the lower density regions are more gas-rich and less chemically-evolved. This is consistent with the traditional inside-out model. The inside-out model posits that central (denser) regions of a galaxy, such as the bulge and inner disk, are formed first through rapid star formation fueled by gas-rich mergers, accretion of gas from the surrounding medium, or other processes.  As the galaxy evolves, star formation activity gradually extends to the outer (less dense) regions, leading to the formation of the outer disk, which is less gas-rich and more chemically-evolved, and the growth of the galaxy as a whole. This indicates that metallicity increases with increasing galactocentric distance. 
   
\subsubsection{Gas-phase metallicity}
Metallicity distributions in Seyfert galaxies provide insights into their star formation history, chemical evolution and interactions with their environment.  Spiral galaxies often exhibit metallicity gradients, with higher metallicities observed in the central regions and decreasing toward the outer disks  \citep[e.g.][]{Searle1971,Dors+15,Stanghellini+15, Nascimento+22}. This gradient reflects the enrichment of  the ISM with heavy elements over cosmic time through processes such as stellar nucleosynthesis and galactic chemical evolution. Seyfert galaxies are often characterized by enhanced levels of metallicity compared to less active galaxies, as they have experienced multiple generations of star formation and enrichment from supernova explosions, stellar winds and accretion onto the central black hole.    
Additionally, interactions with neighbouring galaxies or gas inflows/outflows can lead to mixing and redistribution of metal-enriched material from external sources or within the galaxy, affecting the metallicity distribution within the galaxy. 

Figs.~\ref{fig6} and \ref{fig7}  show the gas-phase metallicity distributions from the strong-line calibrations. The median metallicity values from the strong-line calibrations by  \textcolor{blue}{SB98F1} and \textcolor{blue}{C20} are shown with the grey dashed lines, while the black dashed lines are represented by \oh = 8.69, which is the solar oxygen abundance value derived by \citet{Asplund+21}. The metallicity maps from \textcolor{blue}{SB98F1} show that most regions within Mrk\,607 are sub-solar, while  metallicity maps from \textcolor{blue}{C20} affirm that most regions within Mrk\,607 are solar or slightly above solar. We show the deprojected radial distance of each spaxel in the galacto-centric frame in  Fig.~\ref{fig7}. The error bars correspond to the uncertainties propagated from the individual flux uncertainties by performing 100 Monte Carlo iterations for the emission line ratios used in the metallicity estimations. The deprojected radial distribution of dust and electron density slightly decreases from the central source to the outskirt. 
We find strong positive and negative (i.e. metal-rich outskirts increasing with radius and vice versa) to flat metallicity gradients from individual galaxies (see Figs~\ref{fig6} and \ref{fig7}), consistent with previous results from star-forming galaxies, supernovae remnants and planetary nebulae  \citep[e.g.][]{Lyman+18,Patricio+19, Curti+20, Wang+20, Simons+21,Li+22,Venturi+24}. However, the exact physical origin  of these  trends has become a staple topic of discussion in the context of redistribution of gas-phase metals and dilution of the ISM in
the central regions of massive galaxies as a result of metal-poor and/or metal-rich gas accretion, star formation, galaxy merging and interactions as well as environmental mechanisms such as ram pressure and tidal striping, which influence the chemical history of galaxies  over cosmic time. 

\begin{figure}
\includegraphics[width=1.\columnwidth]{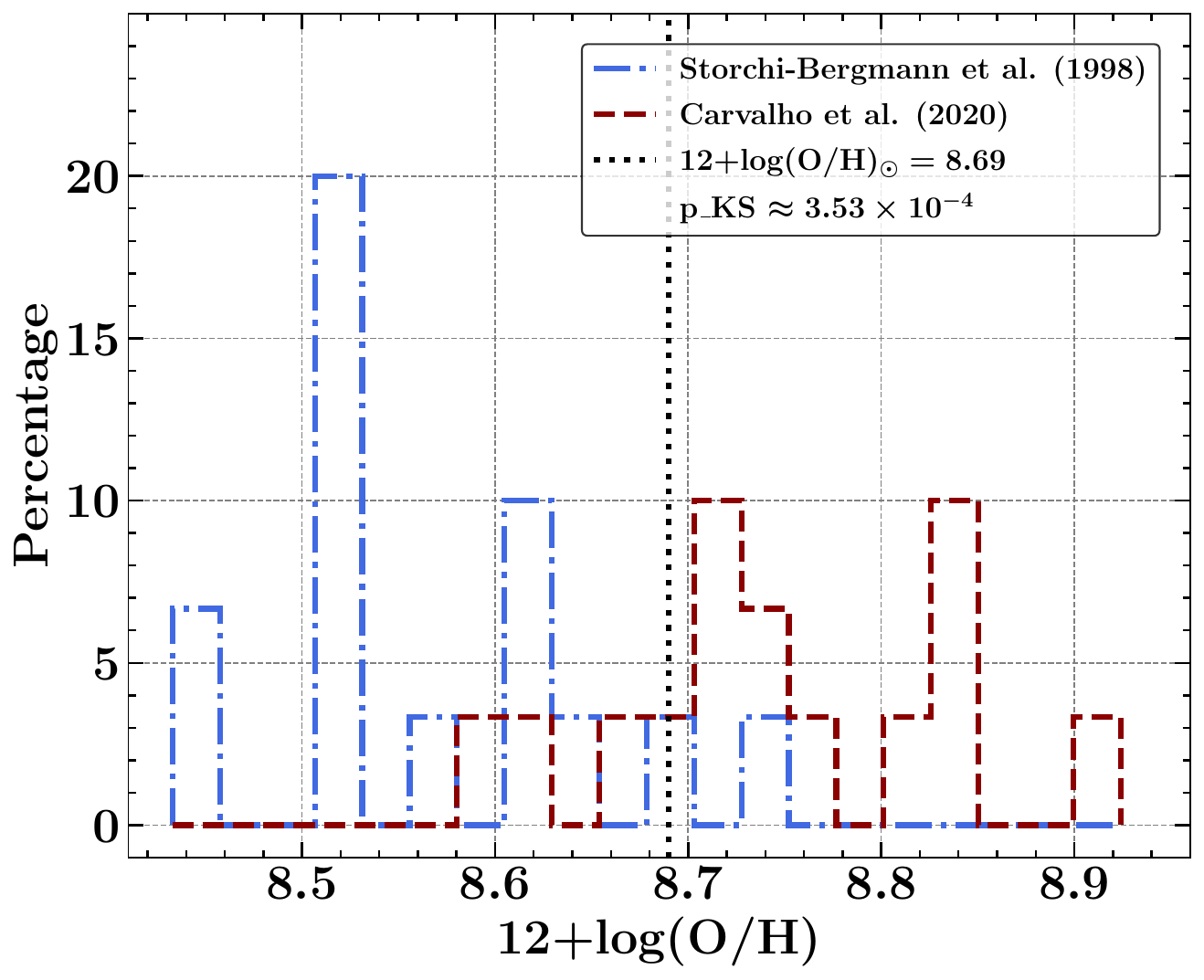}
\caption{Mean metallicity distribution using the calibrations by  \textcolor{blue}{SB98F1}  and   \textcolor{blue}{C20} represented by the blue dash-dot and red dashed lines. For reference, we quote the solar oxygen abundance value \oh = 8.69 \citep{Asplund+21}, denoted by the black dotted line.  A two-sample Kolmogorov--Smirnov (KS) statistical test with {\it p-value},    p\_KS is also shown.  }
\label{oh}
\end{figure}

The \textcolor{blue}{SB98F1} calibration depends on the electron density so it is imperative to  analyse  the potential impact of the electron density on the metallicity estimates.  In order to examine the impact of the electron density on the metallicity, we used a two-sample Kolmogorov--Smirnov (KS) statistical test to examine the metallicity distribution from the two distinct calibrations, as shown in Fig.~\ref{oh}.  We find that the  significant difference between the samples with the {\it p-value},  p\_${\rm KS \approx 3.53\times10^{-4}}$, indicates that, the metallicity estimates via  \textcolor{blue}{SB98F1}
 and   \textcolor{blue}{C20}  are from different distributions following standard
practice by adopting $2\sigma$ (i.e. {\it p-value} = 0.05) as the threshold for a statistically significant difference. We note that this difference is consistent with previous work \citep[see \S~4.3 in][for details]{Armah+23}. Therefore, the electron density has a mild influence on the metallicity derived via the strong-line calibrations by  \textcolor{blue}{SB98F1} consistent with \citet{Armah+23}.

The different strong-line methods, empirical \citep[i.e. based on the $T_{\rm e}$-method, e.g. ][and references therein]{Marino+13,Curti+17}, hybrid or semi-empirical \citep[based on the $T_{\rm e}$-method and photoionization models, e.g.][]{Alloin+79,Denicol+02,Carvalho+20,Dors21,Oliveira+22}, theoretical \citep[based on photoionization models, e.g.][]{McGaugh+91,sb98,Kewley+02,Tremonti+04,Nagao+06,Dopita+16} and Bayesian approach (based on simultaneous fits of most strong emission lines with stellar evolutionary synthesis, e.g.  \texttt{IZI}: \citealt{Blanc+15}; \texttt{BEAGLE}: \citealt{Chevallard+16}; \citealt{VidalGarcia+22}; \texttt{NebulaBayes}: \citealt{Thomas+18};   \texttt{BAGPIPES}: \citealt{Carnall+19} and  \texttt{HCM}: \citealt{PerezMonteiro+19}), as well as metal-recombination lines \citep[e.g.  C{\ii}$\lambda$4267 and O{\ii}$\lambda$4650; see][]{Peimbert+93,Esteban+14},  used in the derivation of metallicities from galaxy spectra, each suffer from its own known limitations, which are likely due to the origins of the discrepancies between the different metallicity calibrations. Nevertheless, the difference ($\sim 0.18$ dex) between the metallicities from the two strong-line calibrations used in this work is in order of  the uncertainty of  abundances via strong line methods
\citep[e.g.][]{Denicol+02, Marino+13}. Moreover, different metallicity calibrations, even when based on the same diagnostics, are usually not consistent with one another irrespective of the ionization mechanism and result in systematic metallicity differences from -0.09 to 0.8~dex  \citep[e.g.][]{Kewley+08,Blanc+15, Bian+17, Dors+20a},  which is consistent with the difference between our estimates. 
We find the nuclear region (AGN-dominated) has lower metallicity than the outer parts, which could possibly be due to the addition of metal-poor gas to the centre of the galaxies.  A low-redshift galaxy is more evolved, indicating that it has undergone a certain amount of chemical enrichment, which
depends principally on its stellar mass, but its current (low) gas-phase metallicity is strongly affected by the inflow of pristine hydrogen gas, which also drives up its star formation.
We attribute this metallicity offset between the inner and outskirts to unusually high gas inflows needed to trigger the star formation, which simultaneously lower their metallicity. 

\begin{figure*}
\includegraphics[width=2.1\columnwidth]{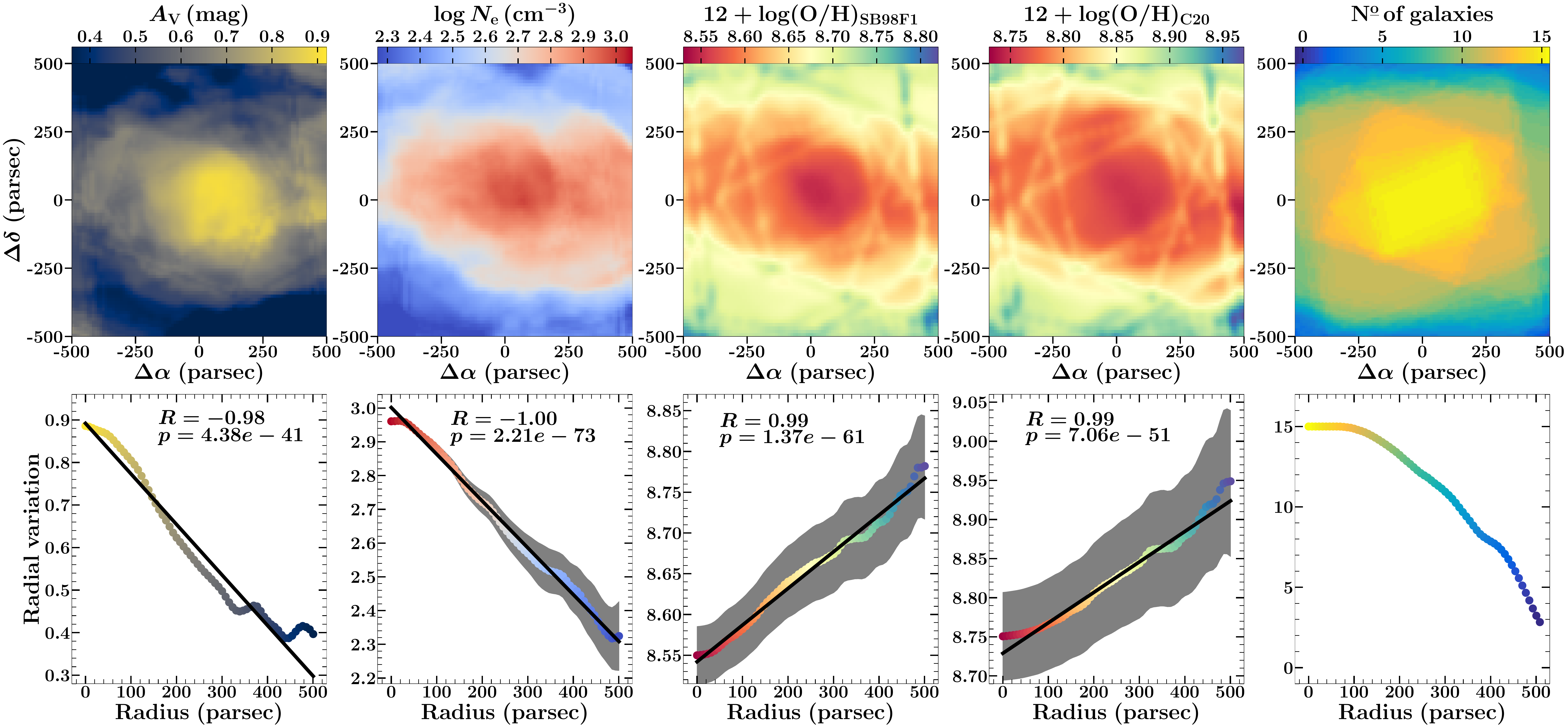}
\caption{Upper panels: from left to right represent averaged properties of dust, electron density and metallicity distributions (using the calibrations by  \textcolor{blue}{SB98F1}
 and   \textcolor{blue}{C20}, respectively) and the number of galaxies contributing to each spatial region. Lower panels: radial profiles from the mean stacked distributions in the upper panels. Each data point represents the average value calculated within a specific annulus for each 10 pc radial bin. The black solid line represents the best-fit regression over all data points using 1000 bootstrap realisations. The curves correspond to the average profiles, and the shaded regions represent 1$\sigma$ uncertainties.  The 
Pearson correlation coefficients with their corresponding {\it p-values} are indicated. The strongly positive (inverted) radial metallicity gradients possibly arise from accretion of pristine gas to the central regions of the galaxies.}
\label{avg}
\end{figure*}

The general positive metallicity radial profile observed in Fig.~\ref{avg}, refers to the trend where the metallicity increases with distance from the galactic centre. Several factors, including but not limited to AGN activity, star formation, gas dynamics and the galaxy environment, contribute to the positive metallicity radial profile observed in Seyfert galaxies. The intense radiation and energetic outflows from the AGN can enrich the ISM with metals. As the radiation and outflows propagate outward from the central black hole, they deposit metals into the surrounding gas, leading to an increase in metallicity with distance from the galactic centre. Seyfert galaxies often exhibit enhanced levels of star formation in their central regions, where AGN activity is most prominent. Stars produce metals through nuclear fusion in their cores and release them into the ISM through stellar winds and supernova explosions. The central concentration of star formation contributes to the higher metallicity observed near the galactic centre.  The radial distribution of metals in Seyfert galaxies can also be influenced by the gravitational potential of the galaxy and dynamical processes such as gas inflows, outflows, and radial mixing. These processes can redistribute metals throughout the galaxy, leading to variations in metallicity as a function of radius. The metallicity radial profile in Seyfert galaxies may be influenced by the surrounding galactic environment, including interactions with neighboring galaxies, gas accretion from the cosmic web, and environmental effects such as ram pressure stripping or galaxy mergers. These external factors can contribute to the enrichment and redistribution of metals within the galaxy.   

The negative and positive radial profiles of  visual dust extinction and gas-phase metallicity (see Fig.~\ref{avg}) respectively, suggest that as metallicity decreases, the amount of dust in the AGN vicinity tends to increase, leading to greater attenuation of emission lines used in metallicity diagnostics. This aligns with the understanding that dust grains primarily form in the atmospheres of evolved stars, which are more prevalent in regions of higher metallicity \citep[e.g.][]{Dwek+98}. However, the correlation suggests a potential link between metallicity and dust production/destruction processes. Higher metallicities could promote dust formation through the availability of heavier elements, while the harsh AGN environment, characterized by intense radiation fields and high-energy phenomena, might lead to increased dust destruction through processes such as sputtering by energetic particles or sublimation by intense radiation \citep[e.g.][]{Feltre+16}.

Similarly, the negative and positive radial profiles of electron density and gas-phase metallicity (see Fig.~\ref{avg}), imply that metallicity decreases with increasing $N_{\rm e}$, which reveals insights into the physical conditions of the gas in the NLR of AGN. Metal-rich gas tends to concentrate in denser regions, possibly due to mechanisms like shock compression or radiation pressure \citep[e.g.][]{Dopita+13}.
Higher electron densities are often associated with more highly ionized gas. The observed correlation suggests that metal-rich gas may be preferentially located in regions with lower ionization levels, potentially due to shielding by dust or variations in the radiation field. The correlation also indicates that the processes responsible for metal enrichment in AGN are less efficient in high-density environments. This can be attributed to the nuclear suppression of star formation or the dilution of enriched material by inflows of lower-metallicity gas.

The observation of nearly constant stellar metallicity within the FoV \citep{DahmerHahn+22}, while gas-phase metallicity increases with distance from the nucleus, presents an intriguing puzzle with several important implications. This decoupling of stellar and gas-phase metallicities suggests that these two components have experienced distinct enrichment histories \citep[e.g.][]{Asari+07,Riffel+24}. Stellar metallicity reflects the chemical composition of stars formed at earlier epochs, while gas-phase metallicity is influenced by ongoing processes like star formation and gas flows.
The increasing gas-phase metallicity with radius indicates that the central AGN is driving metal-enriched outflows \citep[e.g.][]{Davies+14}. Conversely, the constant stellar metallicity suggests that star formation in the central regions has been suppressed or quenched, possibly due to AGN feedback. This feedback could heat or expel the gas, preventing further star formation and maintaining a constant stellar metallicity.
The observed gas-phase metallicity trends can also be influenced by the accretion of metal-poor gas from the circumgalactic medium or the interaction between different gas phases within the galaxy.

\begin{figure}
\includegraphics[width=1.\columnwidth]{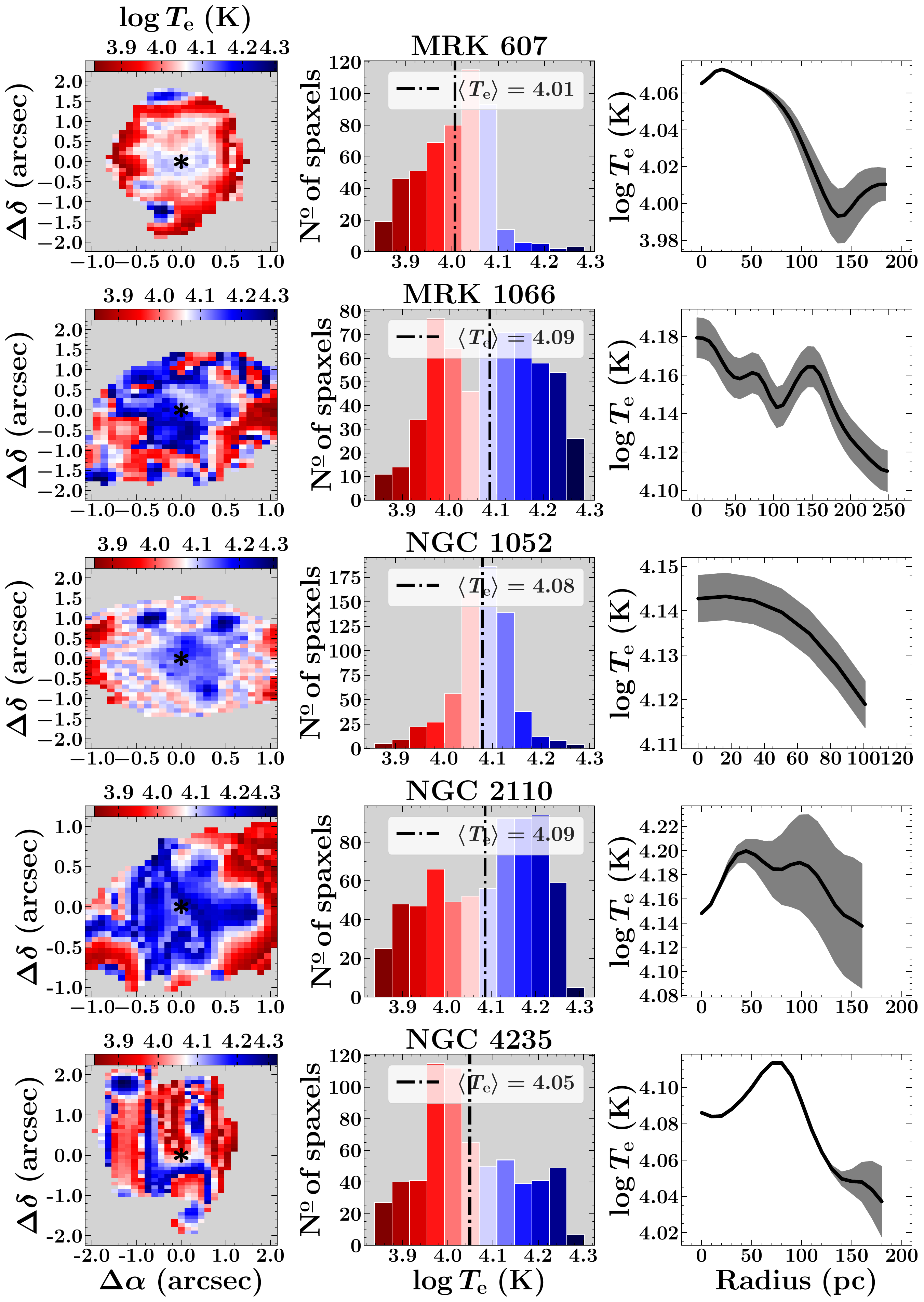}
\caption{The three column panels are maps, histograms, and radial distributions of the electron temperature [$T_{\rm e}\,(\rm K)$ in logarithmic scale] derived from  [O\iii]$\lambda5007/\lambda4363$.  The black solid curves correspond to the average profile. The shaded regions in the radial profiles represent 1$\sigma$ uncertainties. }
\label{all_te}
\end{figure}

\begin{figure}
\includegraphics[width=1.\columnwidth]{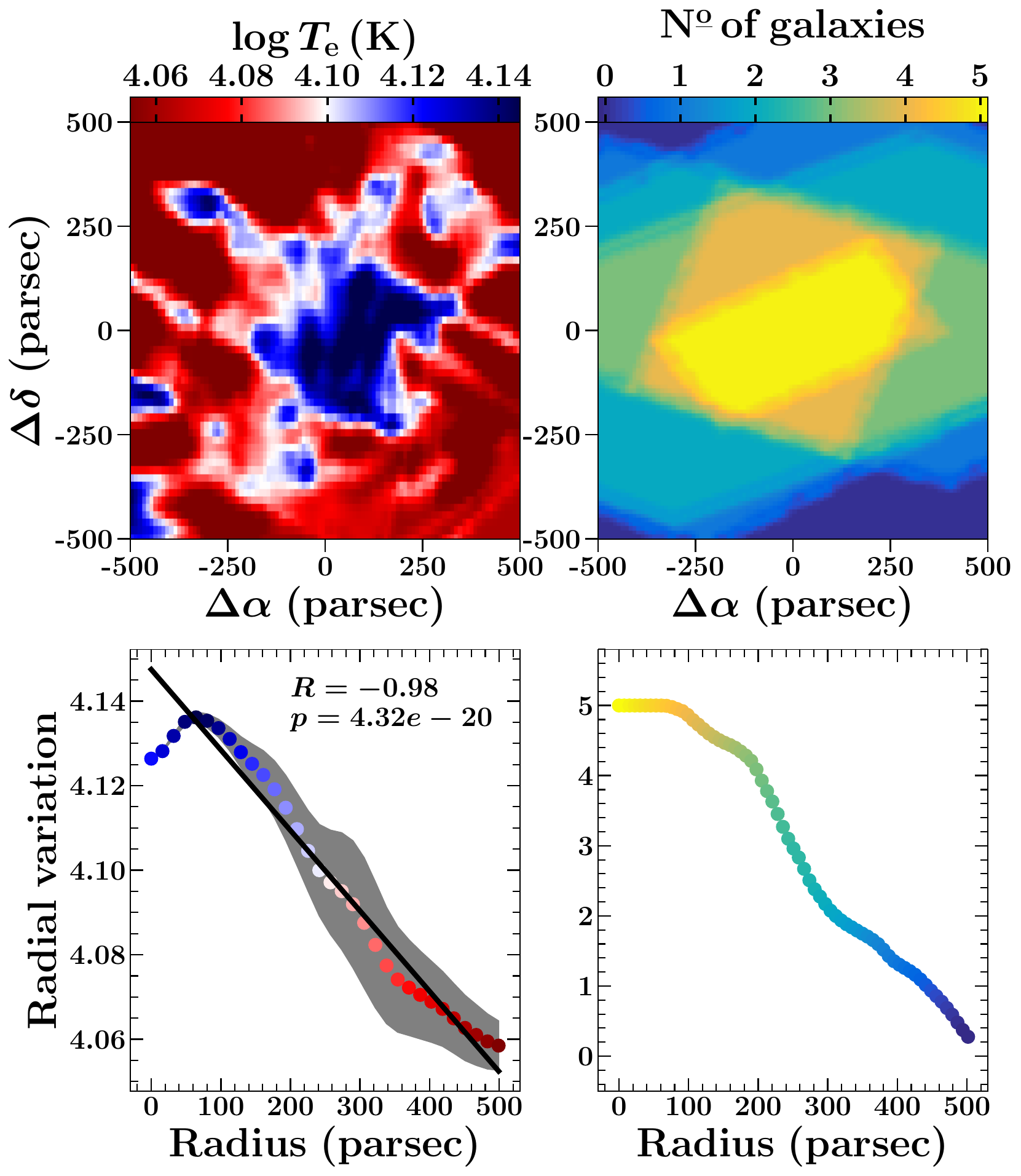}
\caption{Same as Fig.~\ref{avg} but for the electron temperature estimates from Mrk\,607, Mrk\,1066, NGC\,1052, NGC\,2110 and NGC\,4235. }
\label{te_avg}
\end{figure}

To test for the accuracy and reliability of the positive radial metallicity profiles from the strong-line methods, we estimated the electron temperature in 5 (out of 9) targets from the GMOS observations ($7\,000 \lesssim T_{\rm e} {\rm (K)}\lesssim 20\,000$; see Fig.~\ref{all_te}). In Fig.~\ref{te_avg}, we show the stacked and mapped average electron temperature.  The negative radial $T_{\rm e}$ profile has several contributing factors and implications for the gas-phase metallicity via the $T_{\rm e}$-method. The primary source of ionization in Seyfert galaxies is the central AGN, which emits a strong radiation field. As the distance from the central source increases, the radiation field intensity diminishes, leading to lower electron temperatures in the outer regions.  
Lower densities imply reduced collisional excitation and de-excitation rates, further contributing to a decrease in electron temperature. Thus, the decreasing density contributes to the observed negative $T_{\rm e}$ radial profile.  In the case of negative electron density profiles, the lower densities in the outer regions can lead to an overestimation of metallicity when using the $T_{\rm e}$-method. This is because lower densities generally correspond to higher ionization parameters, which can mimic the effects of higher metallicity in the $T_{\rm e}$-method diagnostics. In some cases, shocks and turbulence induced by AGN outflows or other dynamical processes can heat the gas in the central regions, leading to a higher temperature compared to the outer parts.

The negative $T_{\rm e}$ gradient significantly impacts the determination of gas-phase metallicity using the direct method ($T_{\rm e}$-method). The $T_{\rm e}$-method relies on the measurement of temperature-sensitive emission line ratios, which are mildly influenced by electron density. The intensities of the required emission lines for the gas-phase metallicity estimation depend exponentially on the temperature of the gas, therefore, higher electron temperature values lead to lower metallicity estimates and vice versa \citep[e.g.][]{Dors+20b}. The lower electron temperatures in the outer regions favour the production of lower ionization species, typically associated with lower metallicities. If this gradient is not accounted for, metallicity estimations can lead to systematic errors and misinterpretations of the chemical enrichment history of galaxies \citep[e.g.][]{Revalski+18b, RiffelRA+21b}, especially in the outer galactic regions. Therefore, the negative $T_{\rm e}$ radial profile in Fig.~\ref{te_avg} signifies that metallicity via the $T_{\rm e}$-method increases with distance, consistent with the  metallicity profiles via the strong-line methods shown in Fig.~\ref{avg}.  

\subsection{Metallicity and Eddington ratio}
To better understand the evolutionary trajectories of AGNs, we investigate the NLR physics within the context of the co-evolution of SMBHs and galaxies \citep[e.g.][and references therein]{Kormendy+13}. The two primary components of the NLR model are a star-forming disc that continuously creates metals, which are then carried outward by AGN outflows, and stratified cold clouds created by AGN outflows, as seen by the typical blueshift of the high ionization lines such as [O III] \citep[e.g.][]{Komossa+08,Trindade+21}.
The forbidden [O III] emission lines are often regarded as a good tracer of the extended ionized gas since they cannot be produced in the high-density at sub-parsec scales \citep[e.g. $R_{\rm BLR} < 1\,{\rm pc}$;][]{Suganuma+06}--broad line regions of AGNs. The [O III] line profile in AGNs can exhibit a broad and/or blue-wing asymmetry that is typically attributed to an outflowing gas. Several studies have used multicomponent fitting of the [O III] line to describe the outflow in terms of its blue-wing properties \citep[e.g.][and references therein]{RuschelDutra+21,Kakkad+22,Oio+24}. Gas-phase metallicity gradients can be influenced by AGN activity, particularly through AGN-driven outflows. These outflows can expel metal-enriched gas from the central regions of the galaxy into the surrounding ISM. As a result, the metallicity distribution within a galaxy may exhibit variations, with potentially lower metallicities observed in regions affected by AGN outflows. Alternatively, the radiation from AGN can enhance the metallicity of the gas through processes such as photoionization and heating.
It is believed  that gas accretion, which results in a luminous AGN with a significant energy release in the form of radiation, is the main process driving the growth of SMBHs. Additionally, galaxy metallicity is an inherent metric that tracks the evolution of the system, which shows how the ISM becomes more elementally rich over cosmic time.
The evolution and growth of galaxies, as well as their transformation from star-forming to quenched systems, are regulated by the flows of the gas supply. The local cycle of baryons between the galaxy ISM and CGM regulates the star formation processes, provides the supply of new material and distributes the newly formed metals.

\begin{figure*}
\includegraphics[width=2.\columnwidth]{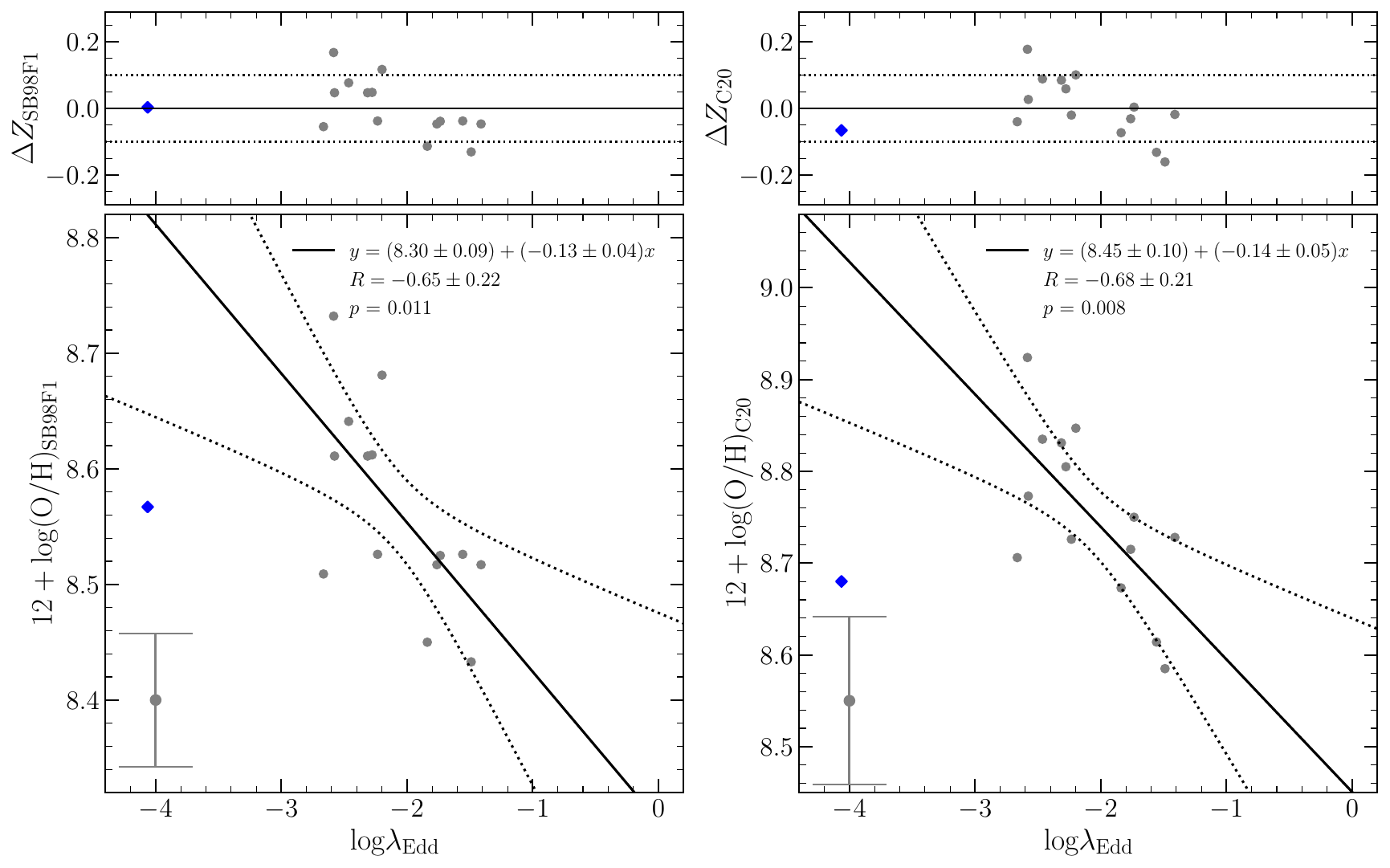}
\caption{Upper  panels: the relation between
the metallicity gradients ($\Delta Z$ from  \textcolor{blue}{SB98F1} and \textcolor{blue}{C20}) and ${\rm \lambda_{Edd}}$, as indicated. Lower panels: 
the relation between the mean metallicity from the calibrations  (\textcolor{blue}{SB98F1} and \textcolor{blue}{C20})  and the Eddington ratio from the intrinsic  X-ray luminosity ($ L_{14-150}$), shown in Table~\ref{tab1}. The error bars in the lower left corners of the lower panels represent the average value of the uncertainty in the y-axis. The solid line represents the best-fit regression over all data points. The 95 per cent
level confidence boundaries around the best-fit are indicated with dashed lines. The correlation parameters  are indicated in each panel.  All our estimates are generally consistent within the 95 per cent level of confidence boundaries, but the NGC\,1052 represented by the blue diamond marker appears to be an outlier. The 
Pearson correlation coefficients and {\it p-values} are indicated in the upper right corners of the lower panels. The  anti-correlation from $Z$-${\rm \lambda_{Edd}}$
relation is consistent with the positive gas-phase metallicity profiles in Fig.~\ref{avg}.}
\label{edd}
\end{figure*}

In order to test for a possible influence of the AGN activity on the metallicity of the host galaxies, we compare the  metallicity with the Eddington ratio, which is a function of luminosity and $M_{\rm BH}$ since there is a similar distribution on the plane log$\lambda_{\rm Edd} \times \log L_{\rm Bol}$ between the two calibrations \citep[see][for details]{Armah+23}.  Fig.~\ref{edd} compares the metallicities and Eddington ratios ($Z$-${\rm \lambda_{Edd}}$), and shows that they are strongly correlated given the regression coefficients ({\it R}$\sim -0.65$ and {\it R}$\sim -0.68$) via the strong-line calibrations by  \textcolor{blue}{SB98F1}
and   \textcolor{blue}{C20},  and the null probability of no correlation ($p$ = 0.011 and $p$ = 0.008)  respectively. The  metallicity and Eddington ratio estimates are generally consistent except for NGC\,1052, which can be attributed to its dual Seyfert/LINER classification, implying that it harbours a less active AGN in comparison with the other objects as noted by \citet{DahmerHahn+22}.  The  correlations do exist, however, larger sample size would be beneficial for a better significant $p$-value. Additionally, in Fig.~\ref{edd}, we show the relation between
the residual of metallicity, which represent the difference between the mean metallicity and the metallicity estimates ($\mathrm{\Delta[12+log(O/H)]}$) and ${\rm \lambda_{Edd}}$.  The  $Z$-${\rm \lambda_{Edd}}$ and $\Delta Z$-${\rm \lambda_{Edd}}$ relations follow the same negative correlation, which arise due to various factors such as the preferential consumption of metal-rich gas by the AGN, dilution of metal-rich gas by newly accreted material with lower metallicity, or differences in the metallicity of gas reservoirs feeding the AGN.

The host galaxy must accrete from the Intergalactic Medium (IGM) to replenish the gas reservoir in its ISM in order to continue its star formation. In turn, AGN activity and star formation create galactic winds that drive metal-rich material out of the system and back into the CGM, possibly also removing it completely from the system, affecting the available gas reservoirs in the ISM and potentially preventing further accretion. Thus, the negative feedback process is initiated by the re-deposition of energy and momentum into the ISM of the galaxy. This cycle of gas plays a crucial role in the processes driving galaxy evolution and star formation. On the other hand, the inflow of metal-poor gas from the IGM and accreted through the CGM onto the galaxy dilutes the existing nuclear metal-rich content, potentially inducing powerful feedback processes that regulate the growth of the host galaxy \citep[e.g.][]{Dekel+09}. These feedback mechanisms can prevent excessive star formation and overcooling of the gas, while also contributing to self-regulation of black hole growth \citep[e.g.][]{Springel+05, Sijacki+07M,Hopkins+11}. This suggests that AGNs can expel metals from their host galaxies to a degree that impacts their chemical evolution. However, it is important to note that if a galaxy is actively forming stars while also losing metals through outflows, a significant drop in metallicity might not be observed as long as the enrichment from star formation counterbalances the metal ejection.  

The most important factors responsible for setting and disrupting metallicity scaling relations are based on different physical processes, including but not limited to: (i) gas consumption resulting from metal-poor or metal-rich gas inflow  from the IGM and accreted through the CGM onto the galaxy; (ii) star-formation enriched material expelled from the galaxy in the form of galactic winds, part of which is re-accreted into the ISM as galactic fountains; (iii) the mixed diffuse gas filling the CGM; and (iv) AGN triggering mechanisms. Each of the aforementioned processes follows its own  unique path. The impact of metal-poor gas accretion from cold filaments strongly depends on the exact location of material deposit--whether gas is deposited in a compact bulge-like or in a more extensive disk region. For materials deposited directly into the centres of galaxies (or radially mixed on quick timescales),  the central metallicity undergoes dilution and flattens metallicity radial profile.  On the other hand, if the metal-poor gas is deposited at the outskirts of galaxies, it should create a negative metallicity radial profile. However, positive metallicity gradients are produced by accretion of external pristine gas towards the central regions of galaxy while creating negative $Z$-${\rm \lambda_{Edd}}$ correlation. Moreover, an inflow of a metal-poor gas from the outskirts of the galaxy or direct accretion of gas from the cosmic web  as a result of AGN activity will still lead to anti-correlation in  $Z$-${\rm \lambda_{Edd}}$ relation. The formation of galactic fountain which takes place when the ejection from stellar feedback is not powerful enough to expel the gas from the CGM of the galaxy and metals ejected into the CGM fall back onto the outskirts of galaxies. In this scenario, a usual flattening of metallicity radial profile will occur as a result of re-accretion of the galactic fountain \citep[e.g.][]{Ma+17}.  Metals in the outer regions of the galaxy originate from the highly enriched regions of the inner part of the galaxy through either radial flows or pristine gas accretion.  When two galaxies merge, their individual black holes are brought together, causing SMBHs to grow. Since galactic fountains acquire angular momentum via mixing of low angular momentum, wind-recycled gas with high angular momentum gas in the CGM, mergers boost gas accretion activity by helping to align the disc and CGM rotation axes, as well as cause the  galactic fountain to lose angular momentum. Therefore, mergers may induce instabilities that can funnel a large amount of gas into the central region of a galaxy, thereby inducing a starburst \citep[e.g.][]{Sanders+88,Pierce+23}. The negative correlation in  $Z$-${\rm \lambda_{Edd}}$ can be linked to the mechanisms triggering AGN activity. Galaxy interactions or mergers might trigger both AGN activity and enhanced star formation, leading to higher $Z$. However, as the AGN becomes more luminous (higher Eddington ratio), it may start to dominate over star formation, leading to a decrease in $Z$ despite the ongoing AGN activity.

We argue that, the inflow of metal-poor gas into the centres of galaxies considered here is an ongoing process. It is plausible to expect that the rate at which metals are re-distributed in galaxies and the ISM is diluted by metal-poor gas accretion would drop with time as the prevalence of all of the disrupting metallicity scaling relation mechanisms such as star formation, gas accretion and mergers decline. 
The AGN accretes more gas at higher Eddington ratios and consumes the available gas reservoir more rapidly. This rapid gas consumption reduces the amount of gas available for star formation, which in turn lowers the metallicity of the remaining gas reservoir due to reduced  enrichment from stellar nucleosynthesis. The anti-correlation in $Z$-${\rm \lambda_{Edd}}$ indicates more luminous AGN corresponding to higher Eddington ratios, are associated with weaker metallicity  enhancement mechanisms. These mechanisms can include the injection of metal-poor gas into the ISM through AGN-driven outflows, as well as reduced enrichment of gas from less efficient supernovae explosion and stellar winds processes. Higher gas-phase metallicities in galaxies hosting more luminous AGNs reflect the evolutionary history of these galaxies. 

\section{Concluding remarks}
\label{concl}
In this work, we present integral field unit spectroscopy of Seyfert galaxies using data from GMOS and MUSE. The data  allowed us to analyse the emission line ratios and BPT diagrams showed the presence of AGN activity or/and shocks as the ionization sources in these galaxies. Additionally, we used the data to derive metallicity via the strong-line methods from the nebular emissions based on regions in the galaxies that are predominantly ionized by AGNs. We derived metallicity distributions
 across the field of view for individual AGN data points.  We summarize our main findings below.
 \begin{enumerate}
  \item  Comparing the strong-line methods used in determining the oxygen abundances for individual spaxels, the metallicity calibration indicators, in particular, the metallicities from [N\ii]$\lambda$$\lambda$6548,6584/H$\alpha$ and [O\iii]$\lambda$$\lambda$4959,5007/H$\beta$ tracers of oxygen abundance were  considered. 
 We estimated the gas-phase metallicity and found mean values for the oxygen dependent ($Z \sim 0.75Z_\odot$) and  nitrogen dependent ($Z \sim 1.14Z_\odot$) calibrations. We find that the metallicity distribution shows excellent agreement  with $\Delta Z \approx 0.19$ dex and $\Delta Z \approx 0.18$ dex between the mean values from the two strong-line calibrations for GMOS and MUSE respectively, which is consistent with  the order of uncertainty in metallicities via the strong-line methods from integrated observations such as long-slit or single fiber.
Our results show that the traditional  strong-line calibration procedures which have been used for the derivation of metallicity from single-aperture or long-slit spectrographs observational data are equally useful for metallicity estimates in spatially resolved IFU data, where it is often impossible to detect weak emission lines or as a result of observational wavelength coverage limitations, especially for the MUSE observation. 
\item  Despite the aforementioned drawback, we found that in almost all cases, the galaxies ($15/15 - 100\,\%$ from \textcolor{blue}{SB98F1} and $13/15 \sim 87\,\%$ from \textcolor{blue}{C20}) in our sample have strong positive gas-phase metallicity gradients.  This result is generally consistent with previous results in star-forming galaxies at high redshift from the literature \citep[$z\gtrsim2$; e.g.][]{Curti+20, Wang+20, Simons+21,Li+22,Venturi+24} and nearby AGN \citep[e.g.][]{Amiri+24} but in marked contrast with nearby ($z\sim0$) star-forming galaxies  \citep[e.g.][]{Belfiore+17} and AGNs \citep[$\sim66\,\%$, $\sim9\,\%$ and $\sim25\,\%$ of AGN hosts exhibiting negative, flat and positive gradients, respectively;][]{Nascimento+22}. 
\item  We found a significant negative correlation  between  metallicity and $\lambda_{\rm Edd}$ (metallicity decreases with increasing Eddington ratio).  We argue that, a drop in metallicity from the centre to the outskirt of a galaxy could be evidence for the influence of either pristine gas inflows or galactic outflows triggered by starbursts. The existence of a negative correlation between  metallicity and $\lambda_{\rm Edd}$ is  driven by the X-ray luminosity, indicating that the AGN is driving the  chemical enrichment of its host galaxy, as a result of the  inflow of pristine gas that is diluting the metal-rich gas, reiterating the nature of the metallicity radial profile. We confirm the metallicity--Eddington ratio relation using analysis of IFU data, which agrees with previous
results based on single-fibre spectroscopic surveys. By successfully demonstrating the power of IFU data to reveal the metallicity distribution in nearby Seyfert galaxies, our finding has yielded invaluable insights into the metallicity distribution cycle in nearby galaxies. 

While gas-phase metallicity gradients provide valuable insights into galaxy evolution, their utility is constrained by model assumptions, observational limitations, and the complex interplay of various physical processes within galaxies. The interpretation of metallicity gradients in the context of galaxy evolution requires careful consideration of various factors (including but not limited to mergers, interactions, gas dynamics, star formation efficiency, and influence of the environment), making it challenging to disentangle the specific contributions of different processes to the observed gradients.

\end{enumerate}

\section*{Acknowledgements}

We express our gratitude to the anonymous referee for their insightful comments and suggestions. MA gratefully acknowledges support from Coordenação de Aperfeiçoamento de Pessoal de Nível Superior (CAPES, Proj. 88887.595469/2020-00). 
RR thanks  Conselho Nacional de Desenvolvimento Cient\'{i}fico e Tecnol\'ogico  (CNPq, Proj. 311223/2020-6,  304927/2017-1 and  400352/2016-8), Funda\c{c}\~ao de amparo \`{a} pesquisa do Rio Grande do Sul (FAPERGS, Proj. 16/2551-0000251-7 and 19/1750-2), Coordena\c{c}\~ao de Aperfei\c{c}oamento de Pessoal de N\'{i}vel Superior (CAPES, Proj. 0001). RAR acknowledges support from Conselho Nacional de Desenvolvimento Cient\'ifico e Tecnol\'ogico (CNPq; Proj. 303450/2022-3, 403398/2023-1, \& 441722/2023-7), Funda\c c\~ao de Amparo \`a pesquisa do Estado do Rio Grande do Sul (FAPERGS; Proj. 21/2551-0002018-0), and CAPES (Proj. 88887.894973/2023-00).  ARA acknowledges partial support for this work from the Conselho Nacional de Desenvolvimento Científico e Tecnológico (CNPq) under grant number 313739/2023-4.

%%%%%%%%%%%%%%%%%%%%%%%%%%%%%%%%%%%%%%%%%%%%%%%%%%
\section*{Data Availability}

The GMOS data underlying this article are available in the Gemini Observatory Archive,\footnote{\href{https://archive.gemini.edu/searchform}{https://archive.gemini.edu/searchform}} and can be accessed with project numbers GN-2013A-Q-61, GS-2013B-Q-20,  GN-2014B-Q-87, GN-2017B-Q-44 and GS-2018A-Q-225. The MUSE data underlying this article are available in the 
European Southern Observatory Science Archive Facility,\footnote{\href{http://archive.eso.org/cms.html}{http://archive.eso.org/cms.html}} and can be accessed with project numbers 094.B-0298(A), 094.B-0321(A), 097.B-0640(A), 098.B-0551(A) and 099.B-0242(B). Post-processed data and analysis codes will be shared on reasonable request with the corresponding author.



\bibliographystyle{mnras}
\bibliography{ref} % if your bibtex file is called example.bib

%%%%%%%%%%%%%%%%%%%%%%%%%%%%%%%%%%%%%%%%%%%%%%%%%%

%%%%%%%%%%%%%%%%% APPENDICES %%%%%%%%%%%%%%%%%%%%%
\newpage
\appendix
\section{Spatially-resolved map properties from gmos and muse observations}
\label{appendix}
\subsection{Individual galaxy results}
\subsubsection{Mrk\,607  (NGC\,1320)}
Mrk\,607 is a  high-ionization Sa galaxy harboring a Seyfert\,2 nucleus. Close to the central region where the AGN is located, the dust distribution is heavily influenced by the intense radiation emitted by the AGN. Dust grains in this region experience high levels of ionization due to the strong radiation field, leading to the emission of infrared radiation and potentially altering their spectral characteristics. As one moves farther away from the AGN, the influence of its radiation on the dust distribution decreases, and the dust properties resemble those found in regions less affected by AGN activity.  The density of the ISM in Mrk\,607 exhibits a gradient with distance from the AGN.  
Close to the AGN, where AGN-driven outflows are strongest, the density of the ISM is higher probably due to compression and shocks induced by the outflows. Mrk\,607 shows ionized outflows and counter rotation stellar and gas disks \citep[e.g.][]{Riffel+17a, Bianchin+22}. These regions may exhibit enhanced emission in certain spectral lines, such as H$\alpha$ or [O III]. Moving away from the AGN, the density of the ISM decreases, although the exact profile of the density gradient depend on factors such as the strength and orientation of the AGN outflows. The  metallicity depicts positive gradients from both \textcolor{blue}{SB98F1}
and   \textcolor{blue}{C20} strong-line calibrations. The metallicity distributions can be classified as moderate to high metallicity, similar to other spiral galaxies.

\subsubsection{Mrk\,1066}
This galaxy is classified as SB0 and  harbours a Seyfert~2 nucleus.  The negative radial dust extinction in Mrk\,1066 suggests that the central region of the galaxy is more obscured by dust compared to the outer regions. This is consistent with typical observations in many galaxies, where the central regions are often more dust-obscured due to the higher concentration of stars and gas. The negative radial electron density gradient indicates that the electron density decreases as we move away from the galactic centre. This trend is consistent with the general expectation that gas density decreases with increasing radius in galaxies.   The  metallicity depicts positive gradients from both \textcolor{blue}{SB98F1}
and   \textcolor{blue}{C20} strong-line calibrations, respectively. 

\subsubsection{NGC\,1052}
NGG\,1052 is a giant E4 galaxy \citep{Forbes+01,Xilouris+04}, with an ambiguous LINER/Seyfert classification, well known for having one of the nearest radio-loud AGN \citep{Heckman+80,Ho+97,Riffel+17a}.  As with other elliptical galaxies, its properties differ from those of spiral galaxies. Elliptical galaxies like NGG\,1052 typically have very little interstellar dust compared to spiral galaxies. The absence of significant dust lanes is a characteristic feature of elliptical galaxies. However, we observe the presence of moderate dust, primarily concentrated in the central regions of the galaxy. The density distribution is typically smooth and symmetric. Moreover, Elliptical galaxies lack the distinct spiral arms seen in spiral galaxies, so their density distribution tends to be more uniform. However, we observe a complex density distribution in NGG\,1052, with a higher concentration at the outskirts and decreasing towards the central regions. Like the dust and density distributions, elliptical galaxies generally exhibit a relatively uniform metallicity throughout their stellar populations. However, there can be variations in metallicity, especially in the central regions where more metal-rich gas may be found due to previous star formation episodes or interactions with other galaxies. The metallicity distribution in NGG\,1052 is uneven, but follows the same aforementioned distribution patterns as the dust and density, which can be attributed to its dual LINER/Seyfert classification, implying that it harbors a less active AGN compared to the rest of the sample.  The  metallicity depicts positive gradients from both \textcolor{blue}{SB98F1}
and   \textcolor{blue}{C20} strong-line calibrations.

\subsubsection{NGC\,2110}
NGC\,2110 is classified as SAB0, harboring a Seyfert\,2 nucleus.  It shows moderate amounts of interstellar dust distribution, which is concentrated in the vicinity of the AGN and along its spiral arms. The density distribution varies, with higher densities observed in the central regions consistent with similar estimate by \citet{Peralta+23}.  The  metallicity depicts positive gradients from both \textcolor{blue}{SB98F1}
and   \textcolor{blue}{C20} strong-line calibrations.
NGC2110 show inflows in molecular gas and outflows in ionized gas \citep{Diniz+15}.

\subsubsection{NGC\,3516}
NGC\,3516 is morphologically classified as SB0. It exhibits the characteristics of  a Seyfert\,1.5 AGN, with intense radiation emitted from the central SMBH. It shows moderate amounts of interstellar dust distribution, which is concentrated in the central regions around the AGN and along its spiral arms. The density distribution vary with higher densities observed in the central bulge and regions surrounding the AGN.  The  metallicity depicts positive gradients from both \textcolor{blue}{SB98F1}
and   \textcolor{blue}{C20} strong-line calibrations.

\subsubsection{NGC\,3786}
This galaxy has SABa classification, with a Seyfert\,1.8 nucleus. It exhibits characteristics typical of barred spiral galaxies, including a central bar structure and spiral arms. The dust in NGC\,3786 is likely concentrated in its spiral arms and central regions, where active star formation occurs. Interstellar dust plays an important role in the formation of new stars by providing the raw material necessary for the process. The density distribution follows the typical pattern observed in barred spiral galaxies, with the usual higher densities concentrated along the central bar and in the spiral arms. The presence of a bar structure can lead to enhanced star formation activity and gas dynamics within the galaxy. 
The metallicity depicts positive and negative gradients from  \textcolor{blue}{SB98F1}
and   \textcolor{blue}{C20} strong-line calibrations, respectively.  The gas-phase metallicity in Mrk\,1066, tends to be highest in the central regions and decreases toward the outer regions in one phase and vice versa.  This radial gradient discrepancy is probably caused by a combination of factors, including but not limited to different metallicity calibrations, variations in the star formation efficiency, electron density, radial mixing of gas, and gas inflows and outflows (see \S~\ref{res_1} for details).

\subsubsection{NGC\,4235}
NGC\,4235 is an SAa galaxy hosting a Seyfert\,1 active nucleus. As with other spiral galaxies, its properties, including dust, density and metallicity distributions, are influenced by its structure and evolutionary history. It contains significant interstellar dust distributed throughout its spiral arms and central regions. Dust play a significant role in the evolution of the galaxy by absorbing and scattering starlight, affecting observations across different wavelengths. The density distribution in NGC\,4235 follows the typical pattern observed in spiral galaxies, with higher densities concentrated in the central bulge and along the spiral arms. Regions of active star formation may exhibit higher densities due to the presence of young massive stars.
Similar to NGC\,3786, the metallicity depicts positive and negative gradients from  \textcolor{blue}{SB98F1}
and   \textcolor{blue}{C20} strong-line calibrations, respectively.

\subsubsection{NGC\,4939}
NGC\,4939 is classified as SAbc galaxy, specifically Seyfert\,1, indicating the presence of an active AGN obscured by significant amounts of gas and dust. While specific information on the dust, density and metallicity distributions in NGC\,4939 may vary depending on the observational data and studies, here we provide details based on optical observational study.  NGC\,4939 contains significant amounts of interstellar dust, specifically in the central regions surrounding the AGN. The distribution of dust is likely influenced by factors such as interactions with the AGN, ongoing star formation and the overall morphology of the galaxy. It shows  higher densities in the central regions due to the presence of the AGN and ongoing star formation.
The  metallicity depicts positive gradients from both \textcolor{blue}{SB98F1} and  \textcolor{blue}{C20} strong-line calibrations.

\subsubsection{NGC\,5899}
NGC\,5899 is a barred spiral galaxy (SABc), hosting a Seyfert\,2 AGN. 1. Dust in galaxies like NGC 5899 tends to trace the distribution of gas and stars. The distribution of dust in NGC\,5899 reflects its spiral structure, with higher concentrations in the arms and lower concentrations in the interarm regions.  The electron density just like the dust distribution, is highest in the spiral arms and regions of ongoing star formation.
Like most of the targets, the  metallicity presents positive gradients from both \textcolor{blue}{SB98F1} and  \textcolor{blue}{C20} strong-line calibrations, respectively. 

\subsubsection{NGC\,1068}
It is classified as Seyfert~2, with an SAb Hubble classification. It is one of the nearest and brightest Seyfert galaxies \citep[e.g.][]{Meyer+04,Murase+16}, known for its active core and intense star formation activity.  A drop in both $A_{\rm V}$ and $N_{\rm e}$ is detected from the nucleus. The fact that we detected a dust extinction drop in the nucleus of this galaxy and others means that the central regions are so heavily obscured that the light observed should be unquenched stellar emission in the leading edge of the host galaxy, however, it is mainly composed of AGN photoionized gas. Therefore, this drop could be due to energy transport as a result of shock effect on the NLR ionization \citep[e.g.][]{Mizumoto+24}.  The density distribution in it exhibits a barred spiral structure, with higher densities observed along the central bar and in the spiral arms. The presence of a bar in the structure of the galaxy can lead to enhanced star formation activity and gas dynamics. The metallicity distribution indicates metal enrichment content compared to less active galaxies.
The  metallicity depicts positive gradients from both \textcolor{blue}{SB98F1} and \textcolor{blue}{C20} strong-line calibrations. The AGN activity has likely contributed to the enrichment processes such as supernova explosions and stellar winds.

\subsubsection{NGC\,1194}
NGC\,1194 is a barred spiral galaxy (SA0), with a Seyfert\,2 AGN, which can influence the distributions of dust, density, and gas-phase metallicity within the galaxy. It contains significant amounts of interstellar dust, particularly in the central regions surrounding the AGN.  The intense radiation emitted by the AGN can heat and ionize the surrounding dust grains, leading to the formation of a warm dust component. This warm dust emission can be observed at infrared wavelengths. AGN-driven outflows can also stir up and redistribute dust within the galaxy, affecting its spatial distribution. In NGC 1194, the radiation and outflows from the AGN activity may lead to variations in the distribution and properties of dust, with higher concentrations closer to the central region hosting the AGN.  AGN activity can influence the density distribution of the ISM in NGC 1194 through various mechanisms. AGN-driven outflows can compress and shock the surrounding gas, leading to regions of enhanced density. These density enhancements may be observed as regions of increased emission in certain spectral lines, such as H$\alpha$ or [O III]. Additionally, AGN feedback processes can regulate star formation activity and gas dynamics, further affecting the density distribution within the galaxy.  The  metallicity depicts positive gradients from both \textcolor{blue}{SB98F1} and \textcolor{blue}{C20} strong-line calibrations.

\subsubsection{NGC\,2992}
NGC\,2992 is an edge-on Sa Seyfert galaxy known for its active galactic nucleus and the complex interplay between its multiphase disk and wind. It has a prominent dust lane extending along its major axis, crossing the nucleus. The dust features are indicative of the complex dynamics within the galaxy, influenced by interactions with its companion, NGC\,2993. Observations suggest that the encounter between these galaxies may have triggered the active nucleus in NGC\,2992, with significant amounts of dust obscuring the nucleus and outflowing material from the core. The dust distribution is closely tied to the gas kinematics, with the dusty molecular outflowing clumps and turbulent ionized gas located at the edges of radio bubbles, suggesting interactions through shocks \citep{Zanchettin+23}.
The electron density in NGC\,2992 has been studied through spatially resolved maps, revealing a clumpy ionized wind distributed in wide-opening angle ionization cones extending up to 7 kpc \citep{Zanchettin+23}. The ionized outflow mass and rate have been inferred based on these electron density maps, providing insights into the AGN-driven kpc-scale ionized wind and its interplay with the multiphase disk \citep{Zanchettin+23}. The electron density variations are crucial for understanding the ionization structure and the mechanisms driving the outflows and inflows within the galaxy.
The gas-phase metallicity in NGC\,2992 is influenced by the multiphase disk and wind interaction. The velocity dispersion of the cold molecular phase suggests that the disk-wind interaction locally boosts the gas turbulence \citep{Zanchettin+23}. The metallicity gradients in galaxies like NGC\,2992 are shaped by various factors, including metal production, transport, consumption, and loss. A new model for the evolution of gas-phase metallicity gradients from first principles shows that these gradients depend on ratios describing the metal equilibration timescale and the competition between radial advection, metal production, and accretion of metal-poor gas \citep[e.g.][]{Sharda+21}. The  metallicity depicts positive gradients from both \textcolor{blue}{SB98F1} and \textcolor{blue}{C20} strong-line calibrations.
The gradients can be altered by radial inflows and outflows, which are likely important processes in setting the gas-phase metallicity gradients \citep{Cheng+24}. 

\subsubsection{NGC\,3081}
NGC\,3081 is classified as SAB0, with a Seyfert\,2 nucleus. The AGN activity in NGC\,3081 can have significant influence on the dust, density, and metallicity distributions within the galaxy.
In most spiral galaxies, including NGC\,3081, dust is often found concentrated in the spiral arms and central regions. These dusty regions are associated with active star formation, where newly formed stars heat up the surrounding dust, causing it to emit infrared radiation. The presence of an AGN in NGC\,3081 may also give rise to a dusty torus structure surrounding the central black hole. This torus can obscure the central region from optical view and contribute to infrared emission.  The distribution of dust in NGC\,3081 can be influenced by various factors, including interactions with nearby galaxies, ongoing star formation activity, and the presence of supernova remnants. The density distribution of gas in NGC\,3081 follows the typical pattern observed in spiral galaxies, with higher densities in the central bulge and spiral arms. The metallicity gradients follow the usual spiral galaxy metallicity distribution, with higher metallicities observed in the central regions and decreasing toward the outer disk. The  metallicity depicts positive gradients from both \textcolor{blue}{SB98F1} and \textcolor{blue}{C20} strong-line calibrations.

\subsubsection{NGC\,3393}
This galaxy is classified as SBa, harboring a Seyfert\,2 nucleus. The spaxel profile  indicates that this object has two distinct spiral arms composed of a combination of old and young stellar populations. Close to the central region where the AGN is located, the dust distribution may be heavily influenced by the intense radiation emitted by the AGN. Dust grains in this region experience high levels of ionization due to the strong radiation field, leading to the emission of infrared radiation and potentially altering their spectral characteristics. As one moves farther away from the AGN, the influence of its radiation on the dust distribution decreases, and the dust properties resemble those found in regions less affected by AGN activity.  The density of the ISM in NGC 3393 exhibit a gradient with distance from the AGN. Close to the AGN, where AGN-driven outflows may be strongest, the density of the ISM is higher due to compression and shocks induced by the outflows. These regions exhibit enhanced emission in certain spectral lines, such as H$\alpha$ or [O III]. Moving away from the AGN, the density of the ISM decreases, although the exact profile of the density gradient will depend on factors such as the strength and orientation of the AGN outflows. The gas-phase metallicity in NGC 3393 also vary with distance from the AGN and close to the AGN, where AGN-driven outflows may be prominent, the metallicity of the gas is influenced by the enrichment from accretion disk of the AGN and supernova explosions triggered by AGN activity. These regions exhibit enhanced metallicity compared to more distant regions. However, the metallicity gradient becomes shallower or in a reverse direction at larger distances from the AGN, as the influence of AGN-driven processes diminishes and other factors, such as star formation and chemical enrichment from older stellar populations, become more dominant.  The metallicities from both \textcolor{blue}{SB98F1} and \textcolor{blue}{C20} strong-line calibrations show positive gradients.

\subsubsection{NGC\,5728}
NGC\,5728 is an SABa galaxy, hosting a Seyfert\,2 AGN. The dust distribution reveals significant amounts of interstellar dust, particularly in its spiral arms and central regions. This dust plays a crucial role in star formation processes within the galaxy, as it provides the raw material from which new stars can form. The density distribution follows the typical pattern observed in barred spiral galaxies, with higher densities concentrated in the central bulge and along the spiral arms. The density decreases gradually away from the centre.
It exhibits varied metallicity distribution across its stellar populations. Similar to most of the targets, the  metallicity shows positive gradients from both \textcolor{blue}{SB98F1} and \textcolor{blue}{C20} strong-line calibrations. The central regions of the galaxy tend to have lower metallicities,  likely due to different star formation history or interactions with neighbouring galaxies. In contrast, the outer regions have higher metallicities, indicating enrichment from previous generations of stars.

\begin{figure*}
\includegraphics[width=2.1\columnwidth]{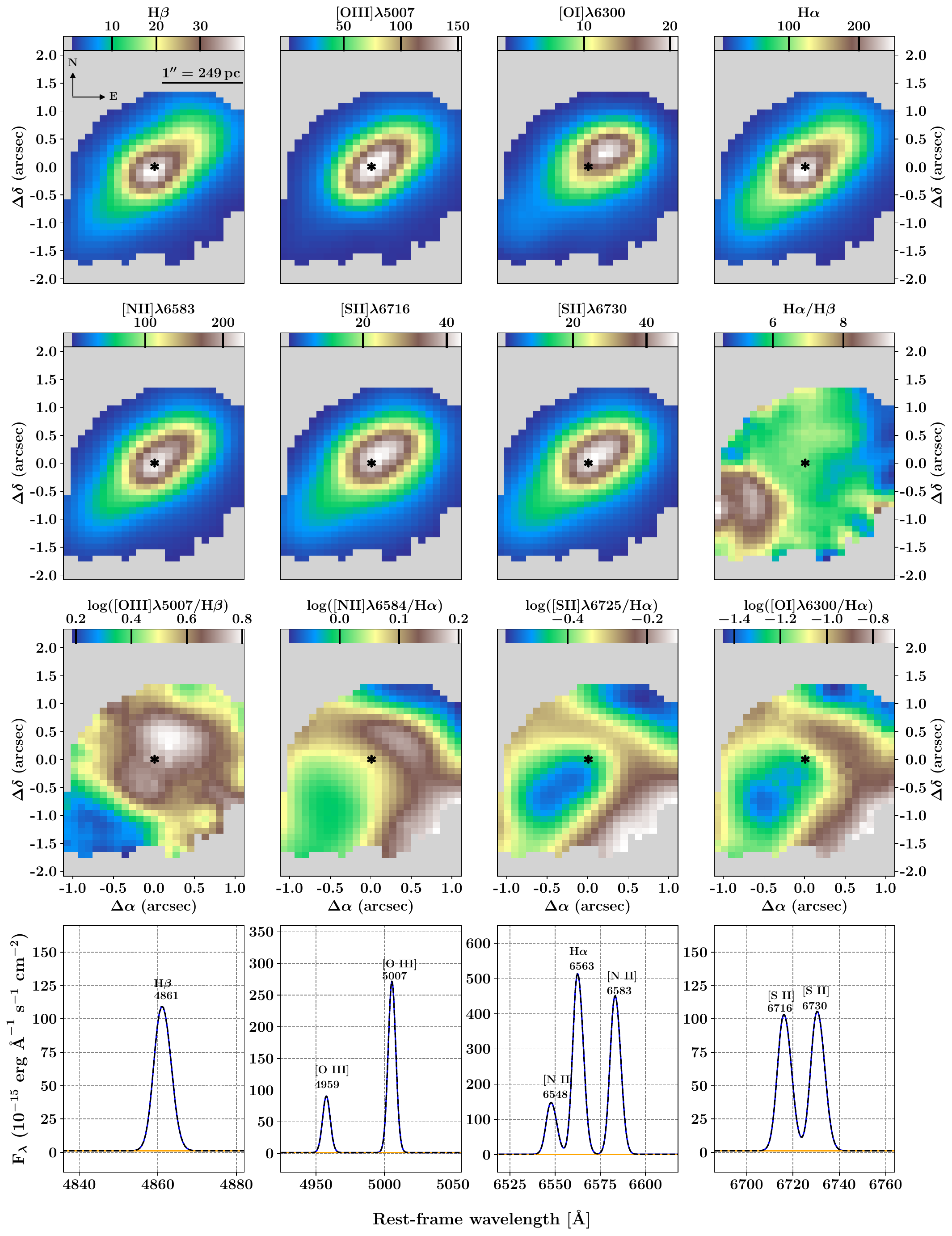}
\caption[]{Same as Fig.~\ref{ratios} but for MRK\,1066.}
\end{figure*}

\begin{figure*}
\includegraphics[width=2.1\columnwidth]{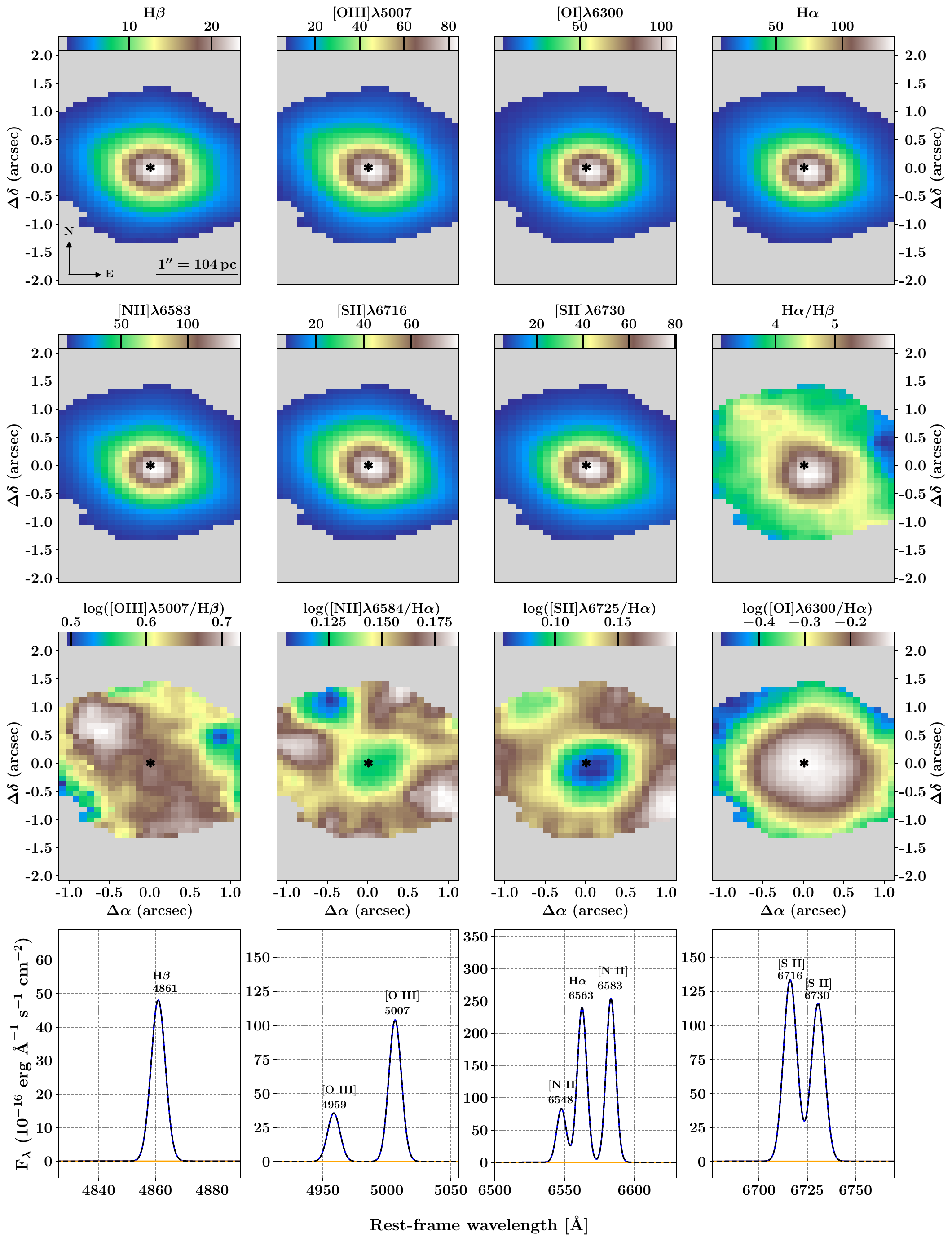}
\caption[]{Same as Fig.~\ref{ratios} but for NGC\,1052.}
\end{figure*}

\begin{figure*}
\includegraphics[width=2.1\columnwidth]{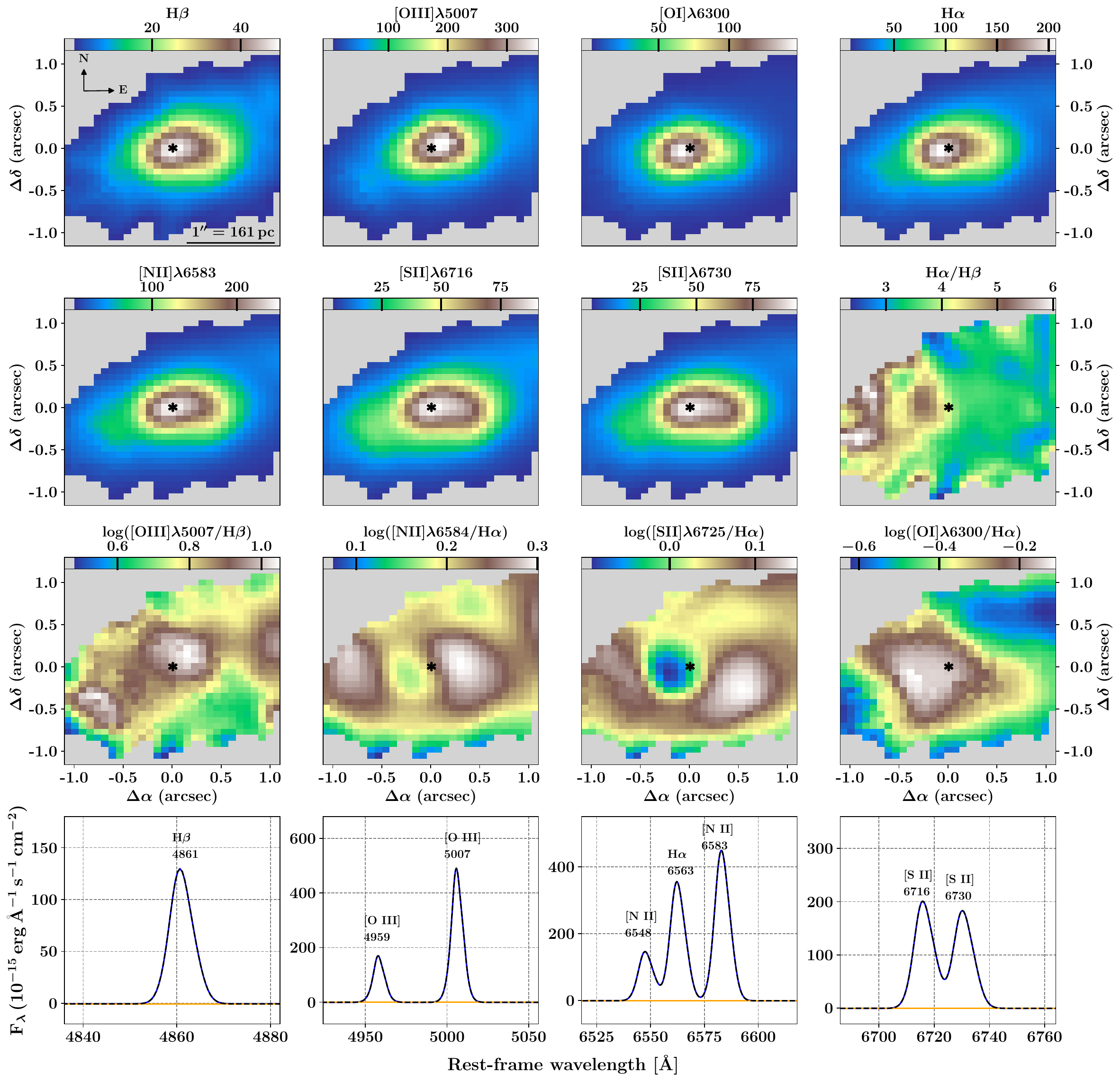}
\caption[]{Same as Fig.~\ref{ratios} but for NGC\,2110.}
\end{figure*}

\begin{figure*}
\includegraphics[width=2.1\columnwidth]{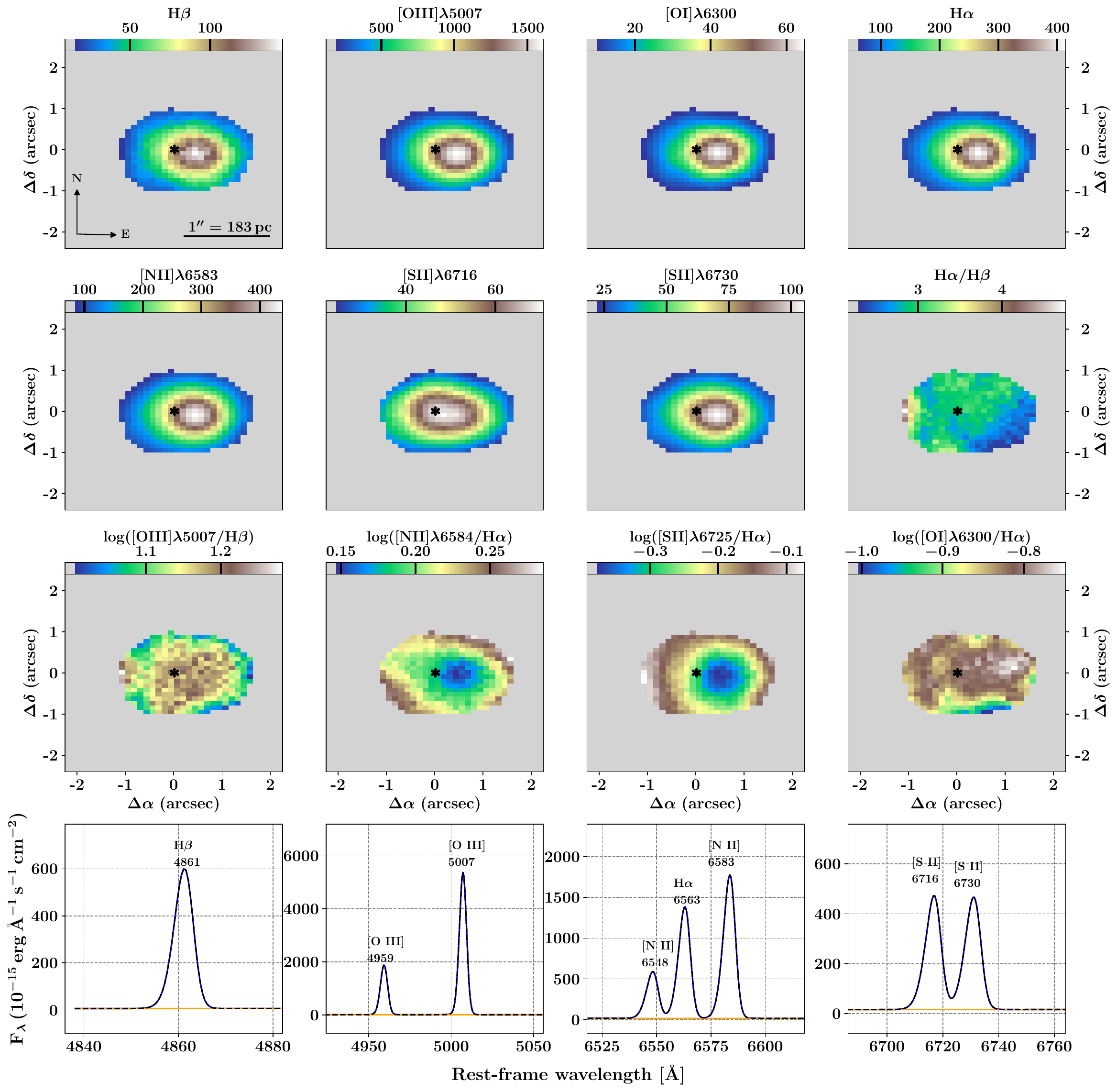}
\caption[]{Same as Fig.~\ref{ratios} but for NGC\,3516.}
\end{figure*}

\begin{figure*}
\includegraphics[width=2.1\columnwidth]{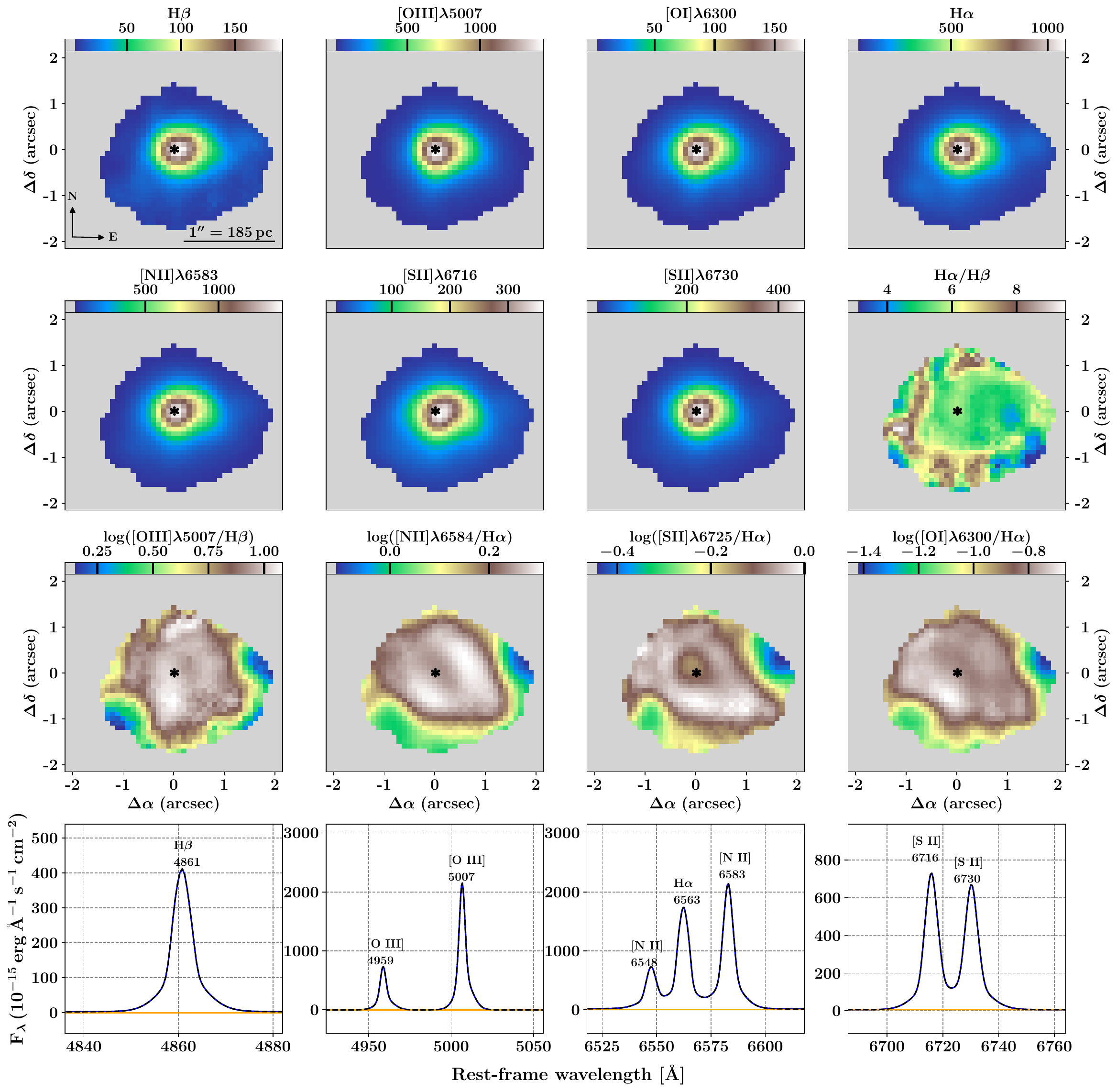}
\caption[]{Same as Fig.~\ref{ratios} but for NGC\,3786.}
\end{figure*}

\begin{figure*}
\includegraphics[width=2.1\columnwidth]{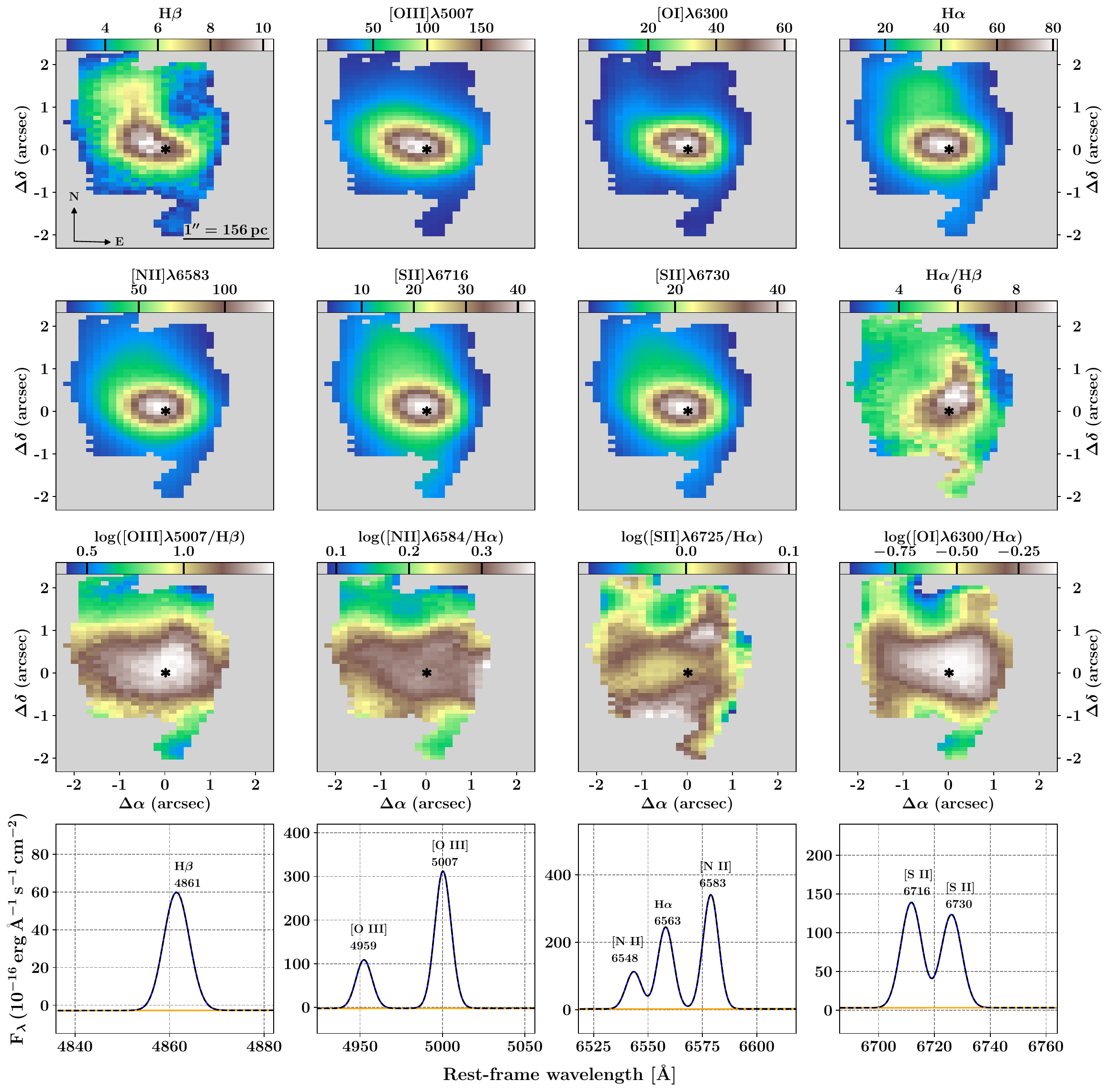}
\caption[]{Same as Fig.~\ref{ratios} but for NGC\,4235.}
\end{figure*}

\begin{figure*}
\includegraphics[width=2.1\columnwidth]{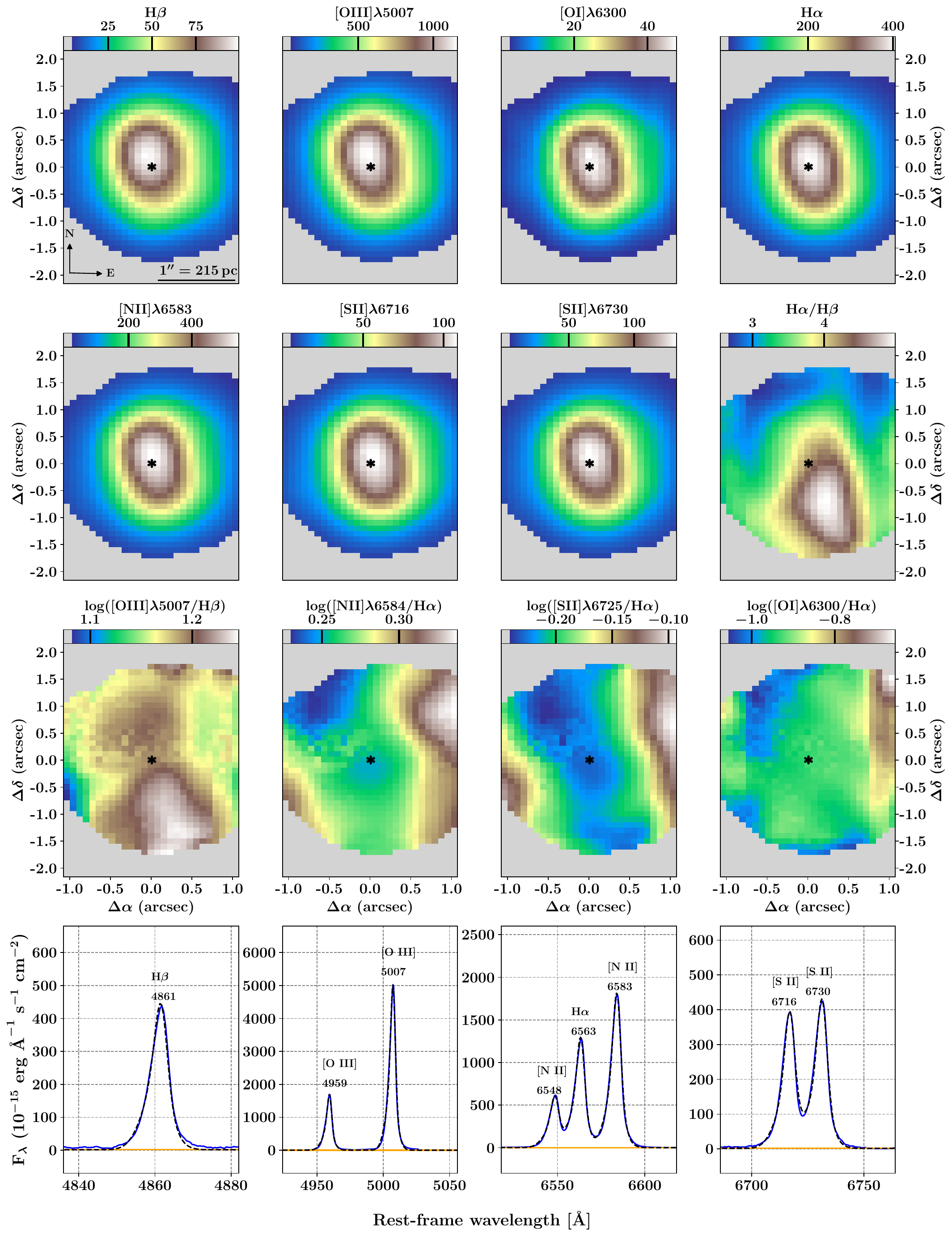}
\caption[]{Same as Fig.~\ref{ratios} but for NGC\,4939.}
\end{figure*}

\begin{figure*}
\includegraphics[width=2.1\columnwidth]{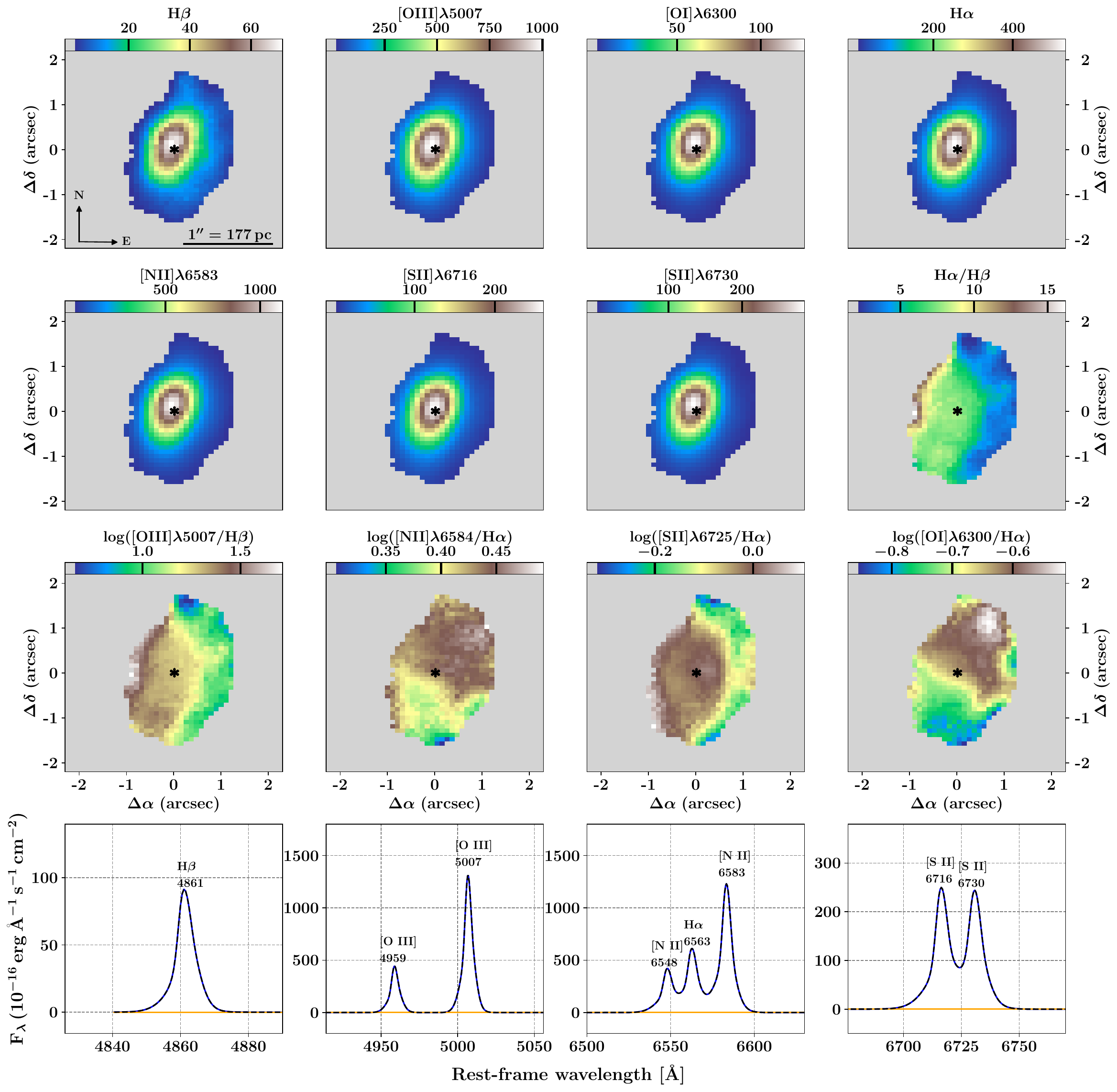}
\caption[]{Same as Fig.~\ref{ratios} but for NGC\,5899.}
\end{figure*}

\begin{figure*}
\includegraphics[width=2.1\columnwidth]{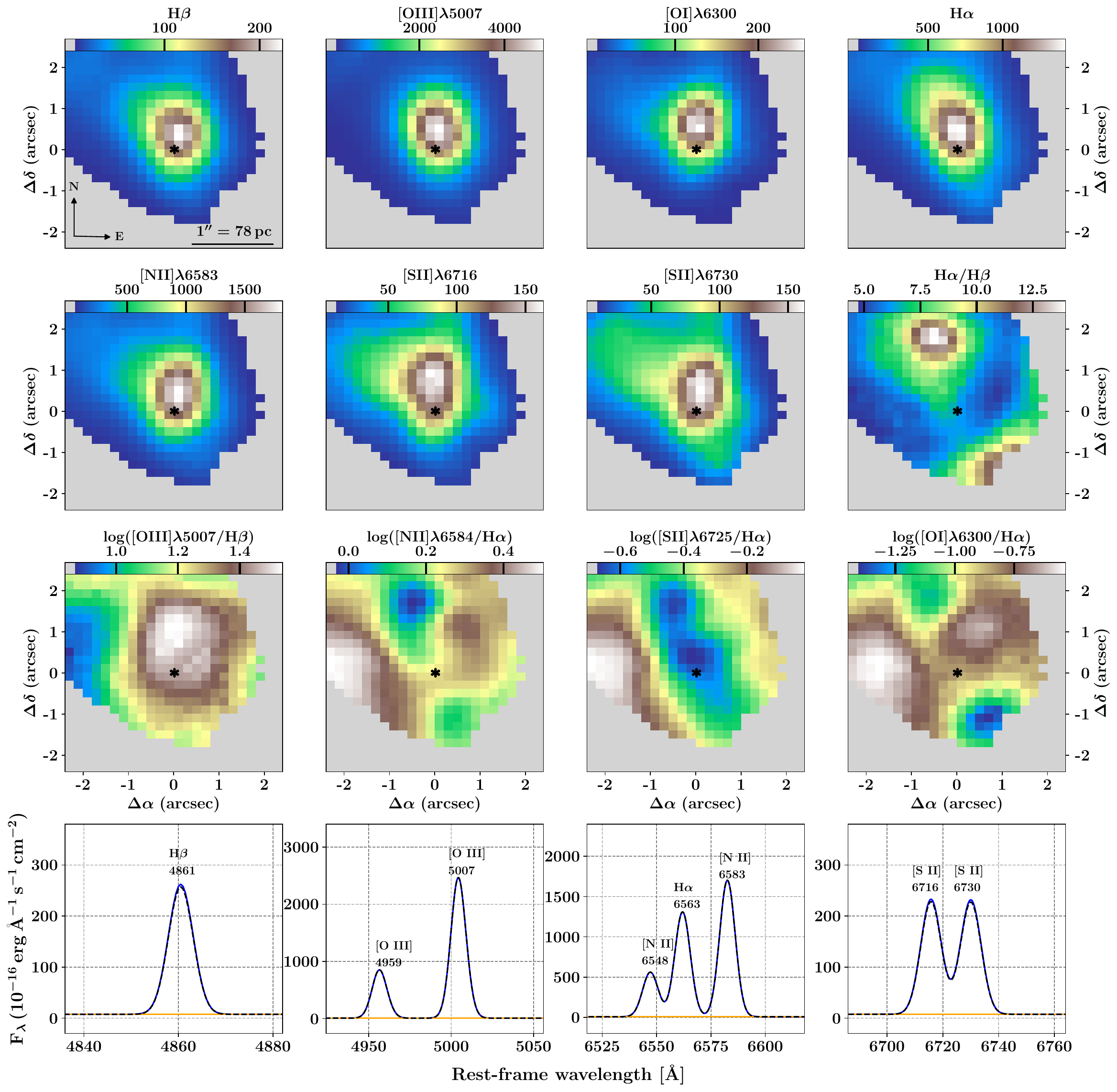}
\caption[]{Same as Fig.~\ref{ratios} but for NGC\,1068.}
\end{figure*}

\begin{figure*}
\includegraphics[width=2.1\columnwidth]{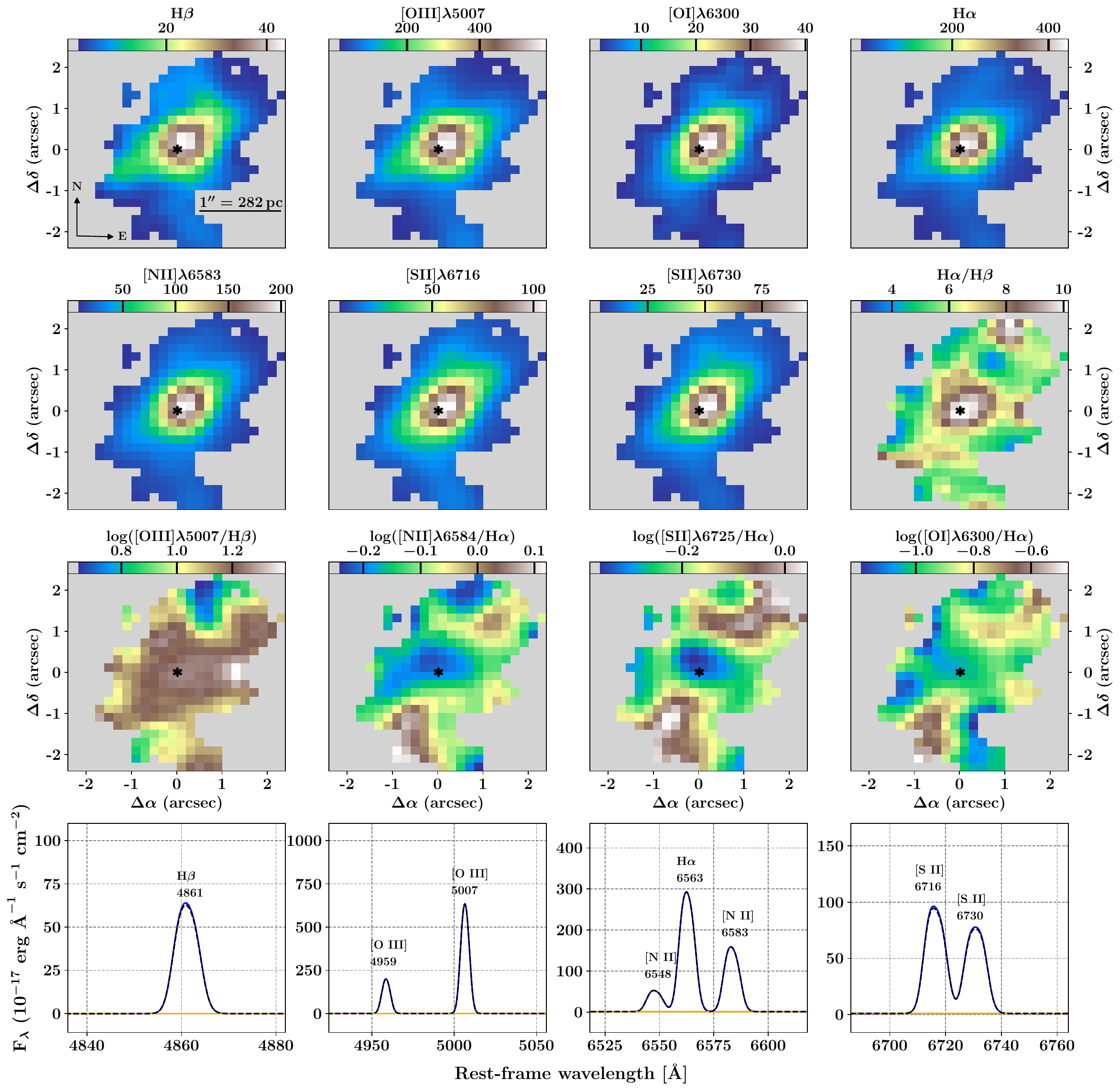}
\caption[]{Same as Fig.~\ref{ratios} but for NGC\,1194.}
\end{figure*}

\begin{figure*}
\includegraphics[width=2.1\columnwidth]{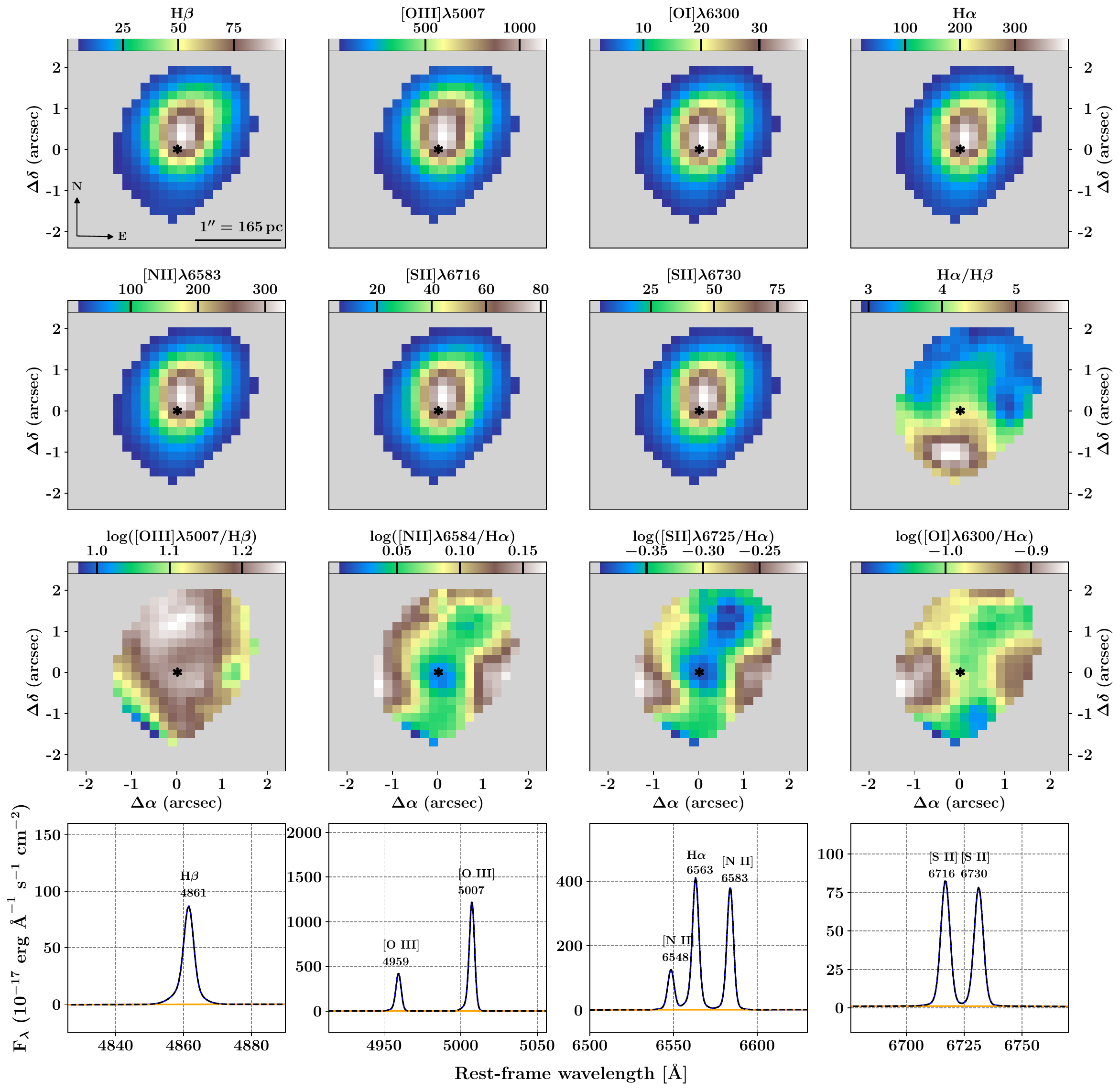}
\caption[]{Same as Fig.~\ref{ratios} but for NGC\,3081.}
\end{figure*}

\begin{figure*}
\includegraphics[width=2.1\columnwidth]{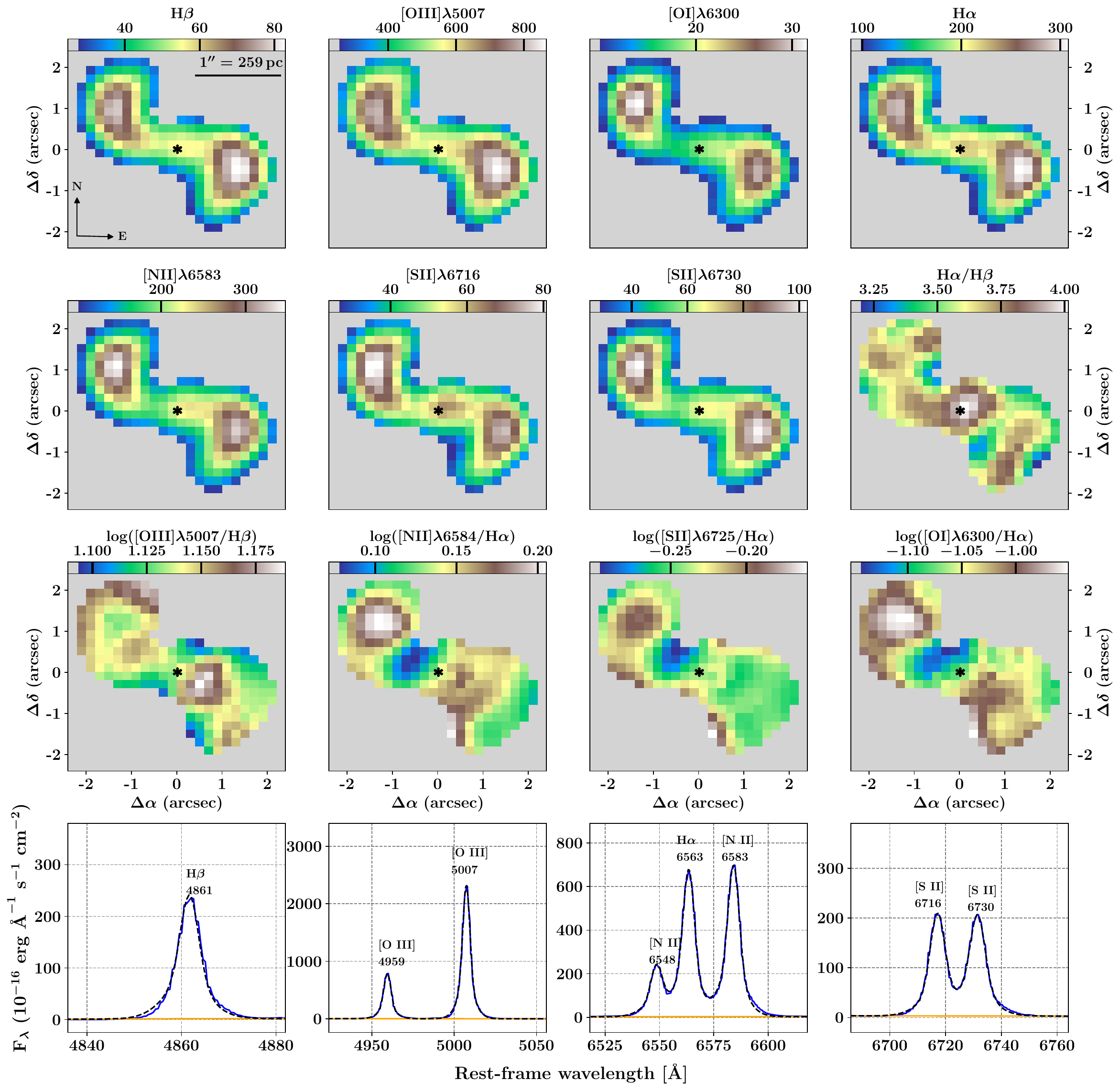}
\caption[]{Same as Fig.~\ref{ratios} but for NGC\,3393.}
\end{figure*}

\begin{figure*}
\includegraphics[width=2.1\columnwidth]{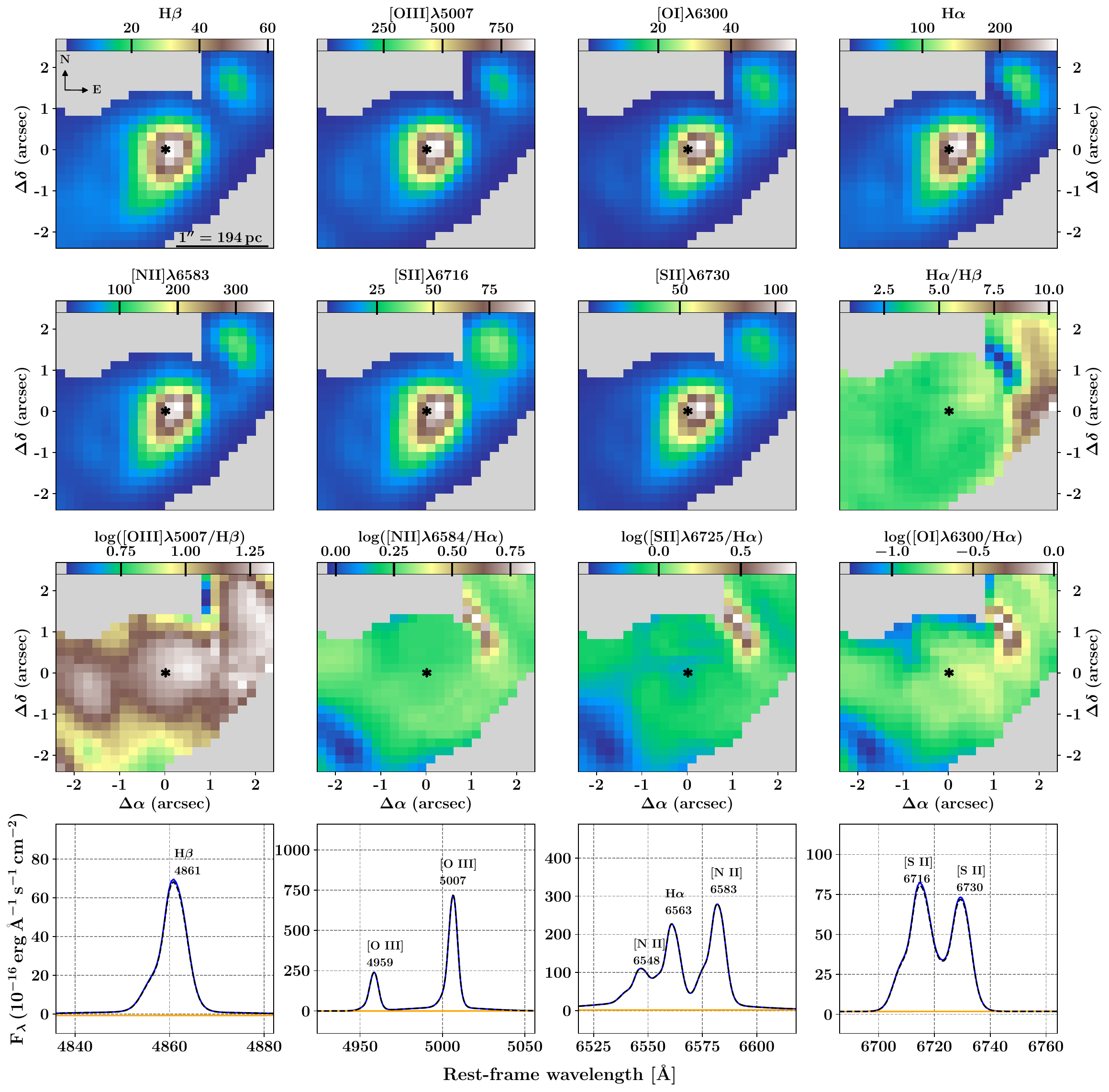}
\caption[]{Same as Fig.~\ref{ratios} but for NGC\,5728.}
\end{figure*}

%%%%%%%%%%%%%%%%%%%%%%%%%%%%%%%%%%%%%%%%%%%%%%%%%%

% Don't change these lines
\bsp	% typesetting comment
\label{lastpage}
\end{document}